\documentclass{jfm}
\usepackage{graphicx}
\usepackage{epstopdf, epsfig}
\usepackage{amsmath}
\usepackage{bm}
\usepackage{color}
\usepackage{hyperref}
\usepackage{url}

\hypersetup{
    colorlinks=true,
    linkcolor=blue,
    filecolor=blue,
    urlcolor=blue,
    citecolor=blue,
}

\shorttitle{{Wall cooling effect on spectra and structures}}
\shortauthor{D. Xu, J. Wang and S. Chen}

\title{{Wall cooling effect on spectra and structures of thermodynamic variables in hypersonic turbulent boundary layers}}

\author{Dehao Xu\aff{1},
	Jianchun Wang\aff{2},
\corresp{\email{wangjc@sustech.edu.cn}}
\and Shiyi Chen \aff{3,2,1}
\corresp{\email{chensy@sustech.edu.cn}}}
		
\affiliation{\aff{1}State Key Laboratory of Turbulence and Complex Systems, College of Engineering, Peking University, Beijing 100871, People$^{'}$s Republic of China
\aff{2}Department of Mechanics and Aerospace Engineering, Southern University of Science and Technology, Shenzhen 518055, People$^{'}$s Republic of China
\aff{3}Eastern Institute for Advanced Study, Ningbo 315200, People$^{'}$s Republic of China
}

\begin{document}

\maketitle

\begin{abstract}
    {The wall cooling effect on the spectra and structures of thermodynamic variables are investigated in hypersonic turbulent boundary layers.} The density and temperature can be divided into the acoustic and entropic modes based on the Kovasznay decomposition. {The intensities of the pressure and the acoustic modes of density and temperature attain the maximum values near the wall, while those of the entropy and the entropic modes of density and temperature achieve their primary peaks near the edge of boundary layer.} {In the near-wall region, the pressure and the acoustic modes of density and temperature are significantly enhanced when the wall is strongly cooled, which can be attributed to the appearance of the travelling-wave-like alternating positive and negative structures. Moreover, the intensities of the entropy and the entropic modes of density and temperature become stronger near the wall as the wall temperature decreases, due to the appearance of the streaky entropic structures.} The streaky entropic structures are mainly caused by the advection effect of the strong positive wall-normal gradient of the mean temperature associated with ejection and sweep events. {It is also found that the profiles of the intensities of the entropy, density and temperature are similar to each other far from the wall, which are mainly due to the reason that the entropic modes are dominant in the fluctuating density and temperature in the far-wall region. The acoustic modes of density and temperature only have significant contributions in the near-wall region.}
\end{abstract}

\section{Introduction}
The mechanisms of the hypersonic turbulent boundary layers are of great importance in aerospace industry due to the direct application to the hypersonic vehicles \citep[]{Smits2006,Gatski2009}. {It has been widely observed that the cold wall can significantly enhance the compressibility effect near the wall in hypersonic turbulent boundary layers} \citep[]{Duan2010,Zhang2017,Zhang2018,Xu2021a,Xu2021b,Xu2022a,Xu2022b,Xu2022c,Huang2022}. Therefore, the systematic investigations of the properties of the physical quantities in the cooled wall hypersonic turbulent boundary layers are extraordinarily critical to better understanding of the underlying mechanisms and more accurate physics-based modelling for this type of flows.

Most of the previous investigations about the hypersonic turbulent boundary layers were concentrated {on} the flow statistics of velocities \citep[]{Duan2010,Duan2011,Lagha2011,Chu2013,Zhang2018,Xu2021a,Xu2021b,Xu2022a,Xu2022b,Xu2022c,Huang2022}. \citet[]{Duan2010} performed direct numerical simulation (DNS) of hypersonic turbulent boundary layers at Mach number 5 with isothermal boundary condition. The ratio of the wall-to-edge temperature is ranging from 1.0 to 5.4. The effect of wall cooling on Morkovin$^{'}$s scaling, Walz$^{'}$s equation, the strong Reynolds analogy (SRA), turbulent kinetic energy budgets, compressibility {effect} and near-wall coherent structures were systematically investigated. They found that many scaling relations for the non-adiabatic hypersonic turbulent boundary layers are similar to those found in adiabatic wall cases, and the compressibility effect is insignificantly enhanced by wall cooling. Furthermore, \citet[]{Zhang2018} developed {DNS databases of} spatially evolving zero-pressure-gradient compressible turbulent boundary layers with nominal free-stream Mach number ranging from 2.5 to 14 and wall-to-recovery temperature ranging from 0.18 to 1.0. They assessed the performance of compressibility transformations, including the Morkovin$^{'}$s scaling and SRA, as well as the mean velocity and temperature scaling. Recently, a series of researches were aiming to reveal the effect of wall cooling on other complicated flow statistics beyond the well-observed compressibility transformations and the mean velocity and temperature scaling \citep[]{Xu2021a,Xu2021b,Xu2022a,Xu2022b,Xu2022c}. \citet[]{Xu2021a,Xu2021b} performed the DNS of hypersonic turbulent boundary layers at Mach numbers 6 and 8 with wall-to-recovery temperature ranging from 0.15 to 0.8. They used the Helmholtz decomposition to divide the fluctuating velocities into the solenoidal and dilatational components. They investigated the interactions among mean and fluctuating fields of kinetic and internal energy \citep[]{Xu2021a} as well as the kinetic energy transfer across different scales \citep[]{Xu2021b}. Furthermore, the flow topology and its effect on the kinetic energy transfer across different scales were also systematically investigated in \citet[]{Xu2022a,Xu2022b}. {In order to explain the possible reasons of the overshoot phenomena of the wall skin friction and wall heat transfer in transitional hypersonic boundary layers, \citet[]{Xu2022c} applied the decomposition method on the wall skin friction and heat transfer coefficients based on the two-fold repeated integration.} Moreover, the effect of the wall cooling on the wall skin friction and heat transfer decomposition in hypersonic turbulent boundary layers was also discussed.

However, most of the previous studies were {focused on} the flow statistics and structures of velocities, while the mechanisms of the thermodynamic statistics in hypersonic turbulent boundary layers were {less} studied \citep[]{Duan2016,Zhang2017,Ritos2019,Zhang2022,Cogo2022}. Recently, \citet[]{Zhang2022} investigated the wall cooling effect on pressure fluctuations in compressible turbulent boundary layers. They explored the generating mechanisms of pressure fluctuations by dividing the pressure fluctuations into five components, among which the rapid pressure, slow pressure and compressible pressure are dominant. {Furthermore, \citet[]{Cogo2022} investigated the high-Reynolds-number effect in hypersonic turbulent boundary layers. They studied the structural properties of the uniform streamwise momentum and uniform temperature regions in the high-speed regime. Furthermore, they also evaluated the accuracy of different compressibility transformations and temperature-velocity relations at moderate-high Reynolds numbers. A revised scaling for the characteristic length scales of the spanwise spectra of the fluctuating velocity and temperature at various wall distances was proposed based on the local mean shear. Nevertheless, the wall cooling effect on the multi-scale properties and the spatial structures of the pressure, density, temperature and entropy in hypersonic turbulent boundary layers need more systematic investigations, for the sake of a better understanding of the underlying mechanisms and more accurate physics-based modelling of the thermodynamic variables.}

    {The goal of this study is to systematically explore the wall cooling effect on the spectra and structures of the thermodynamic variables in hypersonic turbulent boundary layers by direct numerical simulation.} The fluctuating density and temperature are divided into the acoustic and entropic modes based on the Kovasznay decomposition \citep[]{Kovasznay1953,Chassaing2002,Gauthier2017,Wang2019}. The streamwise and spanwise spectra of the thermodynamic variables are systematically studied to figure out the multi-scale properties and spatial structures of thermodynamic variables. The streamwise and spanwise spectra of the fluctuating streamwise velocity are also investigated aiming for comparing with those of the thermodynamic variables. {It is found that the wall cooling effect on the spectra and structures of the thermodynamic variables are much larger than those of the fluctuating streamwise velocity.}

The remainder of the paper is organized as follows. The governing equations and simulation parameters are outlined in Section \ref{sec: n1}. The turbulent intensities of the streamwise velocity and thermodynamic variables are shown in Section \ref{sec: n2}. Section \ref{sec: n3} presented the streamwise and spanwise spectra of the streamwise velocity and thermodynamic variables. Some discussions are given in Section \ref{sec: n4}. Finally, summary and conclusions are given in Section \ref{sec: n5}.

\section{Governing equations and simulation parameters}\label{sec: n1}
The compressible Navier-Stokes equations can be non-dimensionalised by a set of reference scales: the reference length $L_{\infty }$, free-stream density $\rho  _{\infty }$, velocity $U_{\infty }$, temperature $T_{\infty}$, pressure $p_{\infty }= \rho  _{\infty }U_{\infty }^{2}$, energy per unit volume $\rho  _{\infty }U_{\infty }^{2}$, viscosity $\mu _{\infty }$ and thermal conductivity $\kappa _{\infty }$. Therefore, there are three non-dimensional governing parameters, namely the Reynolds number $ Re= \rho _{\infty }U_{\infty }L_{\infty }/\mu _{\infty }$, Mach number $ M= U_{\infty }/c_{\infty }$ and Prandtl number $ Pr= \mu _{\infty } C_{p}/\kappa _{\infty }$. The ratio of specific heat at constant pressure $C_{p}$ to that at constant volume $C_{v}$ is defined as $\gamma = C_{p}/C_{v}=1.4$. The parameter $ \alpha$ is defined as $ \alpha = PrRe\left ( \gamma -1 \right )M^{2}$, where $Pr=0.7$.

The following compressible dimensionless Navier-Stokes equations in the conservative form are solved numerically \citep[]{Liang2015,Xu2021a,Xu2021b,Xu2022a,Xu2022b,Xu2022c}
\begin{equation}
    \frac{\partial \rho }{\partial t}+\frac{\partial \left ( \rho u_{j} \right )}{\partial x_{j}}=0,
\end{equation}
\begin{equation}
    \frac{\partial \left ( \rho u_{i} \right )}{\partial t}+\frac{\partial \left [ \rho u_{i}u_{j}+p\delta _{ij} \right ]}{\partial x_{j}}=\frac{1}{Re}\frac{\partial \sigma _{ij}}{\partial x_{j}},
\end{equation}
\begin{equation}
    \frac{\partial E}{\partial t}+\frac{\partial \left [ \left ( E+p \right )u_{j} \right ]}{\partial x_{j}}=\frac{1}{\alpha }\frac{\partial }{\partial x_{j}}\left ( \kappa \frac{\partial T}{\partial x_{j}} \right )+\frac{1}{Re}\frac{\partial \left ( \sigma _{ij}u_{i} \right )}{\partial x_{j}},
\end{equation}
\begin{equation}
    p=\rho T/\left ( \gamma M^{2} \right ),
\end{equation}
where $\rho $, $u_{i}$, $T$ and $p$ are the density, velocity component, temperature and pressure, respectively. The viscous stress $\sigma _{ij}$ is defined as
\begin{equation}
    \sigma _{ij}=\mu \left ( \frac{\partial u_{i}}{\partial x_{j}}+\frac{\partial u_{j}}{\partial x_{i}} \right )-\frac{2}{3}\mu \theta \delta _{ij},
\end{equation}
where $\theta = \partial u_{k}/\partial x_{k}$ is the velocity divergence, {and the viscosity $\mu $ is determined by the Sutherland$^{'}$s law}. The total energy per unit volume $E$ is
\begin{equation}
    E=\frac{p}{\gamma -1}+\frac{1}{2}\rho \left ( u_{j}u_{j} \right ).
\end{equation}

\begin{figure}\centering
    \includegraphics[width=0.99\linewidth]{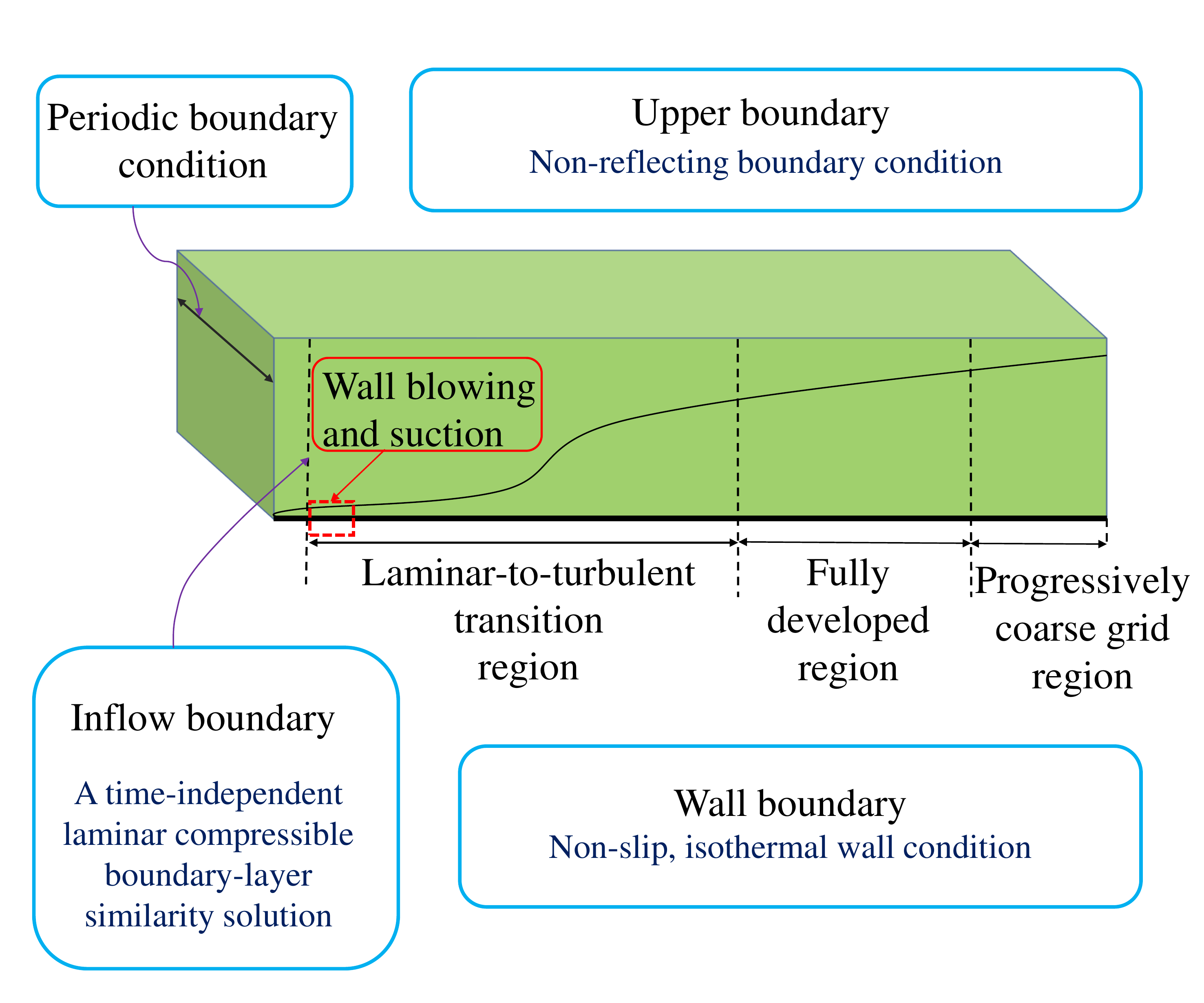}
    \caption{The schematic of the hypersonic transitional and turbulent boundary layers.}
    \label{fig: d1}
\end{figure}

{The convection terms of the compressible governing equations are discretized by a hybrid scheme. In order to judge on the local smoothness of the numerical solution, the modified Jameson sensor \citep[]{Jameson1981} is used in the hybrid scheme, which can be given by \citep[]{Dang2022}}
\begin{equation}
    \begin{split}
        &{\phi _{i}=\frac{\left | -p_{i-1}+2p_{i}-p_{i+1} \right |}{p_{i-1}+2p_{i}+p_{i+1}},} \\
        &{\phi _{j}=\frac{\left | -p_{j-1}+2p_{j}-p_{j+1} \right |}{p_{j-1}+2p_{j}+p_{j+1}}, }\\
        &{\phi _{k}=\frac{\left | -p_{k-1}+2p_{k}-p_{k+1} \right |}{p_{k-1}+2p_{k}+p_{k+1}},}
    \end{split}
\end{equation}
\begin{equation}
    {\Theta =\phi _{i}+\phi _{j}+\phi _{k}.}
\end{equation}
{The threshold $\Theta _{1}$ is set to 0.02 \citep[]{Dang2022}. When $\Theta \leq \Theta _{1}$, the eighth-order central difference scheme is used; when $\Theta > \Theta _{1}$, the seventh-order weighted essentially non-oscillatory scheme \citep[]{Balsara2000} is applied.} Furthermore, the viscous terms are approximated by an eighth-order central difference scheme. A third-order total variation diminishing type of Runge-Kutta method is utilized for time advancing \citep{shu1988}. The compressible governing equations are numerically solved by the OPENCFD code, which has been widely validated in compressible transitional and turbulent wall-bounded flows \citep[]{Liang2015,Xu2021a,Xu2021b,Xu2022a,Xu2022b,Xu2022c,Dang2022}. The schematic of the hypersonic transitional and turbulent boundary layers is shown in figure \ref{fig: d1}. The spatially evolving hypersonic transitional and turbulent boundary layer is numerically simulated under the inflow and outflow boundary conditions, a wall boundary condition, an upper far-field boundary condition, and a periodic boundary condition in the spanwise direction. A time-independent laminar compressible boundary-layer similarity solution is applied at the inflow boundary. The laminar flow is disturbed by the wall blowing and suction region, and then transitioned to the fully developed turbulent state. Moreover, in order to inhibit the reflection of disturbance due to the numerical treatment of the outflow boundary condition, a progressively coarse grid is implemented in the streamwise direction near the outflow boundary condition. The non-slip and isothermal boundary conditions are applied for the wall boundary, and the non-reflecting boundary condition is imposed for the upper boundary. More detailed descriptions can refer to \citet[]{Pirozzoli2004,Liang2015,Xu2021a,Xu2021b,Xu2022a,Xu2022b,Xu2022c}.

In this study, $\overline{f}$ denotes the Reynolds average (spanwise and time average) of flow field $f$, and the fluctuating component of the Reynolds average is ${f}'=f-\overline{f}$. Furthermore, $\widetilde{f}=\overline{\rho f}/\bar{\rho }$ represents the Favre average of $f$, and the fluctuating component is ${f}''=f-\widetilde{f}$.

The DNS of three hypersonic transitional and turbulent boundary layers at Mach number 8 with different wall temperatures are performed and the fundamental parameters of the database are listed in table \ref{tab: tab1}. {The free-stream temperature $T_{\infty }$ is prescribed to be $T_{\infty }=169.44K$.} Temperature $T_{w}$ is the wall temperature, and the recovery temperature $T_{r}$ can be defined as $T_{r}=T_{\infty }\left ( 1+r\left ( \left ( \gamma -1 \right )/2 \right )M_{\infty }^{2} \right )$ with recovery factor $r=0.9$ \citep[]{Duan2010,Xu2022c}. The coordinates along the streamwise, wall-normal and spanwise directions are represented by $x$, $y$ and $z$ respectively. The computational domains $L_{x}$, $L_{y}$ and $L_{z}$ are nondimensionalized by the inflow boundary layer thickness $\delta _{in}$, and the symbols $N_{x}$, $N_{y}$ and $N_{z}$ represent the grid resolutions along the streamwise, wall-normal and spanwise directions respectively.

    {It should be pointed out that the no-ideal gas effect, as well as the non-equilibrium and radiative effects are neglected in the present DNS databases. The reasons are as follows. It is noted that the free-stream temperature $T_{\infty }$ is prescribed to be $T_{\infty }=169.44K$ in the present study. The largest temperature in the highest wall temperature case M8T08 is approximately 1700K. In the previous studies, the maximum temperatures of cases ``M7'', ``M8'' and ``M12'' in \citet{Duan2011} and $M_{\infty }=7.5$, 10, 15 and 20 in \citet{Lagha2011} are much larger than that of M8T08, and the no-ideal gas effect, as well as the non-equilibrium and radiative effects were also neglected in their study. Therefore, the neglect of the no-ideal gas, non-equilibrium and heat radiative effects is a reasonable simplification for the present study of the hypersonic turbulent boundary layers. The influence of the no-ideal gas, non-equilibrium and radiative effects in the hypersonic turbulent boundary layers will be considered in the future.}

\begin{table}
    \begin{center}
        \def~{\hphantom{0}}
        \begin{tabular}{ccccccc}
            Case   & $M$ & $Re$               & $T_{w}/T_{\infty }$ & $T_{w}/T_{r}$ & $L_{x}/\delta _{in}\times L_{y}/\delta _{in}\times L_{z}/\delta _{in}$ & $N_{x}\times N_{y}\times N_{z}$ \\ [3pt]
            M8T015 & 8   & $5.2\times 10^{4}$ & 1.9                 & 0.15          & $1000 \times 95 \times 42$                                             & $6000\times500\times600$        \\
            M8T04  & 8   & $2.6\times 10^{5}$ & 5.0                 & 0.4           & $1500\times 97 \times 47$                                              & $7000\times400\times500$        \\
            M8T08  & 8   & $8.0\times 10^{5}$ & 10.03               & 0.8           & $1700\times 103 \times 51$                                             & $7000\times400\times400$        \\
        \end{tabular}
        \caption{Summary of computational parameters for the three DNS database at Mach number 8 with different wall temperatures.}
        \label{tab: tab1}
    \end{center}
\end{table}

{Three} sets of data in a small streamwise window of $ \left [ x_{a}-0.5\delta ,x_{a}+0.5\delta  \right ]$ extracted from the fully developed region of the above three transitional and hypersonic turbulent boundary layers are used for following statistical analysis, where $x_{a}$ is the reference streamwise location selected for statistical analysis, and $\delta $ is the boundary layer thickness {at the streamwise location $x_{a}$}. It is noted that a similar technique has been used by the previous studies of \citet[]{Pirozzoli2011}, \citet[]{Zhang2018} and \citet[]{Huang2022}, and the width of the streamwise window in this study is consistent with that of \citet[]{Huang2022}. The fundamental parameters of the {three} sets of data are listed in table \ref{tab: tab2}. The friction Reynolds number $Re_{\tau}$ is defined as $Re_{\tau}=\bar{\rho}_{w}u_{\tau}\delta/\bar{\mu}_{w}$, where $\bar{\rho}_{w}$ and $\bar{\mu}_{w}$ are the mean wall density and wall viscosity respectively, and $u_{\tau }=\sqrt{\tau _{w}/\bar{\rho} _{w}}$ and $\tau _{w}=\left (\mu \partial \bar{u}/\partial y  \right ) _{y=0}$ are the friction velocity and the wall shear stress respectively. {Furthermore, $\Delta x^{+}=\Delta x /\delta _{\nu}$, $\Delta y_{w}^{+}=\Delta y_{w} /\delta _{\nu}$, $\Delta y_{e}^{+}=\Delta y_{e} /\delta _{\nu}$ and $\Delta z^{+}=\Delta z /\delta _{\nu}$ are the normalized spacing of the streamwise direction, the first point off the wall, the wall-normal grid at the edge of the boundary layer and the spanwise direction respectively}, where $\delta _{\nu}=\bar{\mu }_{w}/(\bar{\rho} _{w}u_{\tau })$ is the viscous length scale. {The semi-local lengthscale is defined as $\delta _{\nu}^{*}=\bar{\mu }/\left ( \bar{\rho }u_{\tau }^{*} \right )$, where $u_{\tau }^{*}=\sqrt{\tau _{w}/\bar{\rho }}$ \citep[]{Huang1995}. The semi-local Reynolds number can be defined as $Re_{\tau }^{*}=\delta /\left (\delta _{\nu}^{*}  \right )_{e}$. It is noted that the grid resolutions $\Delta x^{+}$, $\Delta y^{+}_{w}$, $\Delta y^{+}_{e}$ and $\Delta z^{+}$ in three cases are comparable and even smaller than many previous investigations including \citet{Duan2010}, \citet{Pirozzoli2013}, \citet{Zhang2018} and \citet{Huang2022}, indicating that the grid resolutions of the present DNS study are fine enough. Furthermore, the accuracy of the DNS cases in this study are validated via comparisons with the available DNS database in \citet[]{Zhang2018} in Appendix A.}

\begin{table}
    \begin{center}
        \def~{\hphantom{0}}
        \begin{tabular}{ccccccccc}
            Case   & $x_{a}/\delta _{in}$ & $\Delta x^{+}$ & $\Delta y^{+}_{w}$ & $\Delta y^{+}_{e}$ & $\Delta z^{+}$ & $Re_{\tau}$ & $Re_{\tau }^{*}$ & $\delta /\delta _{in}$ \\ [3pt]
            M8T015 & 870                  & 8.3            & 0.50               & 6.4                & 4.1            & 920         & 1715             & 15.4                   \\
            M8T04  & 1100                 & 8.0            & 0.51               & 5.9                & 4.4            & 700         & 3885             & 14.3                   \\
            M8T08  & 1580                 & 8.4            & 0.50               & 6.1                & 5.2            & 700         & 7888             & 16.7                   \\
            \setlength{\tabcolsep}{12mm}
        \end{tabular}
        \caption{The fundamental parameters of the three sets of data.}
        \label{tab: tab2}
    \end{center}
\end{table}

\begin{figure}\centering
    \includegraphics[width=0.99\linewidth]{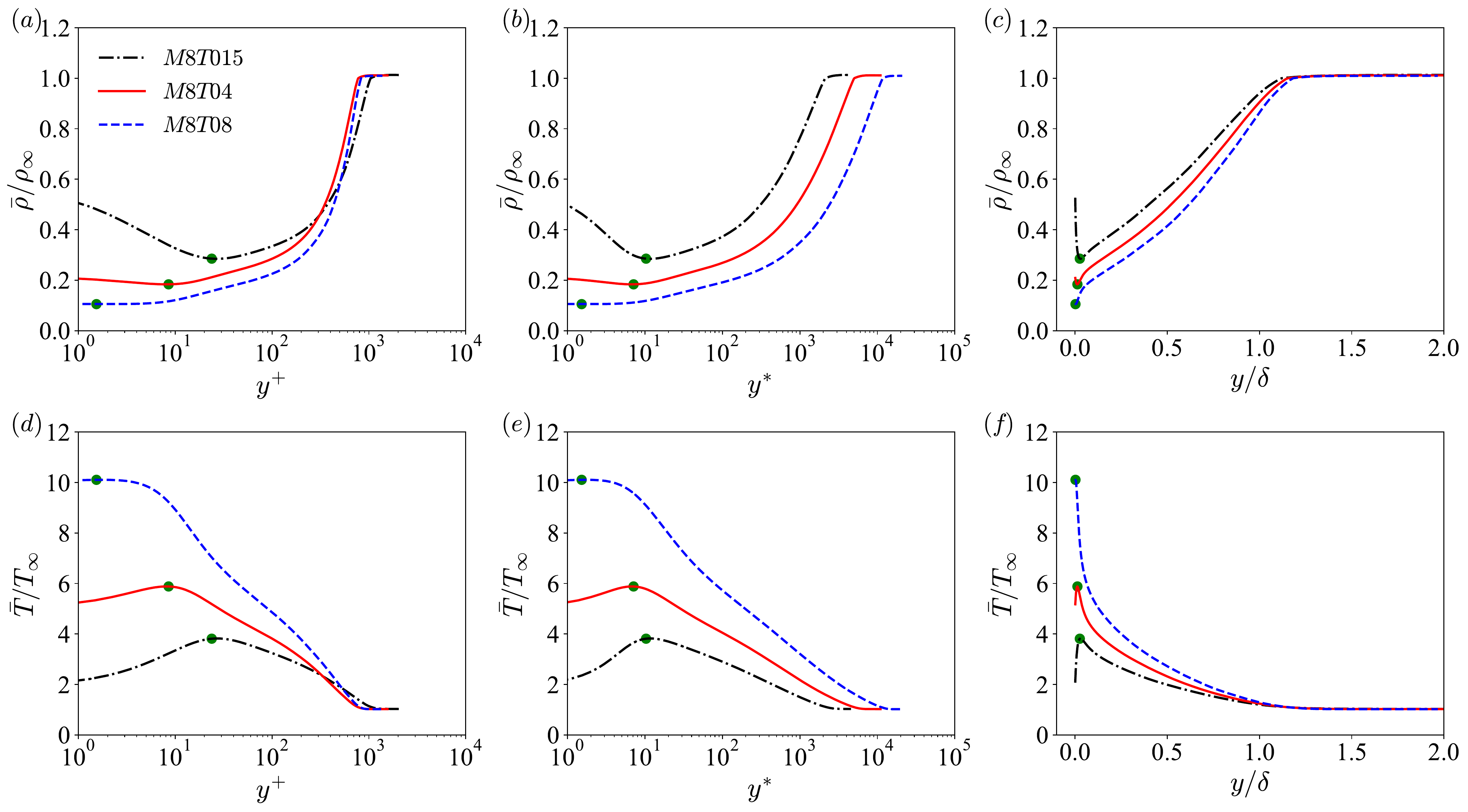}
    \caption{(a) (b) and (c): The mean density profile along wall-normal direction plotted against (a) wall unit scaling ($y^{+}$), (b) semi-local scaling ($y^{*}$) and (c) outer scaling ($y/\delta$). (d) (e) and (f): The mean temperature profile along wall-normal direction plotted against (d) wall unit scaling ($y^{+}$), (e) semi-local scaling ($y^{*}$) and (f) outer scaling ($y/\delta$). Here the ``turning points'' marked by the green circles represent the minimum values of the mean density in (a), (b) and (c), and the maximum values of the mean temperature in (d), (e) and (f).}
    \label{fig: d2}
\end{figure}

The mean density and temperature profiles along wall-normal direction are shown in figure \ref{fig: d2}. Here the ``turning points'' marked by the green circles represent the minimum values of the mean density in figure \ref{fig: d2} (a), (b) and (c), and the maximum values of the mean temperature in figure \ref{fig: d2} (d), (e) and (f). The wall unit scaling is defined as $y^{+}=y/\delta _{\nu}$, and the semi-local scaling is defined as $y^{*}=y/\delta _{\nu}^{*}$. It is noted that the wall unit scaling $y^{+}$ and the semi-local scaling $y^{*}$ are the inner scaling in order to reveal the scaling relation in the near-wall region, while the outer scaling $y/\delta$ shows the statistical behaviour in the far-wall region.

The mean density and temperature are significantly influenced by the wall temperature. When the wall temperature is close to the recovery temperature $T_{r}$ (``M8T08'' case), the mean temperature is nearly constant near the wall, and then decreases drastically when $y^{+}>4$. {However, in ``M8T04'' and ``M8T015'', the mean temperature initially increases near the wall. After reaching the maximum value at the turning point, the mean temperature then decreases as $y$ increases. It is also found that as the wall temperature decreases, the positive values of the wall-normal gradient of the mean temperature become larger below the wall-normal location of the turning point, while the negative values of the wall-normal gradient of the mean temperature become smaller above it. Furthermore, when the wall temperature becomes cooler, the maximum value of the mean temperature decreases, while the wall-normal location $y^{+}$ of the turning point increases under the wall unit scaling.} {The semi-local scaling $y^{*}$ can significantly decrease the discrepancy of the wall-normal locations of turning points in different wall temperature cases. It should be noted that the mean density profiles reveal the opposite variation trends compared with the mean temperature profiles.}

\begin{figure}\centering
    \includegraphics[width=0.99\linewidth]{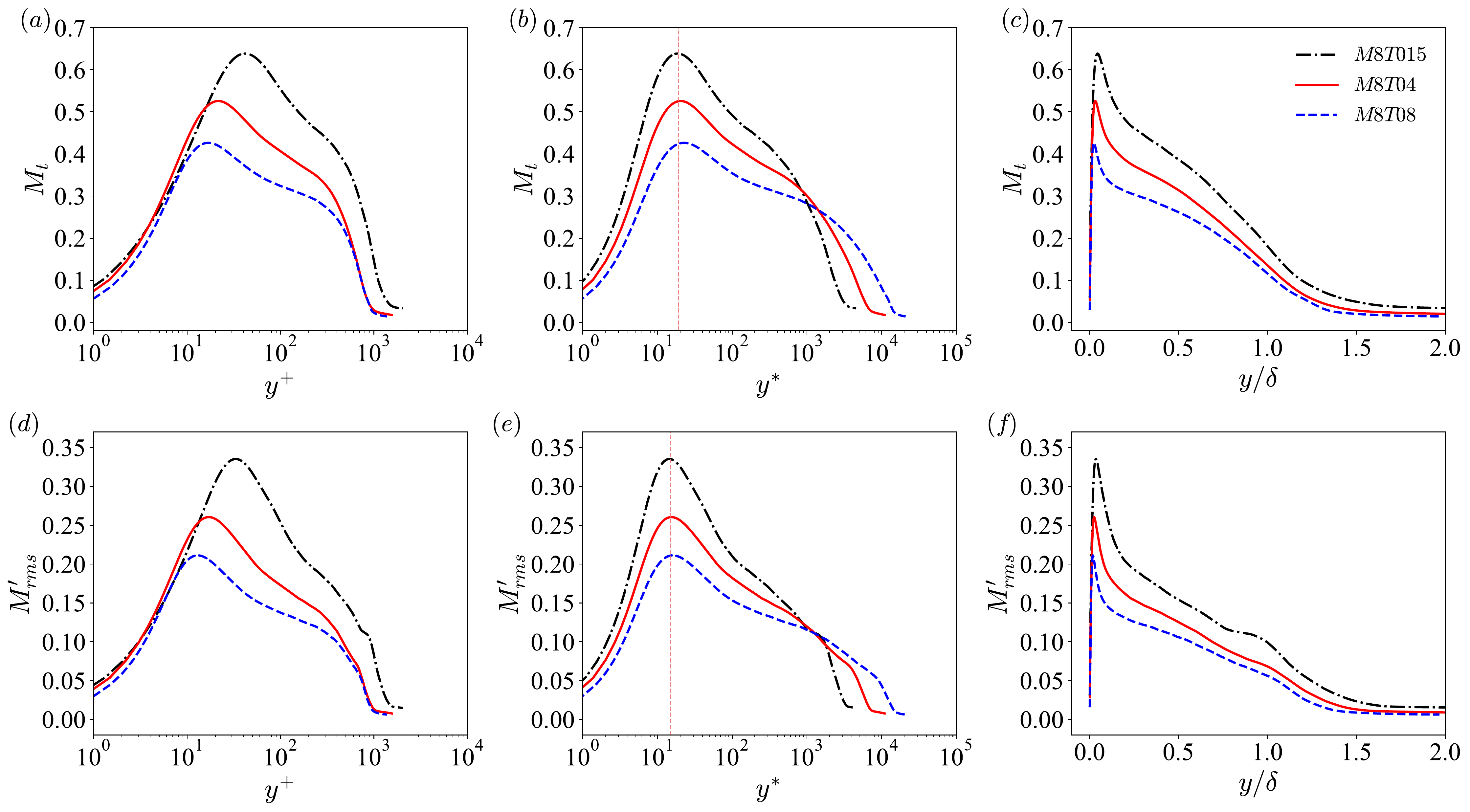}
    \caption{(a) (b) and (c): The turbulent Mach number $M_{t}$ along wall-normal direction plotted against (a) wall unit scaling ($y^{+}$), (b) semi-local scaling ($y^{*}$) and (c) outer scaling ($y/\delta$). The vertical dashed line represents $y^{*}=19$ in (b). (d) (e) and (f): The r.m.s values of the local Mach number ${M}'_{rms}$ along wall-normal direction plotted against (d) wall unit scaling ($y^{+}$), (e) semi-local scaling ($y^{*}$) and (f) outer scaling ($y/\delta$). The vertical dashed line represents $y^{*}=15$ in (e).}
    \label{fig: d3}
\end{figure}

In order to investigate the effect of wall temperature on the compressibility effect, the turbulent Mach number $M_{t}=\sqrt{\overline{{u}''_{i}{u}''_{i}}}/\bar{c}$ and the root mean square (r.m.s) values of the local Mach number based on the fluctuating velocities ${M}'_{rms}$ are evaluated in figure \ref{fig: d3}. Here the local Mach number based on the fluctuating velocities $M'$ is defined as ${M}'=\sqrt{{u}'_{i}{u}'_{i}}/c$, and $c$ represents the local sound speed. As the wall temperature decreases, the peak values of $M_{t}$ and ${M}'_{rms}$ increase, indicating that the cooling wall can enhance the compressibility effect, which is consistent with many previous studies including \citet[]{Zhang2018}, \citet[]{Xu2021b} and \citet[]{Zhang2022}. Moreover, {it is shown in figure \ref{fig: d3} (a) and (d) that the wall-normal locations $y^{+}$ of the peak values of $M_{t}$ and ${M}'_{rms}$ increase under the wall unit scaling as the wall temperature decreases, while it is found in figure \ref{fig: d3} (b) and (e) that the $M_{t}$ and ${M}'_{rms}$ profiles against the semi-local scaling attain their peaks at almost the same values of $y^{*}$ in different wall temperature cases.}

\section{The turbulent intensities of the streamwise velocity and the thermodynamic variables}\label{sec: n2}
\begin{figure}\centering
    \includegraphics[width=0.99\linewidth]{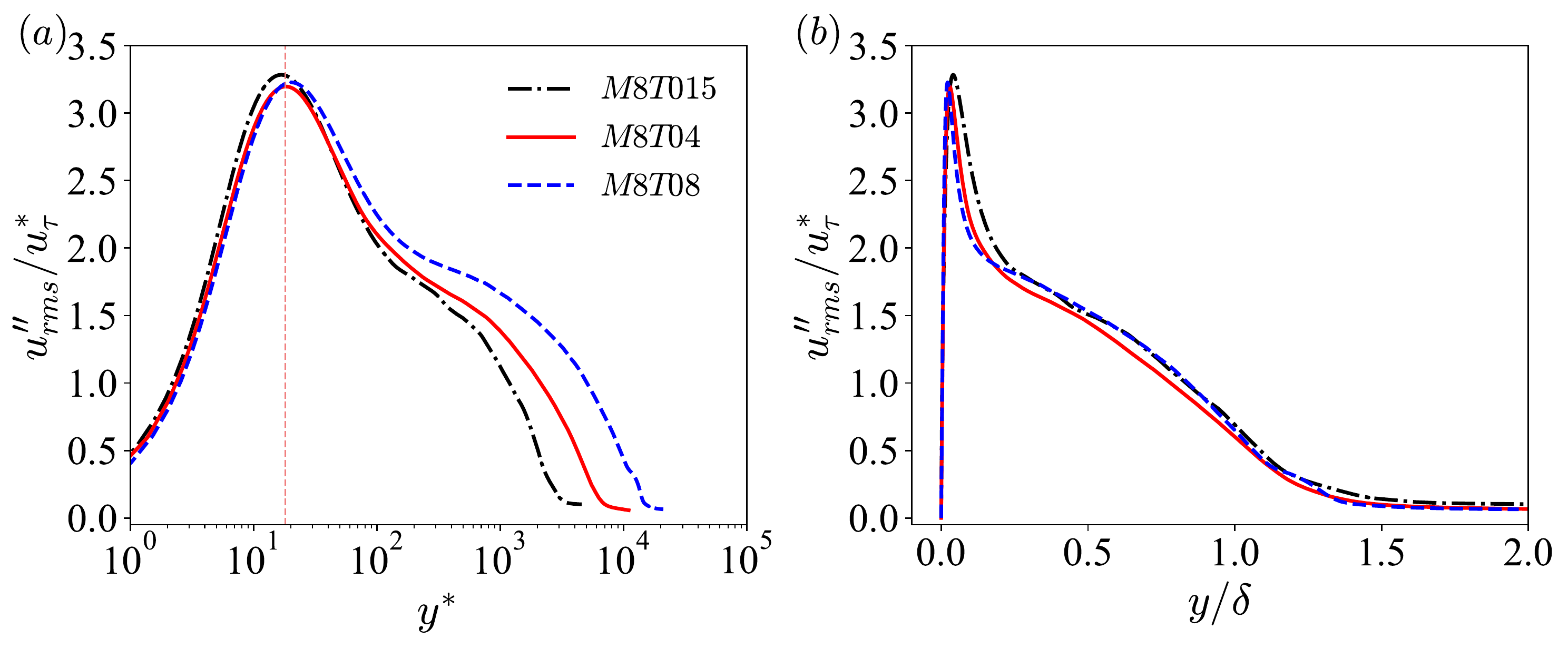}
    \caption{The normalised turbulent intensity of the streamwise velocity $u _{rms}^{\prime \prime}/u_{\tau}^{*}$ plotted against (a) semi-local scaling ($y^{*}$) and (b) outer scaling ($y/\delta$). The vertical dashed line represents $y^{*}=18$ in (a).}
    \label{fig: d4}
\end{figure}

The normalised turbulent intensity of the streamwise velocity $u _{rms}^{\prime \prime}/u_{\tau}^{*}$ along wall-normal direction is shown in figure \ref{fig: d4}. {It is found that $u _{rms}^{\prime  \prime}/u_{\tau}^{*}$ attains its peak in the buffer layer (approximately $y^{*}\approx 18$). Furthermore, the peak values of $u _{rms}^{\prime  \prime}/u_{\tau}^{*}$ are similar in ``M8T04'' and ``M8T08'' cases. However, the peak value is slightly larger in ``M8T015'', which can be ascribed to the strongly colder wall temperature and slightly larger friction Reynolds number $Re_{\tau}$.}

\begin{figure}\centering
    \includegraphics[width=0.99\linewidth]{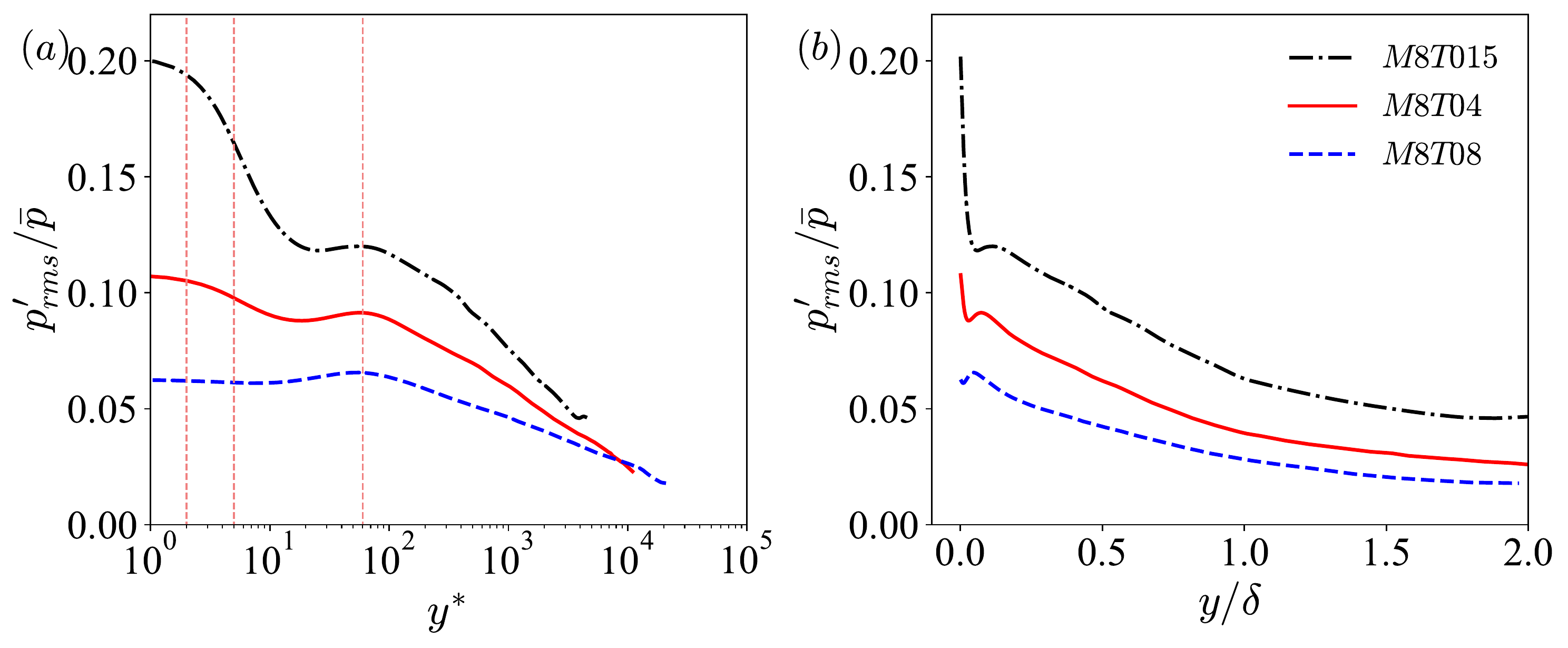}
    \caption{The normalised turbulent intensity of the pressure $p _{rms}^{\prime}/\bar{p}$ plotted against (a) semi-local scaling ($y^{*}$) and (b) outer scaling ($y/\delta$). The vertical dashed lines represent $y^{*}=2, 5, 60$ in (a).}
    \label{fig: d5}
\end{figure}

{The normalised turbulent intensity of the pressure $p _{rms}^{\prime}/\bar{p}$ along wall-normal direction is plotted in figure \ref{fig: d5}. It is found that the $p _{rms}^{\prime}/\bar{p}$ profile in nearly adiabatic wall case (``M8T08'' case) reaches the maximum value at $y^{*} \approx 60$, and then decreases rapidly among the boundary layer to a flat platform. As the wall temperature decreases, the intensity of pressure $p _{rms}^{\prime}/\bar{p}$ significantly increases among the whole boundary layer. The profiles of $p _{rms}^{\prime}/\bar{p}$ in ``M8T04'' and ``M8T015'' have secondary peaks at $y^{*} \approx 60$, which are consistent with the wall-normal location of the primary peak in ``M8T08''. A special phenomenon is observed that the intensities of $p _{rms}^{\prime}/\bar{p}$ are significantly enhanced near the wall when the wall is strongly cooled, which further result in the fact that the $p _{rms}^{\prime}/\bar{p}$ profiles attain their primary peaks at the wall in ``M8T04'' and ``M8T015''. The significant enhancement of $p _{rms}^{\prime}/\bar{p}$ near the wall can be ascribed to the appearance of a special acoustic structure (``the travelling-wave-like alternating positive and negative structures'' or TAPNS) in the near-wall region, which will be specifically discussed in Section \ref{sec: n3}.}

\begin{figure}\centering
    \includegraphics[width=0.99\linewidth]{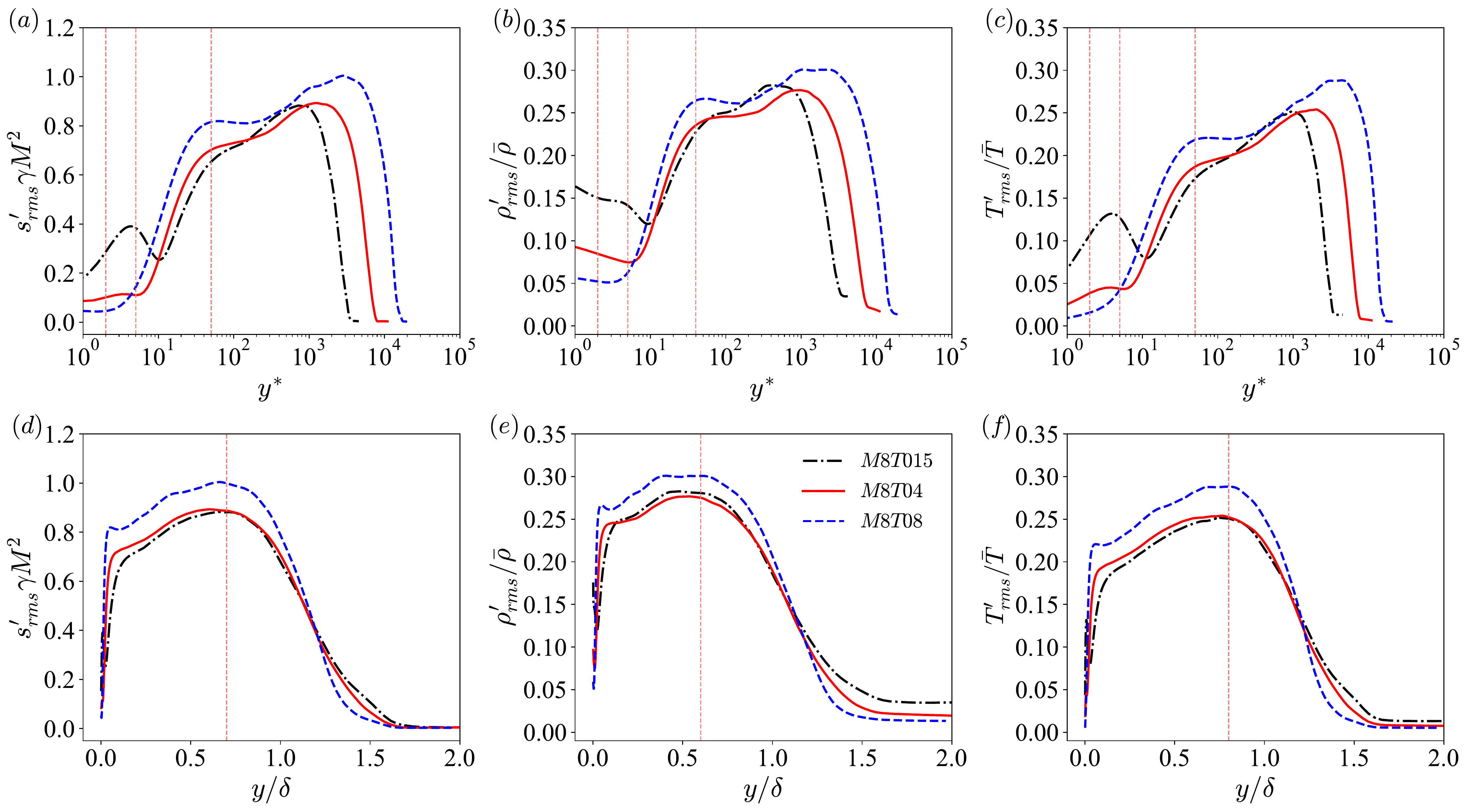}
    \caption{(a) and (d): The normalised turbulent intensity of the entropy $s _{rms}^{\prime}\gamma M^{2}$ along wall-normal direction against (a) semi-local scaling ($y^{*}$) and (d) outer scaling ($y/\delta$). The vertical dashed lines represent $y^{*}=2, 5, 50$ in (a) and $y/\delta=0.7$ in (d) respectively. (b) and (e): The normalised turbulent intensity of the density $\rho _{rms}^{\prime}/\bar{\rho }$ along wall-normal direction against (b) semi-local scaling ($y^{*}$) and (e) outer scaling ($y/\delta$). The vertical dashed lines represent $y^{*}=2, 5, 40$ in (b) and $y/\delta=0.6$ in (e) respectively. (c) and (f): The normalised turbulent intensity of the temperature $T _{rms}^{\prime}/\bar{T}$ along wall-normal direction against (c) semi-local scaling ($y^{*}$) and (f) outer scaling ($y/\delta$). The vertical dashed lines represent $y^{*}=2, 5, 50$ in (c) and $y/\delta=0.8$ in (f) respectively.}
    \label{fig: d6}
\end{figure}

{The dimensionless entropy per unit mass $s$ can be defined as $s=C_{v}log\left ( T/\rho ^{\gamma -1} \right )$ \citep[]{Gerolymos2014,Wang2019}. The normalised turbulent intensities $s _{rms}^{\prime}\gamma M^{2}$, $\rho _{rms}^{\prime}/\bar{\rho }$ and $T _{rms}^{\prime}/\bar{T}$ along wall-normal direction are shown in figure \ref{fig: d6}. It is found in figure \ref{fig: d6} (a) and (d) that the normalised turbulent intensity of the entropy $s _{rms}^{\prime}\gamma M^{2}$ attains its primary peak near the edge of the boundary layer ($y/\delta \approx 0.7$). Furthermore, a secondary peak is observed at $y^{*} \approx 50$ in ``M8T08'', and this local secondary peak gradually disappears as the wall temperature decreases. It is also found that as the wall temperature decreases, the intensity $s _{rms}^{\prime}\gamma M^{2}$ decreases in the region between $y^{*} > 10$ and $y/\delta < 0.7$, while $s _{rms}^{\prime}\gamma M^{2}$ is significantly enhanced in the near-wall region ($y^{*} < 10$). An interesting phenomenon is found that a strong local secondary peak is observed at $y^{*} \approx 5$ in ``M8T015''. It is noted that the local secondary peaks also appear in the near-wall region in ``M6Tw025'' and ``M14Tw018'' cases in \citet[]{Zhang2018}. However, to the best of our knowledge, previous studies have not provided a physical explanation for the strong local secondary peak of $s _{rms}^{\prime}\gamma M^{2}$ in the near-wall region when the wall is strongly cooled. In Section \ref{sec: n3}, it is found that the above phenomenon can be attributed to the appearance of a special entropic structure (``the streaky entropic structures'' or SES) near the wall.}

{It is also found that the intensities $\rho _{rms}^{\prime}/\bar{\rho }$ and $T _{rms}^{\prime}/\bar{T}$ have similar behaviours with $s _{rms}^{\prime}\gamma M^{2}$ among most regions of the boundary layer, except for the much larger values of $\rho _{rms}^{\prime}/\bar{\rho }$ in the vicinity of the wall ($y^{*} < 5$). The similarity among the intensities $\rho _{rms}^{\prime}/\bar{\rho }$, $T _{rms}^{\prime}/\bar{T}$ and $s _{rms}^{\prime}\gamma M^{2}$ can be explained based on the Kovasznay decomposition \citep[]{Kovasznay1953,Chassaing2002,Gauthier2017,Wang2019}. It is noted that the Kovasznay decomposition can decompose the thermodynamic variables into the acoustic modes and the entropic modes \citep[]{Kovasznay1953,Chassaing2002,Gauthier2017,Wang2019}.}

In compressible turbulent flow, the acoustic modes of the thermodynamic variables can be defined as \citep[]{Chassaing2002,Gauthier2017,Wang2019}
\begin{equation}
    p'_{I}=p-\bar{p},
\end{equation}
\begin{equation}
    {\rho }'_{I}=\frac{\bar{\rho }p'_{I}}{\gamma \bar{p}},
\end{equation}
\begin{equation}
    {T}'_{I}=\frac{\left ( \gamma -1 \right )\bar{T}p'_{I}}{\gamma \bar{p}};
\end{equation}
and the entropic modes can be given by \citep[]{Chassaing2002,Gauthier2017,Wang2019}
\begin{equation}
    p'_{E}=0,
\end{equation}
\begin{equation}
    {\rho }'_{E}=\rho -\bar{\rho }-{\rho }'_{I},
\end{equation}
\begin{equation}
    {T}'_{E}=T -\bar{T}-{T}'_{I}.
\end{equation}
Therefore, the acoustic mode of the pressure $p'_{I}$ is consistent with the fluctuating pressure $p^{\prime}$, and the fluctuating density and temperature can be divided into the acoustic modes and the entropic modes respectively: ${\rho }'={\rho }'_{I}+{\rho }'_{E}$; ${T}'={T}'_{I}+{T}'_{E}$. It is noted that the correlation coefficients between variables ${\varphi }'$ and ${\psi }'$ can be defined as
\begin{equation}
    R\left ({\varphi }',{\psi }'  \right )=\frac{\overline{{\varphi }'{\psi }'}}{\sqrt{\overline{{\varphi }^{\prime 2}}}\sqrt{\overline{{\psi }^{\prime 2}}}}.
\end{equation}
The correlation coefficient $R\left ({\varphi }',{\psi }'  \right )=1$ indicates that the variables ${\varphi }'$ and ${\psi }'$ are positively linearly correlated with each other; while $R\left ({\varphi }',{\psi }'  \right )=-1$ suggests that the variables ${\varphi }'$ and ${\psi }'$ are negatively linearly correlated with each other.

\begin{figure}\centering
    \includegraphics[width=0.99\linewidth]{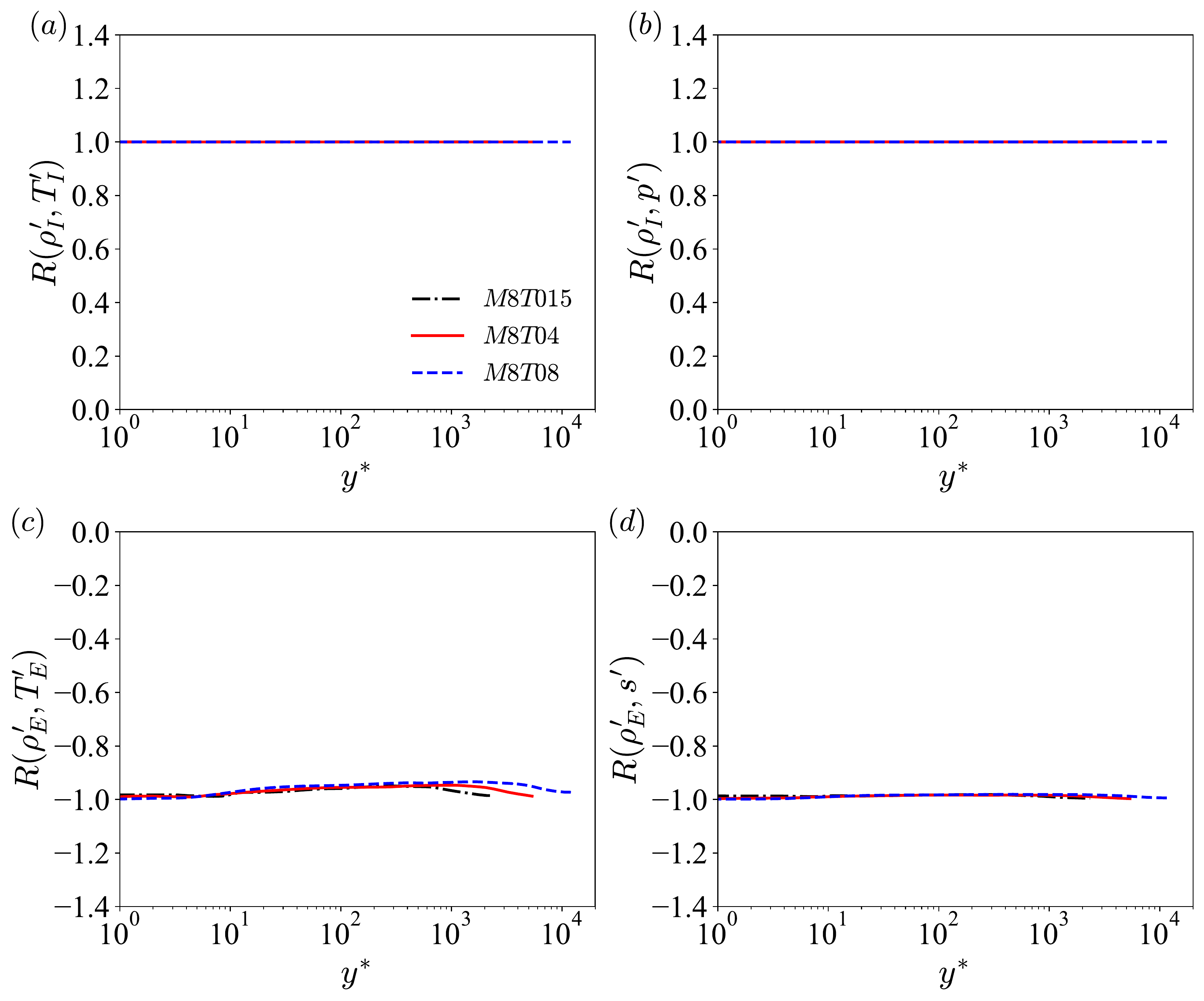}
    \caption{The correlation coefficients (a) $R\left ( \rho _{I} ^{\prime},T _{I}^{\prime} \right )$, (b) $R\left ( \rho _{I} ^{\prime},p^{\prime} \right )$, (c) $R\left ( \rho _{E} ^{\prime},T _{E}^{\prime} \right )$ and (d) $R\left ( \rho _{E} ^{\prime},s^{\prime} \right )$ along wall-normal direction plotted against semi-local scaling ($y^{*}$).}
    \label{fig: d9}
\end{figure}

The correlation coefficients $R\left ( \rho _{I} ^{\prime},T _{I}^{\prime} \right )$, $R\left ( \rho _{I} ^{\prime},p^{\prime} \right )$, $R\left ( \rho _{E} ^{\prime},T _{E}^{\prime} \right )$ and $R\left ( \rho _{E} ^{\prime},s^{\prime} \right )$ along wall-normal direction are shown in figure \ref{fig: d9}. It is found that $R\left ( \rho _{I} ^{\prime},T _{I}^{\prime} \right )=1$ and $R\left ( \rho _{I} ^{\prime},p^{\prime} \right )=1$ along wall-normal direction, confirming that $\rho _{I} ^{\prime}$, $T _{I}^{\prime}$ and $p^{\prime}$ are positively linearly correlated with each other. Furthermore, $R\left ( \rho _{E} ^{\prime},T _{E}^{\prime} \right ) \approx -1$ and $R\left ( \rho _{E} ^{\prime},s^{\prime} \right )\approx -1$, suggesting that $\rho _{E} ^{\prime}$ is almost negatively linearly correlated with $T _{E}^{\prime}$ and $s^{\prime}$ respectively. Accordingly, the acoustic modes of density and temperature $\rho _{I} ^{\prime}$ and $T _{I}^{\prime}$ are linearly correlated with the fluctuating pressure $p^{\prime}$, and the entropic modes of density and temperature $\rho _{E} ^{\prime}$ and $T _{E}^{\prime}$ are almost linearly correlated with the fluctuating entropy $s^{\prime}$.

\begin{figure}\centering
    \includegraphics[width=0.99\linewidth]{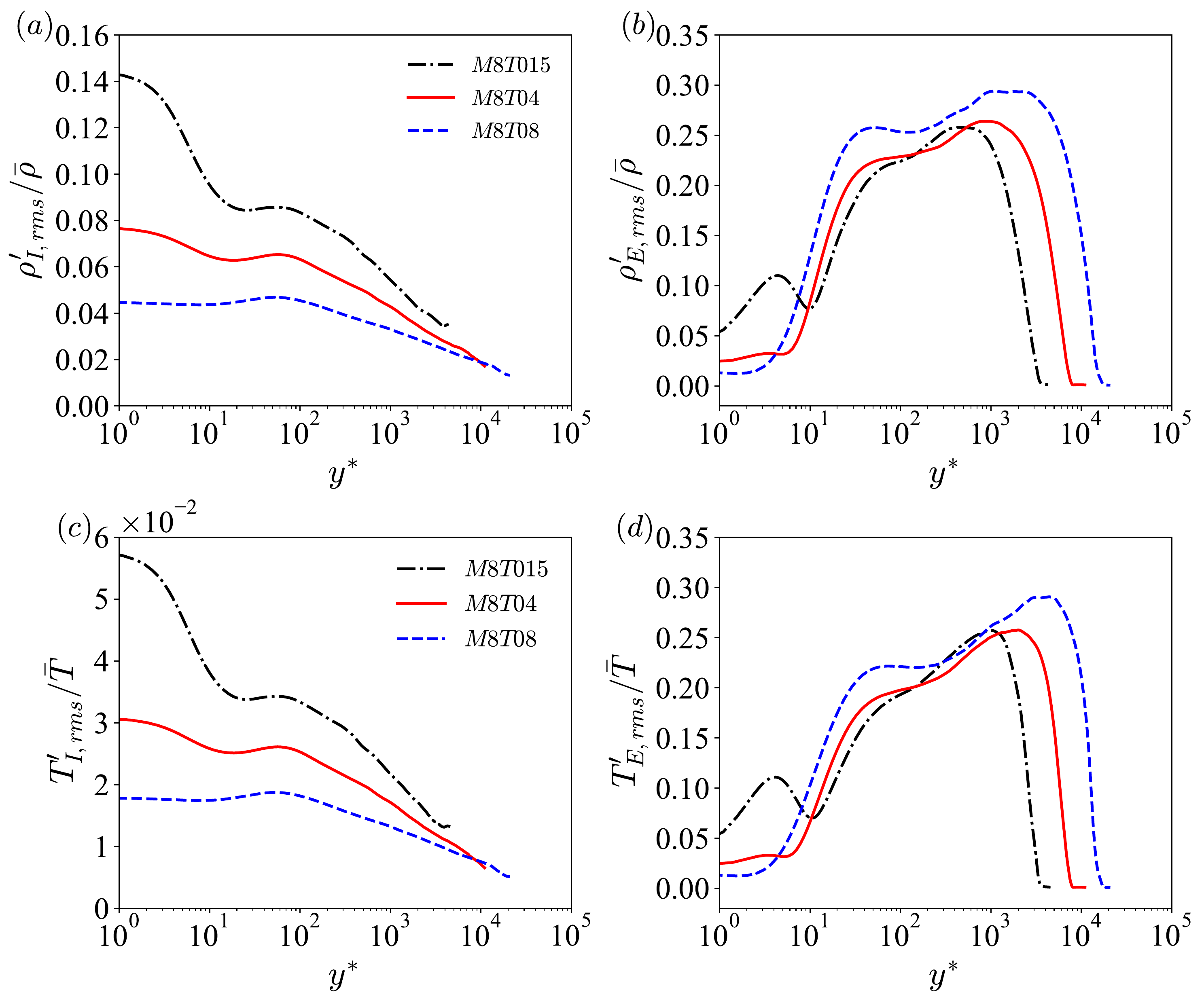}
    \caption{(a) and (b): The normalised turbulent intensity of (a) the acoustic mode of density $\rho _{I,rms}^{\prime}/\bar{\rho }$ and (b) the entropic mode of density $\rho _{E,rms}^{\prime}/\bar{\rho }$ along wall-normal direction. (c) and (d): The normalised turbulent intensity of (c) the acoustic mode of temperature $T _{I,rms}^{\prime}/\bar{T}$ and (d) the entropic mode of temperature $T _{E,rms}^{\prime}/\bar{T}$ along wall-normal direction.}
    \label{fig: d10}
\end{figure}

{The normalised turbulent intensities of the acoustic and entropic modes of density and temperature along wall-normal direction are shown in figure \ref{fig: d10}. It is found that the profiles of the intensities of the acoustic modes of density and temperature (figure \ref{fig: d10} (a) (c)) are similar to that of the fluctuating pressure (figure \ref{fig: d5} (a)), and the intensities of the entropic modes of density and temperature (figure \ref{fig: d10} (b) (d)) also have similar behaviours with that of the fluctuating entropy (figure \ref{fig: d6} (a)).}

\begin{figure}\centering
    \includegraphics[width=0.99\linewidth]{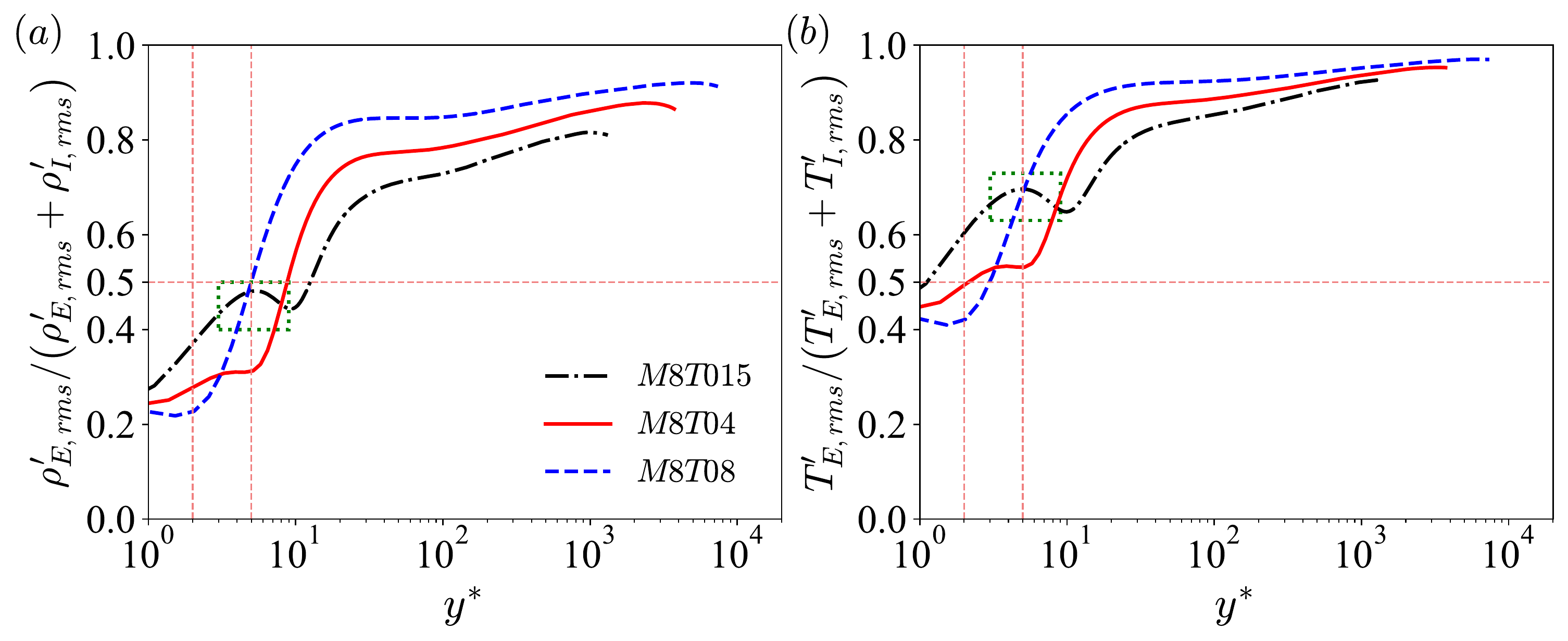}
    \caption{The relative contributions (a) $\rho _{E,rms}^{\prime}/\left (\rho _{E,rms}^{\prime}+ \rho _{I,rms}^{\prime} \right )$ and (b) $T _{E,rms}^{\prime}/\left (T _{E,rms}^{\prime}+T _{I,rms}^{\prime}  \right )$ along wall-normal direction. The vertical dashed lines represent $y^{*}=2, 5$ in (a) and (b).}
    \label{fig: d12}
\end{figure}

The relative contributions $\rho _{E,rms}^{\prime}/\left (\rho _{E,rms}^{\prime}+ \rho _{I,rms}^{\prime} \right )$ and $T _{E,rms}^{\prime}/\left (T _{E,rms}^{\prime}+T _{I,rms}^{\prime}  \right )$ along wall-normal direction are shown in figure \ref{fig: d12}. It is noted that if the relative contributions $\rho _{E,rms}^{\prime}/\left (\rho _{E,rms}^{\prime}+ \rho _{I,rms}^{\prime} \right )$ and $T _{E,rms}^{\prime}/\left (T _{E,rms}^{\prime}+T _{I,rms}^{\prime}  \right )$ are lower than 0.5, the acoustic modes of density and temperature are predominant over their entropic modes; on the contrary, if the relative contributions are larger than 0.5, the entropic modes are dominant. {It is found that the relative contributions $\rho _{E,rms}^{\prime}/\left (\rho _{E,rms}^{\prime}+ \rho _{I,rms}^{\prime} \right )$ and $T _{E,rms}^{\prime}/\left (T _{E,rms}^{\prime}+T _{I,rms}^{\prime}  \right )$ increase as the wall-normal location increases. Furthermore, the relative contribution $\rho _{E,rms}^{\prime}/\left (\rho _{E,rms}^{\prime}+ \rho _{I,rms}^{\prime} \right )$ is totally smaller than $T _{E,rms}^{\prime}/\left (T _{E,rms}^{\prime}+T _{I,rms}^{\prime}  \right )$ among the boundary layer, suggesting that the entropic mode has a much larger contribution in fluctuating temperature than that in fluctuating density. In the near-wall region, the fluctuating density is mainly dominated by its acoustic mode, and the acoustic mode of temperature is slightly predominant over the entropic mode of temperature. However, in the far-wall region, the fluctuating density and temperature are mainly dominated by their entropic modes, which further lead to the similarity between the profiles of $s _{rms}^{\prime}\gamma M^{2}$, $\rho _{rms}^{\prime}/\bar{\rho }$ and $T _{rms}^{\prime}/\bar{T}$ shown in figure \ref{fig: d6}.}

It is also found that the wall temperature has a significant influence on the relative contributions. In the near-wall region, as the wall temperature decreases, the acoustic modes and the entropic modes of density and temperature are all enhanced, but the amounts of the growth of the entropic modes are slightly larger than those of the acoustic modes, which result in the increase of the relative contributions $\rho _{E,rms}^{\prime}/\left (\rho _{E,rms}^{\prime}+ \rho _{I,rms}^{\prime} \right )$ and $T _{E,rms}^{\prime}/\left (T _{E,rms}^{\prime}+T _{I,rms}^{\prime}  \right )$ with colder wall temperature. {In the far-wall region, when the wall temperature becomes colder, the acoustic modes are enhanced while the entropic modes are reduced, which lead to the decrease of the relative contributions of the entropic modes in strongly cooled wall case.} Therefore, as the wall temperature decreases, the increase of $\rho _{rms}^{\prime}/\bar{\rho }$ and $T _{rms}^{\prime}/\bar{T}$ near the wall is mainly due to the contributions of both the acoustic and entropic modes, while the decrease of $\rho _{rms}^{\prime}/\bar{\rho }$ and $T _{rms}^{\prime}/\bar{T}$ far from the wall can be ascribed to the major contributions of the entropic modes.

    {It has been noted previously in figure \ref{fig: d6} (b) that $\rho _{rms}^{\prime}/\bar{\rho }$ has much larger values than those of $s _{rms}^{\prime}\gamma M^{2}$ and $T _{rms}^{\prime}/\bar{T}$ in the vicinity of the wall ($y^{*} < 5$), especially in ``M8T04'' and ``M8T015''. This can be ascribed to the dominant contribution of the acoustic mode to the fluctuating density near the wall (figure \ref{fig: d12} (a)), and the fact that the acoustic mode of density attains the primary peak at the wall in ``M8T04'' and ``M8T015'' (figure \ref{fig: d10} (a)). Moreover, it is also found in figure \ref{fig: d6} (e) that the primary peak value of $\rho _{rms}^{\prime}/\bar{\rho }$ at $y/\delta \approx 0.6$ in ``M8T015'' is slightly larger than that in ``M8T04'', which is opposite to the behaviours of $s _{rms}^{\prime}\gamma M^{2}$ and $T _{rms}^{\prime}/\bar{T}$. This can be attributed to the facts that the intensity of the acoustic mode of density in ``M8T015'' is much larger than that in ``M8T04'' (figure \ref{fig: d10} (a)), and the acoustic mode of density in ``M8T015'' has a much larger contribution to fluctuating density than that in ``M8T04'' near the edge of the boundary layer (figure \ref{fig: d12} (a)).}

\section{The streamwise and spanwise spectra of the streamwise velocity and thermodynamic variables}\label{sec: n3}
In order to quantitatively characterize the characteristic length scales of the energetic structures of the streamwise velocity and thermodynamic variables, the premultiplied streamwise and spanwise spectra are further investigated. It is noted that $k_{x}$ and $k_{z}$ are the wavenumber in the streamwise and spanwise directions respectively. ${\lambda}_{x}$ and ${\lambda}_{z}$ are the corresponding wavelength in the streamwise and spanwise directions respectively. Furthermore, ${\lambda}_{x}^{+}={\lambda}_{x}/\delta _{\nu }$ and ${\lambda}_{x}^{*}={\lambda}_{x}/\delta _{\nu }^{*}$.

\begin{figure}\centering
    \includegraphics[width=0.99\linewidth]{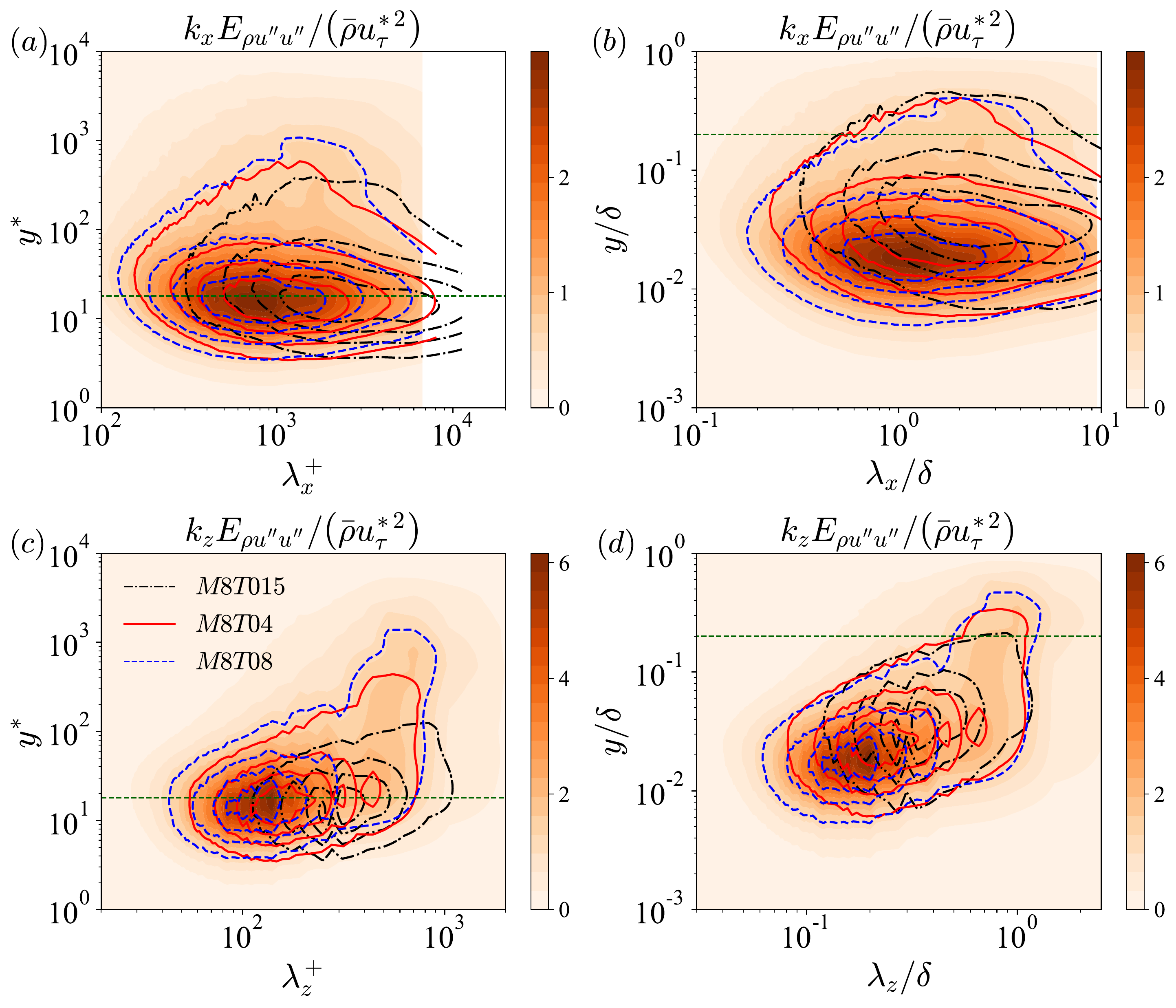}
    \caption{(a) and (b): The normalised premultiplied streamwise spectra of the fluctuating streamwise velocity $k_{x}E _{\rho u^{\prime \prime }u^{\prime \prime }}/\left (\bar{\rho }{u_{\tau }^{*}}^{2}  \right )$ in (a) inner scaling and (b) outer scaling. (c) and (d): The normalised premultiplied spanwise spectra of the fluctuating streamwise velocity $k_{z}E _{\rho u^{\prime \prime }u^{\prime \prime }}/\left (\bar{\rho }{u_{\tau }^{*}}^{2}  \right )$ in (c) inner scaling and (d) outer scaling. The {filled} contour represents the normalised premultiplied spectra in ``M8T08''. The line contour levels are (0.2, 0.4, 0.6, 0.8) times the peak values. The horizontal dashed lines represent $y^{*}=18$ in (a) (c) and $y/\delta=0.2$ in (b) (d) respectively.}
    \label{fig: d13}
\end{figure}

{The normalised premultiplied streamwise and spanwise spectra of the fluctuating streamwise velocity $k_{x}E _{\rho u^{\prime \prime }u^{\prime \prime }}/\left (\bar{\rho }{u_{\tau }^{*}}^{2}  \right )$ and $k_{z}E _{\rho u^{\prime \prime }u^{\prime \prime }}/\left (\bar{\rho }{u_{\tau }^{*}}^{2}  \right )$ are shown in figure \ref{fig: d13}.} It is found in figure \ref{fig: d13} (a) and (c) that the premultiplied streamwise and spanwise spectra of the fluctuating streamwise velocity achieve their {primary} peaks at $y^{*} \approx 18$, which are consistent with the {primary} peak location of $u _{rms}^{\prime \prime}/u_{\tau}^{*}$ (figure \ref{fig: d4} (a)).  {The peak of $u _{rms}^{\prime \prime}/u_{\tau}^{*}$ corresponds to the cycle of the near-wall streak generation} \citep[]{Jimenez1999,Jimenez2013,Hutchins2007b,Monty2009,Pirozzoli2013,Huang2022}. Furthermore, $k_{z}E _{\rho u^{\prime \prime }u^{\prime \prime }}/\left (\bar{\rho }{u_{\tau }^{*}}^{2}  \right )$ has a weak outer peak at nearly $y/\delta \approx 0.2$ (figure \ref{fig: d13} (d)), indicating the long streaky motion in the outer region (i.e. LSMs or VLSMs) \citep[]{Hutchins2007a,Monty2009,Pirozzoli2011,Pirozzoli2013,Hwang2016,Huang2022}.  However, the outer peak of $k_{x}E _{\rho u^{\prime \prime }u^{\prime \prime }}/\left (\bar{\rho }{u_{\tau }^{*}}^{2}  \right )$ is not evident (figure \ref{fig: d13} (b)), mainly due to the {relatively low Reynolds number of the DNS database} \citep[]{Hwang2016}.

\begin{figure}\centering
    \includegraphics[width=0.99\linewidth]{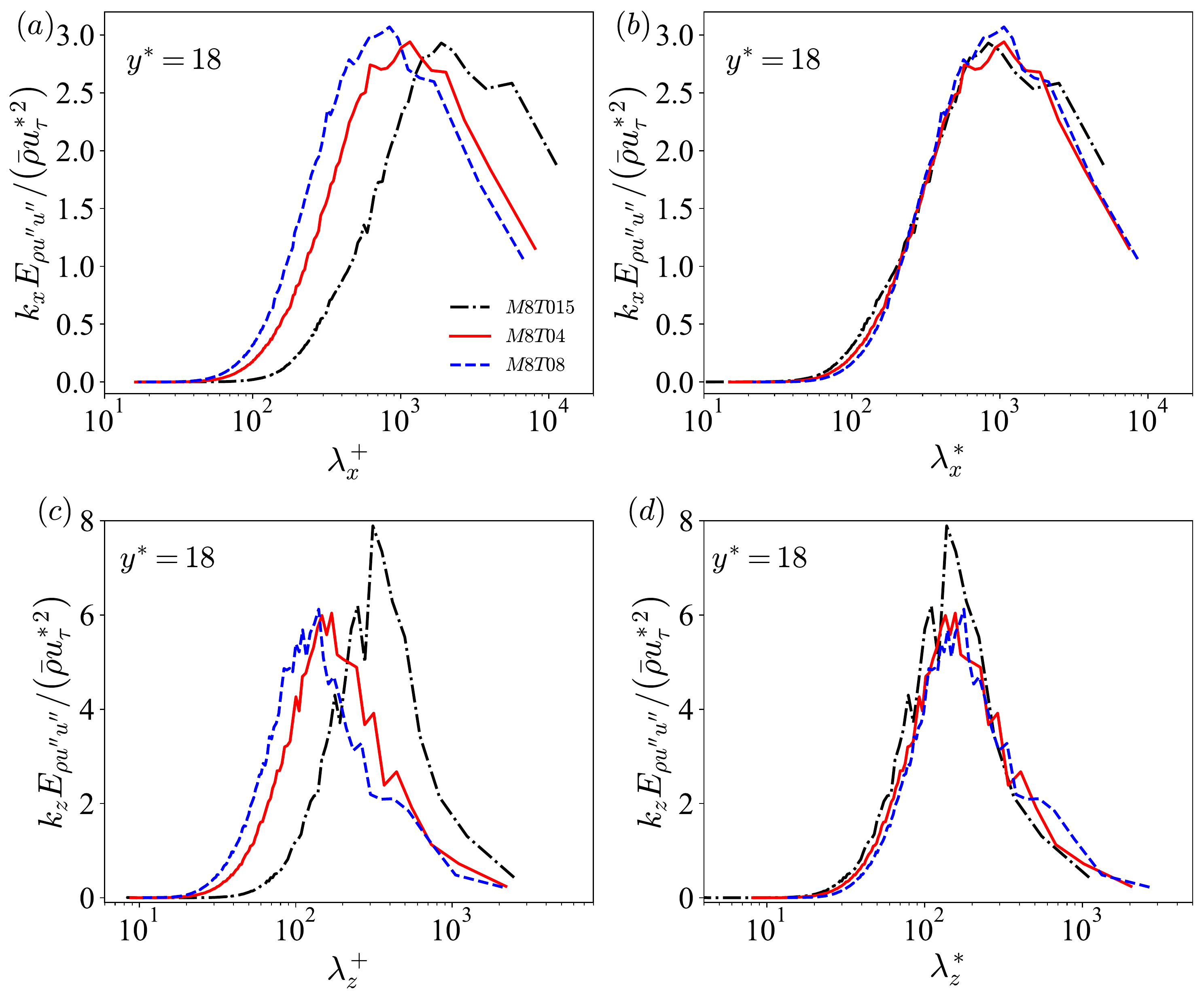}
    \caption{(a) and (b): The normalised premultiplied streamwise spectra of the fluctuating streamwise velocity $k_{x}E _{\rho u^{\prime \prime }u^{\prime \prime }}/\left (\bar{\rho }{u_{\tau }^{*}}^{2}  \right )$ at $y^{*}=18$ plotted against (a) ${\lambda}_{x}^{+}$ and (b) ${\lambda}_{x}^{*}$. (c) and (d): The normalised premultiplied spanwise spectra of the fluctuating streamwise velocity $k_{z}E _{\rho u^{\prime \prime }u^{\prime \prime }}/\left (\bar{\rho }{u_{\tau }^{*}}^{2}  \right )$ at $y^{*}=18$ plotted against (c) ${\lambda}_{z}^{+}$ and (d) ${\lambda}_{z}^{*}$.}
    \label{fig: d14}
\end{figure}

\begin{figure}\centering
    \includegraphics[width=0.99\linewidth]{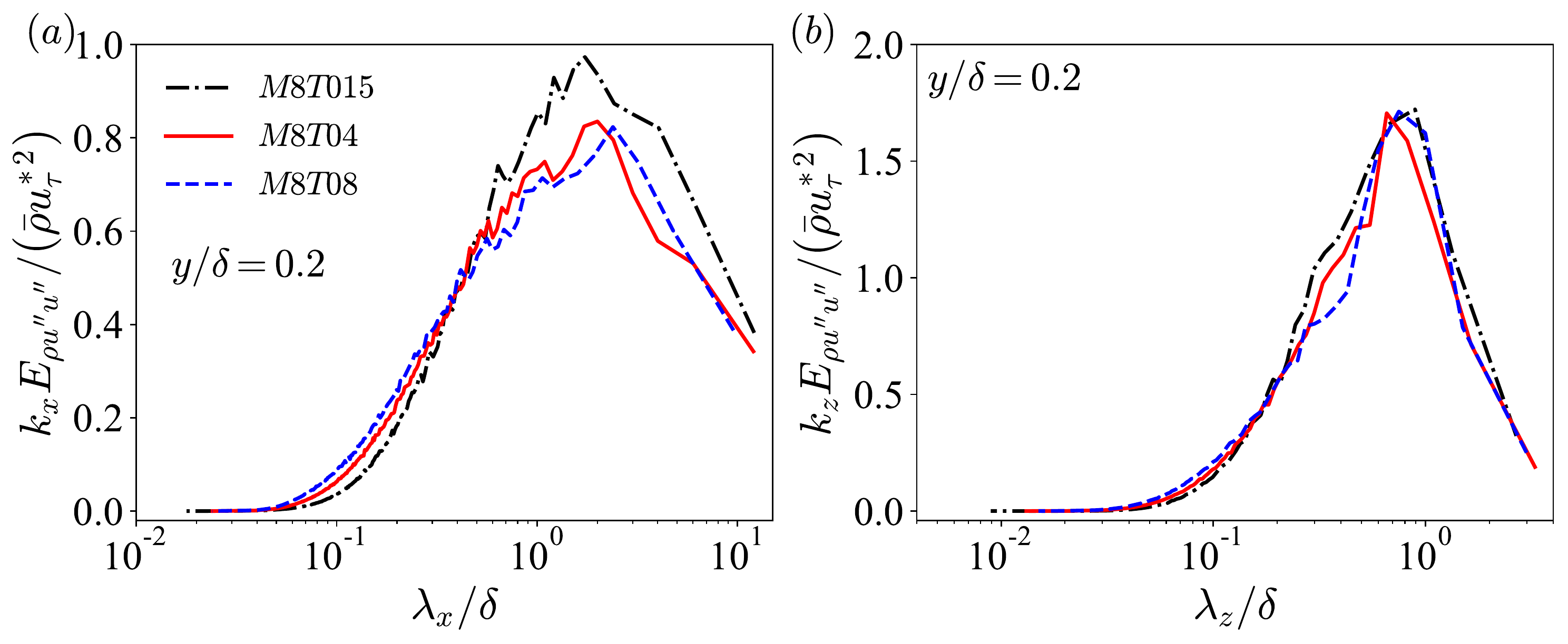}
    \caption{(a) The normalised premultiplied streamwise spectra of the fluctuating streamwise velocity $k_{x}E _{\rho u^{\prime \prime }u^{\prime \prime }}/\left (\bar{\rho }{u_{\tau }^{*}}^{2}  \right )$ at $y/\delta=0.2$ plotted against ${\lambda}_{x}/\delta $. (b) The normalised premultiplied spanwise spectra of the fluctuating streamwise velocity $k_{z}E _{\rho u^{\prime \prime }u^{\prime \prime }}/\left (\bar{\rho }{u_{\tau }^{*}}^{2}  \right )$ at $y/\delta=0.2$ plotted against ${\lambda}_{z}/\delta $.}
    \label{fig: d15}
\end{figure}

The normalised premultiplied streamwise and spanwise spectra of the fluctuating streamwise velocity $k_{x}E _{\rho u^{\prime \prime }u^{\prime \prime }}/\left (\bar{\rho }{u_{\tau }^{*}}^{2}  \right )$ and $k_{z}E _{\rho u^{\prime \prime }u^{\prime \prime }}/\left (\bar{\rho }{u_{\tau }^{*}}^{2}  \right )$ at $y^{*}=18$ and $y/\delta=0.2$ are shown in figure \ref{fig: d14} and figure \ref{fig: d15} respectively. It is found in figure \ref{fig: d14} (a) and (c) that the peak locations of the premultiplied streamwise and spanwise spectra of the fluctuating streamwise velocity in wall unit $\left ( {\lambda}_{x}^{+},{\lambda}_{z}^{+} \right )$ increase significantly as the wall temperature decreases, especially in strongly cooled wall case ``M8T015''. These peak locations represent the characteristic streamwise length and spanwise spacing of the near-wall streaks. However, the semi-local scaling $\left ( {\lambda}_{x}^{*},{\lambda}_{z}^{*} \right )$ can significantly reduce the disparity between the peak locations of spectra (figure \ref{fig: d14} (b) (d)), yielding the characteristic streamwise length ${\lambda}_{x}^{*} \approx 10^{3}$ and spanwise spacing ${\lambda}_{z}^{*} \approx 150$ in different wall temperature cases. Similar values of the characteristic streamwise length and spanwise spacing of the near-wall streaks have also been found in incompressible boundary layers \citep[]{Hutchins2007b,Monty2009} and compressible boundary layers \citep[]{Pirozzoli2013,Huang2022}. Furthermore, it is shown in figure \ref{fig: d15} that the streamwise and spanwise spectra achieve their peak values at ${\lambda}_{x}/\delta \approx 2$ and ${\lambda}_{z}/\delta \approx 1$ respectively at $y/\delta=0.2$, which represent the characteristic streamwise length and spanwise spacing of the long streaky motions in the outer region. {Similar values of the characteristic spanwise spacing at $y/\delta=0.2$ was also obtained in previous study of hypersonic turbulent boundary layers \citep[]{Cogo2022}.}

\begin{figure}\centering
    \includegraphics[width=0.99\linewidth]{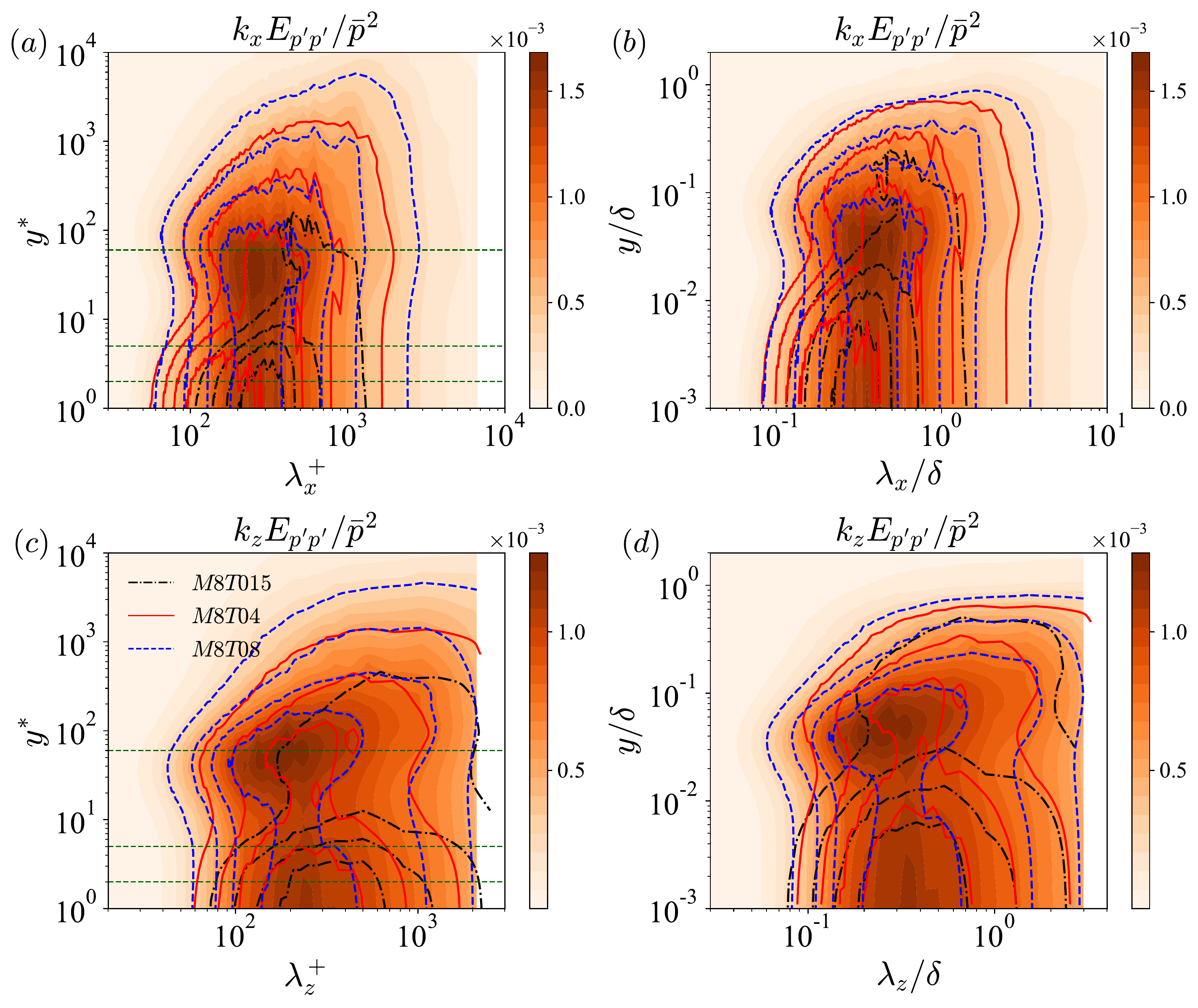}
    \caption{(a) and (b): The normalised premultiplied streamwise spectra of the fluctuating pressure $k_{x}E_{ p^{\prime }p^{\prime }}/\bar{p}^{2}$ in (a) inner scaling and (b) outer scaling. (c) and (d): The normalised premultiplied spanwise spectra of the fluctuating pressure $k_{z}E_{ p^{\prime }p^{\prime }}/\bar{p}^{2}$ in (c) inner scaling and (d) outer scaling. The {filled} contour represents the normalised premultiplied spectra in ``M8T08''. The line contour levels are (0.2, 0.4, 0.6, 0.8) times the peak values. The horizontal dashed lines represent $y^{*}=2, 5, 60$ in (a) (c).}
    \label{fig: d16}
\end{figure}

The normalised premultiplied streamwise and spanwise spectra of the fluctuating pressure $k_{x}E_{ p^{\prime }p^{\prime }}/\bar{p}^{2}$ and $k_{z}E_{ p^{\prime }p^{\prime }}/\bar{p}^{2}$ are shown in figure \ref{fig: d16}. {The $k_{x}E_{ p^{\prime }p^{\prime }}/\bar{p}^{2}$ and $k_{z}E_{ p^{\prime }p^{\prime }}/\bar{p}^{2}$ in the nearly adiabatic wall case ``M8T08'' achieve their primary peaks at $y^{*} \approx 60$, while the pressure spectra of ``M8T04'' and ``M8T015'' achieve their primary peaks at the wall. These observations are consistent with the primary peak locations of $p _{rms}^{\prime}/\bar{p}$ (figure \ref{fig: d5} (a)).}

\begin{figure}\centering
    \includegraphics[width=0.99\linewidth]{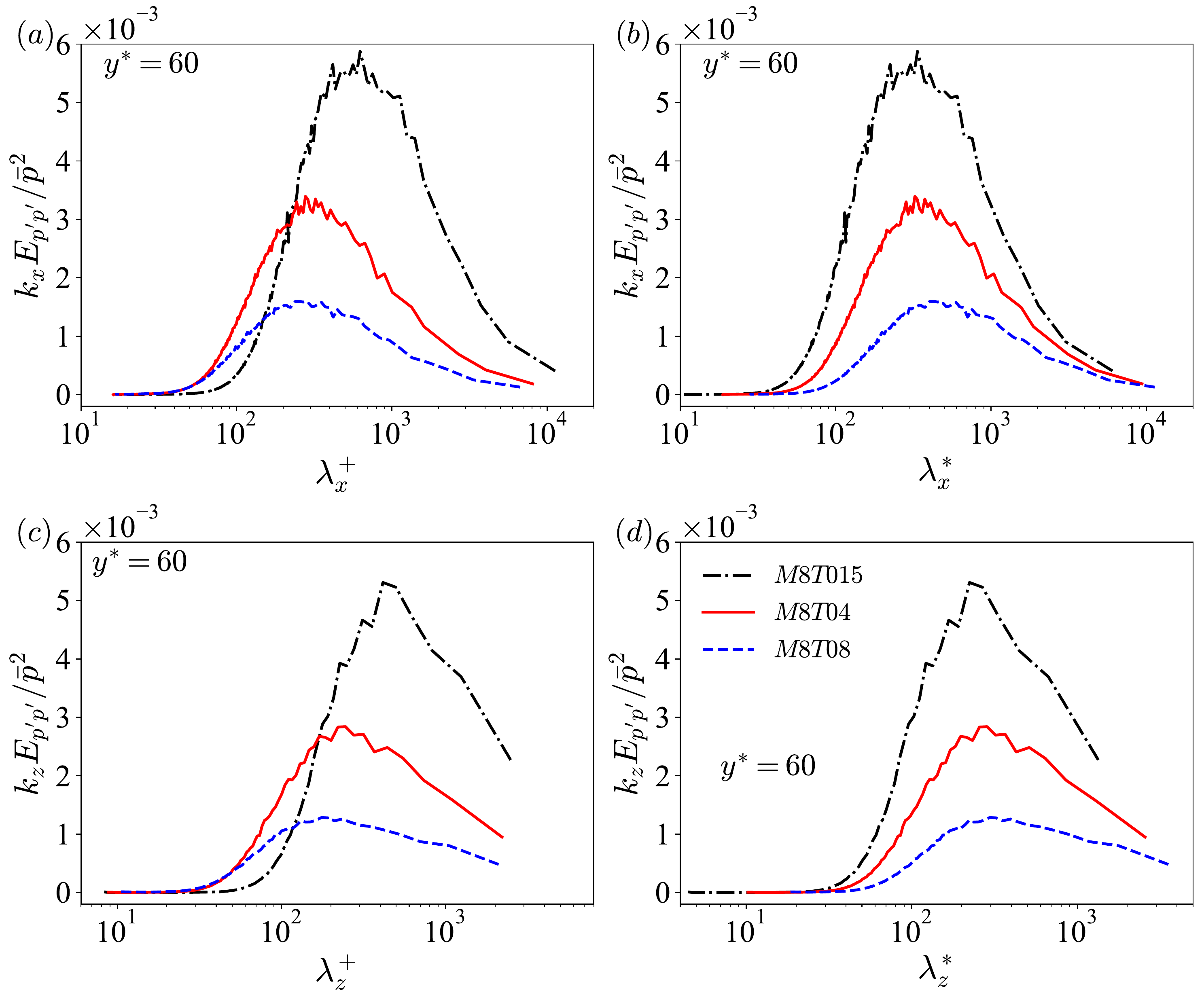}
    \caption{(a) and (b): The normalised premultiplied streamwise spectra of the fluctuating pressure $k_{x}E_{ p^{\prime }p^{\prime }}/\bar{p}^{2}$ at $y^{*}=60$ plotted against (a) ${\lambda}_{x}^{+}$ and (b) ${\lambda}_{x}^{*}$. (c) and (d): The normalised premultiplied spanwise spectra of the fluctuating pressure  $k_{z}E_{ p^{\prime }p^{\prime }}/\bar{p}^{2}$ at $y^{*}=60$ plotted against (c) ${\lambda}_{z}^{+}$ and (d) ${\lambda}_{z}^{*}$.}
    \label{fig: d18}
\end{figure}

\begin{figure}\centering
    \includegraphics[width=0.99\linewidth]{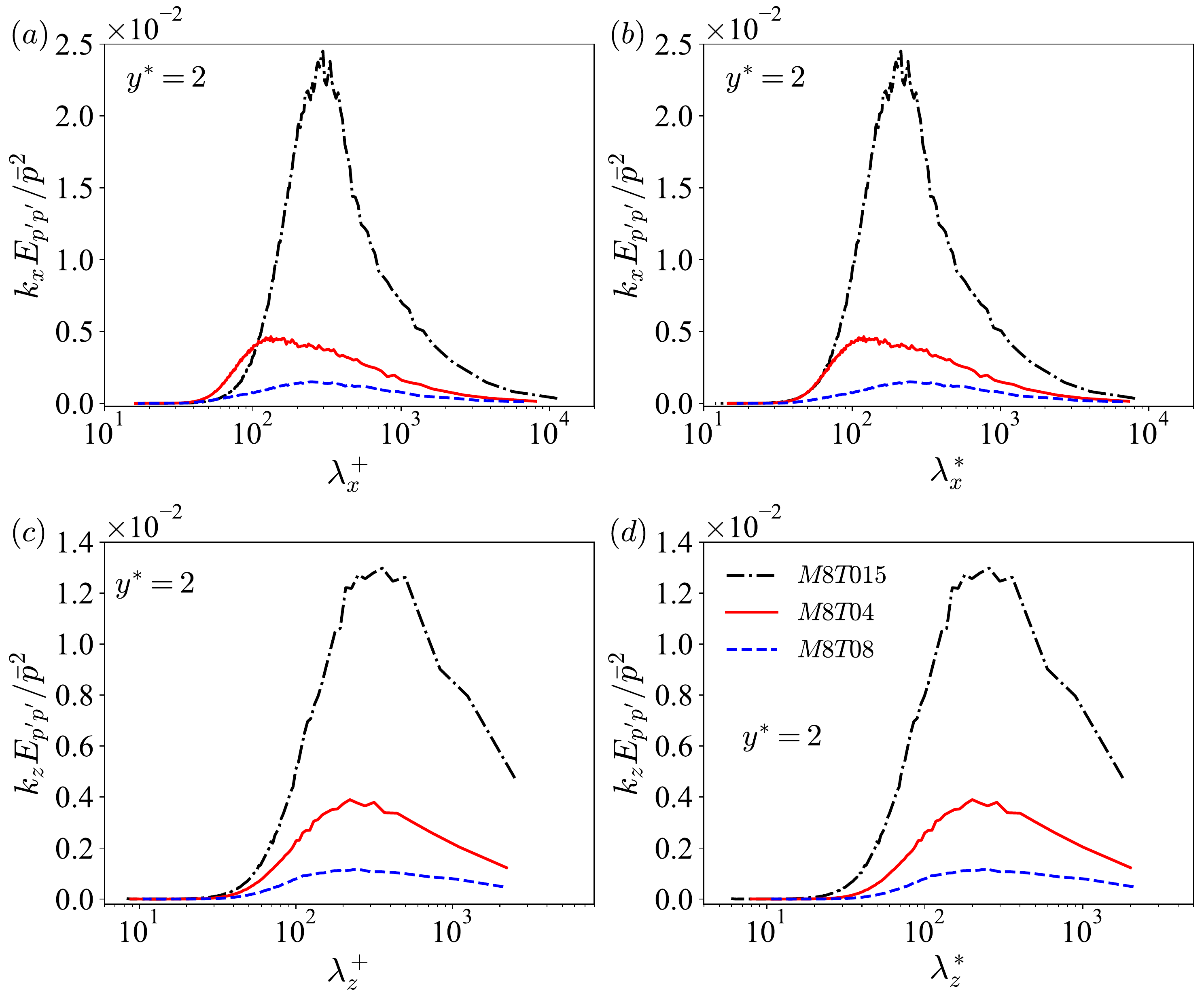}
    \caption{(a) and (b): The normalised premultiplied streamwise spectra of the fluctuating pressure $k_{x}E_{ p^{\prime }p^{\prime }}/\bar{p}^{2}$ at $y^{*}=2$ plotted against (a) ${\lambda}_{x}^{+}$ and (b) ${\lambda}_{x}^{*}$. (c) and (d): The normalised premultiplied spanwise spectra of the fluctuating pressure  $k_{z}E_{ p^{\prime }p^{\prime }}/\bar{p}^{2}$ at $y^{*}=2$ plotted against (c) ${\lambda}_{z}^{+}$ and (d) ${\lambda}_{z}^{*}$.}
    \label{fig: d19}
\end{figure}

\begin{figure}\centering
    \includegraphics[width=0.99\linewidth]{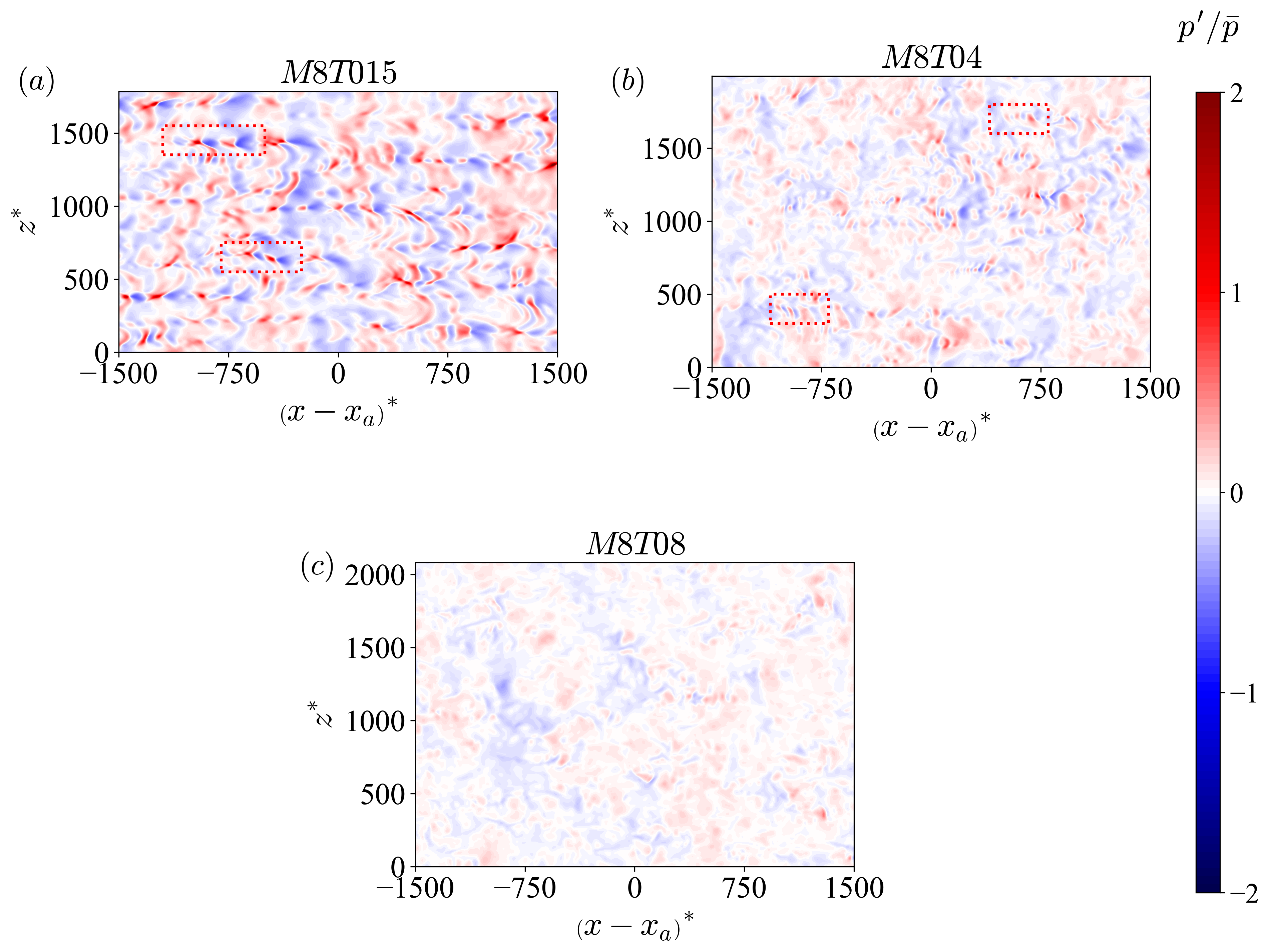}
    \caption{The instantaneous fields of the normalised fluctuating pressure $p^{\prime }/\bar{p}$ at $y^{*}=2$ in (a) ``M8T015'', (b) ``M8T04'' and (c) ``M8T08''.}
    \label{fig: d17}
\end{figure}

\begin{figure}\centering
    \includegraphics[width=0.99\linewidth]{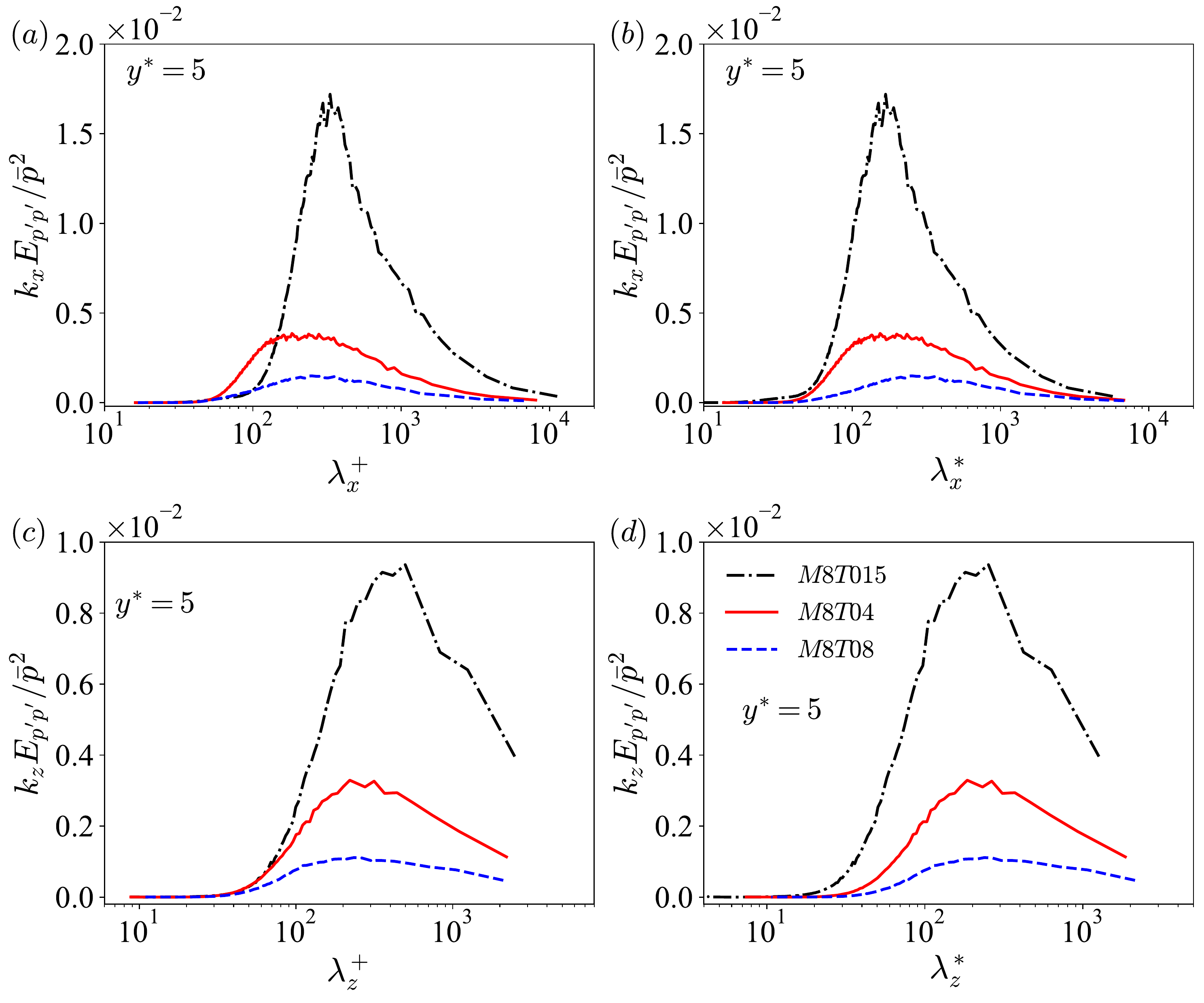}
    \caption{(a) and (b): The normalised premultiplied streamwise spectra of the fluctuating pressure $k_{x}E_{ p^{\prime }p^{\prime }}/\bar{p}^{2}$ at $y^{*}=5$ plotted against (a) ${\lambda}_{x}^{+}$ and (b) ${\lambda}_{x}^{*}$. (c) and (d): The normalised premultiplied spanwise spectra of the fluctuating pressure  $k_{z}E_{ p^{\prime }p^{\prime }}/\bar{p}^{2}$ at $y^{*}=5$ plotted against (c) ${\lambda}_{z}^{+}$ and (d) ${\lambda}_{z}^{*}$.}
    \label{fig: d20}
\end{figure}

The normalised premultiplied streamwise and spanwise spectra of the fluctuating pressure $k_{x}E_{ p^{\prime }p^{\prime }}/\bar{p}^{2}$ and $k_{z}E_{ p^{\prime }p^{\prime }}/\bar{p}^{2}$ at $y^{*}=60$, $y^{*}=2$ and $y^{*}=5$ are shown in figure \ref{fig: d18}, figure \ref{fig: d19} and figure \ref{fig: d20} respectively. It is shown in figure \ref{fig: d18} that as the wall temperature decreases, the peak values of $k_{x}E_{ p^{\prime }p^{\prime }}/\bar{p}^{2}$ and $k_{z}E_{ p^{\prime }p^{\prime }}/\bar{p}^{2}$ are drastically enhanced, and the peak locations also increase under the wall unit $({\lambda}_{x}^{+}, {\lambda}_{z}^{+})$ at $y^{*}=60$. The semi-local scaling $({\lambda}_{x}^{*}, {\lambda}_{z}^{*})$ significantly reduces the disparity between the peak locations, and $k_{x}E_{ p^{\prime }p^{\prime }}/\bar{p}^{2}$ and $k_{z}E_{ p^{\prime }p^{\prime }}/\bar{p}^{2}$ achieve their peaks at ${\lambda}_{x}^{*} \approx 400$ and ${\lambda}_{z}^{*} \approx 250$ respectively at $y^{*}=60$.

However, in the near-wall region ($y^{*}=2$), the behaviours of the pressure spectra are quite different. It is found in figure \ref{fig: d19} that peak locations of $k_{x}E_{ p^{\prime }p^{\prime }}/\bar{p}^{2}$ and $k_{z}E_{ p^{\prime }p^{\prime }}/\bar{p}^{2}$ in the nearly adiabatic wall case ``M8T08'' at $y^{*}=2$ are significantly different from those in cooled wall cases ``M8T04'' and ``M8T015''. {This difference can be explained according to the instantaneous fields of the normalised fluctuating pressure $p^{\prime }/\bar{p}$ at $y^{*}=2$ as shown in figure \ref{fig: d17}. It is shown in figure \ref{fig: d17} (a) and (b) that the special structures marked by red dashed boxes are observed, and these structures are named as ``the travelling-wave-like alternating positive and negative structures'' (TAPNS). These structures are well organized as wavelike alternating  positive and negative patterns along the streamwsie direction, and have also been found in compressible channel flows \citep[]{Yu2019,Tang2020} and turbulent boundary layers \citep[]{Xu2021b,Zhang2022}. Moreover, the TAPNS disappear in nearly adiabatic wall case ``M8T08'', and are strongly enhanced as wall temperature decreases. Particularly, in the strongly cooled wall case ``M8T015'' (figure \ref{fig: d17} (a)), the TAPNS are prevalent in the whole field. The extreme positive and negative values of $p^{\prime }/\bar{p}$ are mainly located among the TAPNS, which further lead to the significant enhancement of $p _{rms}^{\prime}/\bar{p}$ near the wall (figure \ref{fig: d5} (a)).} The peak locations of $k_{x}E_{ p^{\prime }p^{\prime }}/\bar{p}^{2}$ and $k_{z}E_{ p^{\prime }p^{\prime }}/\bar{p}^{2}$ in the nearly adiabatic wall case ``M8T08'' are ${\lambda}_{x}^{*} \approx 300$ and ${\lambda}_{z}^{*} \approx 200$ respectively at $y^{*}=2$, which have similar aspect ratio ${\lambda}_{x}^{*}/{\lambda}_{z}^{*}$ to the characteristic scales at $y^{*}=60$. Nevertheless, the pressure spectra achieve their peaks at ${\lambda}_{x}^{*} \approx 120$ and ${\lambda}_{z}^{*} \approx 200$ in ``M8T04'', and ${\lambda}_{x}^{*} \approx 210$ and ${\lambda}_{z}^{*} \approx 250$ in ``M8T015''. As the wall temperature decreases from ``M8T04'' to ``M8T015'', the characteristic streamwise length scale ${\lambda}_{x}^{*}$ is significantly increased, while the characteristic spanwise spacing scale ${\lambda}_{z}^{*}$ is slightly enhanced, {which is similar to the behaviour of the instantaneous wavelike alternating positive and negative structures shown in figure \ref{fig: d17} (a) and (b).} Therefore, it can be inferred that the characteristic streamwise length scale ${\lambda}_{x}^{*}$ and spanwise spacing scale ${\lambda}_{z}^{*}$ of the cooled wall cases ``M8T04'' and ``M8T015'' represent the scales of the TAPNS. The TAPNS is short and fat (i.e. ${\lambda}_{x}^{*} < {\lambda}_{z}^{*})$. As the wall temperature decreases, the aspect ratio ${\lambda}_{x}^{*}/{\lambda}_{z}^{*}$ of TAPNS increases from almost 0.6 in ``M8T04'' to 0.84 in ``M8T015'', which indicates that the strongly cooled wall prefers to increase the streamwise length scale compared with the spanwise spacing scale of TAPNS. Moreover, the intensity of the TAPNS is strongly enhanced as the wall temperature decreases. It is noted that {these} wavelike alternating positive and negative patterns have also been reported in the fluctuating dilatation ${\theta }''\equiv \partial {u}''_{k}/\partial x_{k}$ in figure 6 of \citet[]{Xu2021b}. Therefore, it can be deduced that as the wall temperature decreases, the peak values of the pressure spectra and the turbulent intensities of the pressure are significantly enhanced in the near-wall region, which further lead to the enhanced compressibility near the wall \citep[]{Duan2010,Zhang2018,Xu2021a,Xu2021b,Zhang2022}. {Moreover, it is shown above in figure \ref{fig: d5} (a) that the maximum values of $p^{\prime }/\bar{p}$ appear at the wall in ``M8T04'' and ``M8T015''. This phenomenon is attributed to the appearance of TAPNS near the wall in the cooled wall cases.}

As the wall-normal location moves further away from the wall (at $y^{*}=5$), it is shown in figure \ref{fig: d20} that most of the peak locations of the streamwise and spanwise spectra are similar to those at $y^{*}=2$, except for the peak location of $k_{x}E_{ p^{\prime }p^{\prime }}/\bar{p}^{2}$ in ``M8T04''. It is found in figure \ref{fig: d20} (b) that the peak of $k_{x}E_{ p^{\prime }p^{\prime }}/\bar{p}^{2}$ in ``M8T04'' becomes much wider than that at $y^{*}=2$, and the characteristic streamwise length scale of $k_{x}E_{ p^{\prime }p^{\prime }}/\bar{p}^{2}$ in ``M8T08'' (${\lambda}_{x}^{*} \approx 300$) is also found in the streamwise spectra in ``M8T04''. Here for better description, the pressure structures with the characteristic streamwise length scale in ``M8T08'' are named as ``the base acoustic structures''. The above observation indicates that the base acoustic structures also exist in the cooled wall cases ``M8T04'' and ``M8T015''. If the wall is significantly cooled and the wall-normal location $y^{*}$ is {close} to the wall, the strength of the TAPNS is much larger than that of the base pressure structures, and only the characteristic streamwise length scale ${\lambda}_{x}^{*}$ of TAPNS is dominant in $k_{x}E_{ p^{\prime }p^{\prime }}/\bar{p}^{2}$ (i.e. ``M8T04'' and ``M8T015'' in figure \ref{fig: d19} (b) and ``M8T015'' in figure \ref{fig: d20} (b)). As the wall temperature increases and $y^{*}$ {moves} away from the wall, the intensity of the TAPNS becomes weaker. {Therefore, the TAPNS and the base acoustic structures have similar intensities in these situations}, which result in the wide peak in ``M8T04'' in figure \ref{fig: d20} (b). When the wall temperature is nearly adiabatic (i.e. ``M8T08''), the TAPNS disappear and only the base acoustic structures are dominant in the near-wall region.

\begin{figure}\centering
    \includegraphics[width=0.99\linewidth]{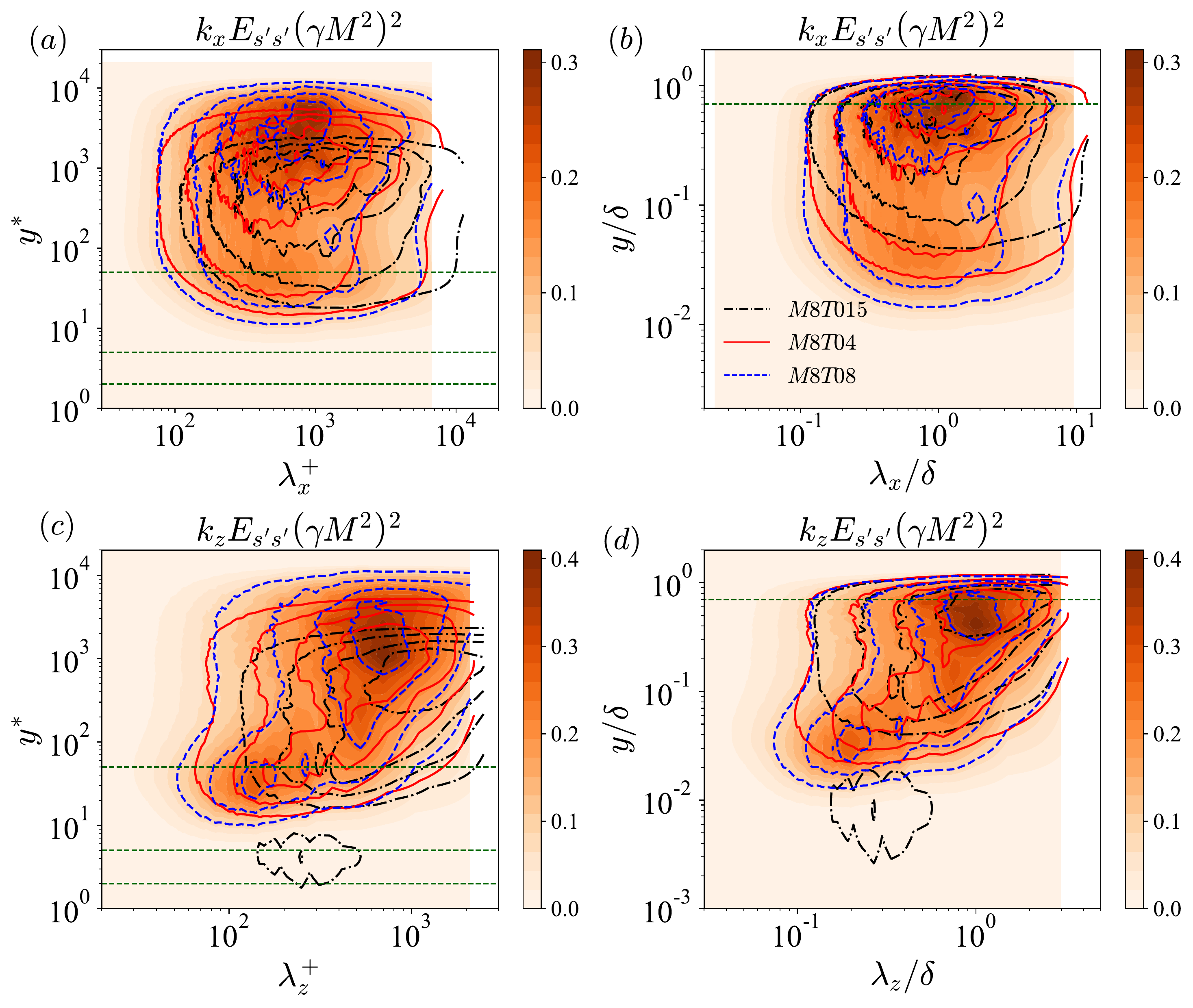}
    \caption{(a) and (b): The normalised premultiplied streamwise spectra of the fluctuating entropy $k_{x}E_{ s^{\prime }s^{\prime }}\left ( \gamma M^{2} \right )^{2}$ in (a) inner scaling and (b) outer scaling. (c) and (d): The normalised premultiplied spanwise spectra of the fluctuating entropy $k_{z}E_{ s^{\prime }s^{\prime }}\left ( \gamma M^{2} \right )^{2}$ in (c) inner scaling and (d) outer scaling. The {filled} contour represents the normalised premultiplied spectra in ``M8T08''. The line contour levels are (0.2, 0.4, 0.6, 0.8) times the peak values. The horizontal dashed lines represent $y^{*}=2, 5, 50$ in (a) (c) and $y/\delta=0.7$ in (b) (d) respectively.}
    \label{fig: d21}
\end{figure}

\begin{figure}\centering
    \includegraphics[width=0.99\linewidth]{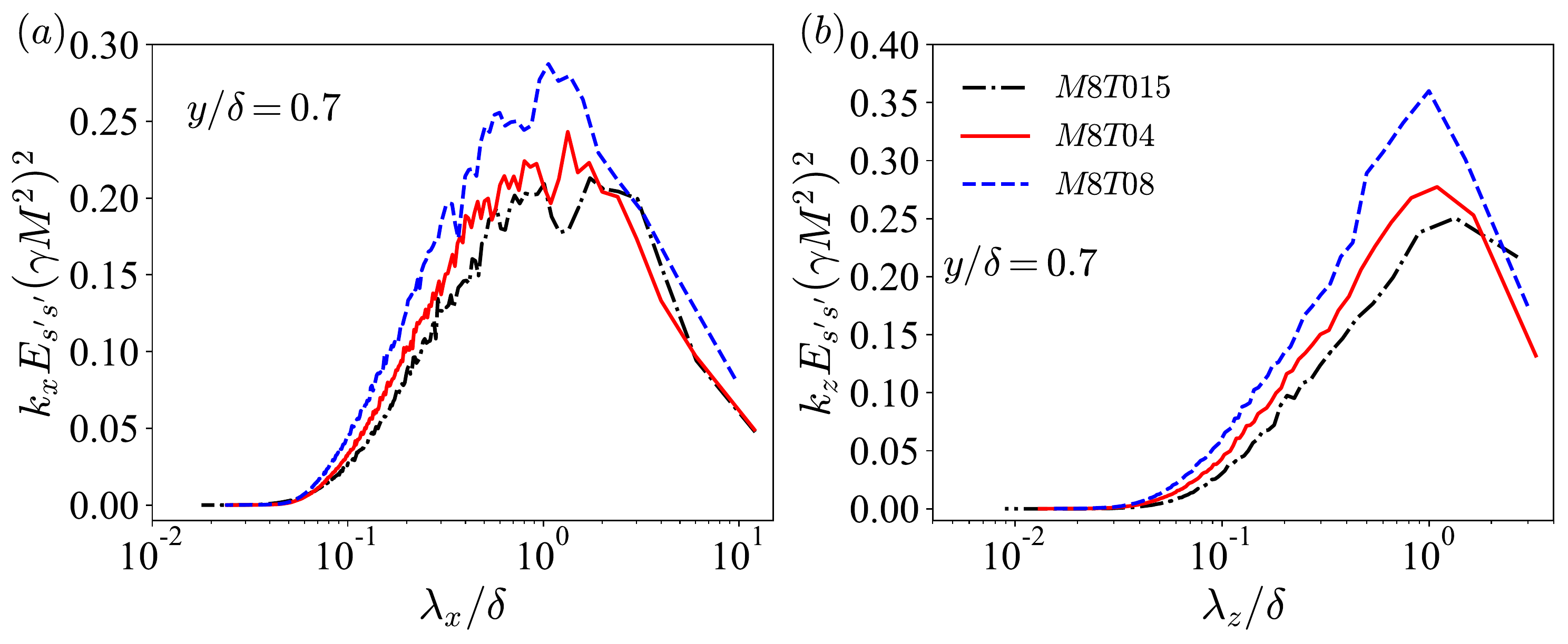}
    \caption{(a) The normalised premultiplied streamwise spectra of the fluctuating entropy $k_{x}E_{ s^{\prime }s^{\prime }}\left ( \gamma M^{2} \right )^{2}$ at $y/\delta=0.7$ plotted against ${\lambda}_{x}/\delta $. (b) The normalised premultiplied spanwise spectra of the fluctuating entropy $k_{z}E_{ s^{\prime }s^{\prime }}\left ( \gamma M^{2} \right )^{2}$ at $y/\delta=0.7$ plotted against ${\lambda}_{z}/\delta $.}
    \label{fig: d22}
\end{figure}

\begin{figure}\centering
    \includegraphics[width=0.99\linewidth]{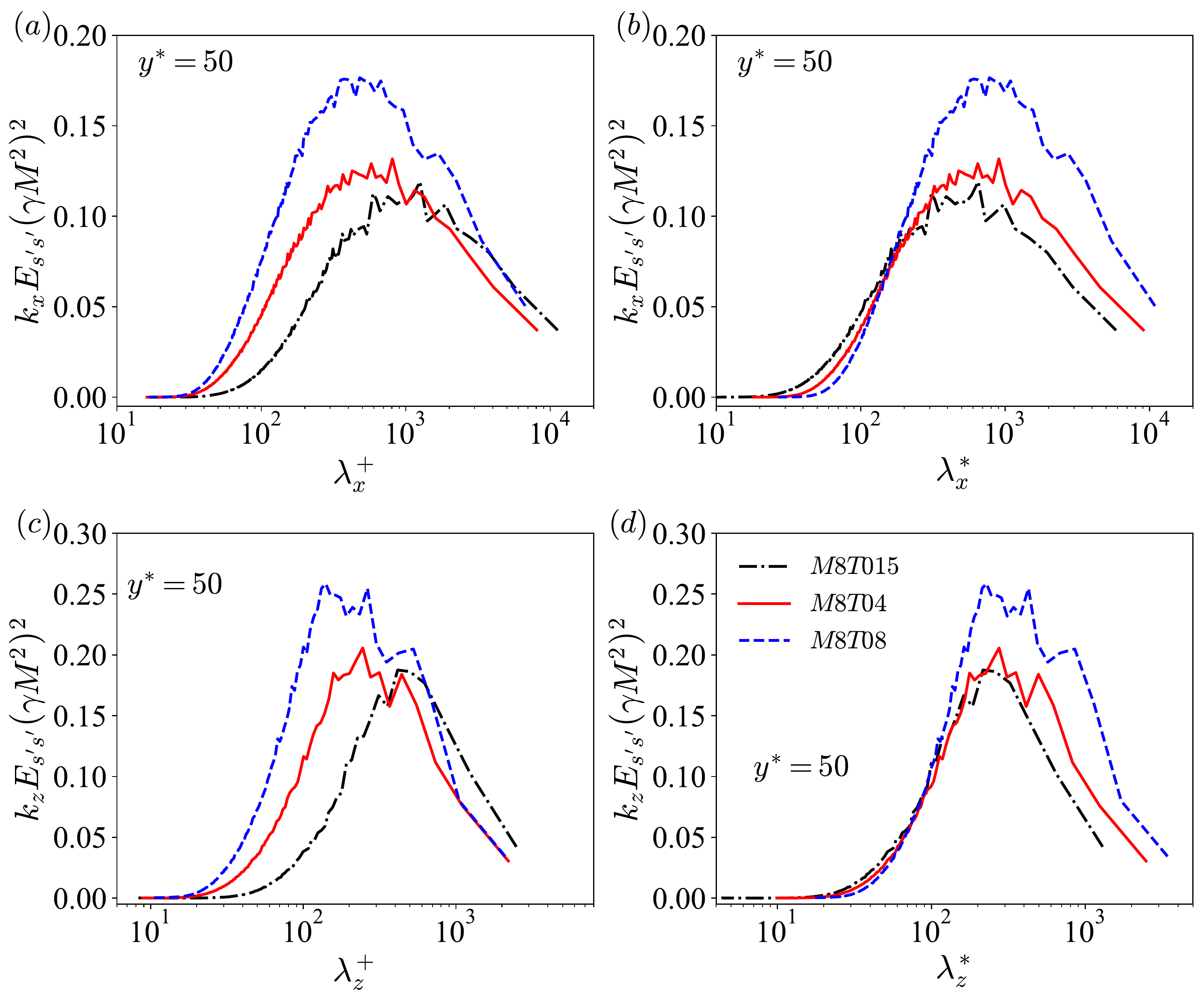}
    \caption{(a) and (b): The normalised premultiplied streamwise spectra of the fluctuating entropy $k_{x}E_{ s^{\prime }s^{\prime }}\left ( \gamma M^{2} \right )^{2}$  at $y^{*}=50$ plotted against (a) ${\lambda}_{x}^{+}$ and (b) ${\lambda}_{x}^{*}$. (c) and (d): The normalised premultiplied spanwise spectra of the fluctuating entropy $k_{z}E_{ s^{\prime }s^{\prime }}\left ( \gamma M^{2} \right )^{2}$ at $y^{*}=50$ plotted against (c) ${\lambda}_{z}^{+}$ and (d) ${\lambda}_{z}^{*}$.}
    \label{fig: d23}
\end{figure}

\begin{figure}\centering
    \includegraphics[width=0.99\linewidth]{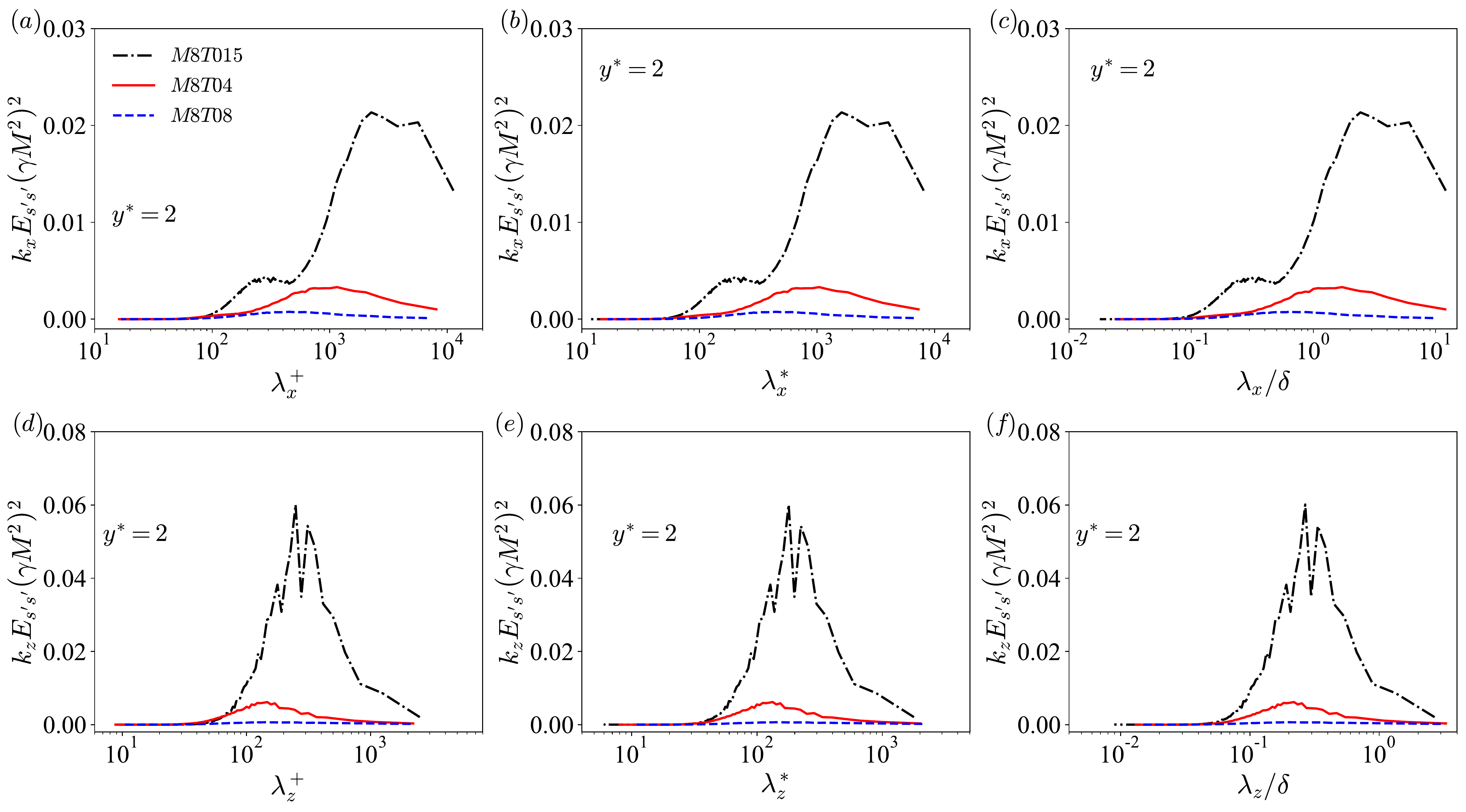}
    \caption{(a), (b) and (c): The normalised premultiplied streamwise spectra of the fluctuating entropy $k_{x}E_{ s^{\prime }s^{\prime }}\left ( \gamma M^{2} \right )^{2}$  at $y^{*}=2$ plotted against (a) ${\lambda}_{x}^{+}$, (b) ${\lambda}_{x}^{*}$ and (c) ${\lambda}_{x}/\delta $. (d), (e) and (f): The normalised premultiplied spanwise spectra of the fluctuating entropy $k_{z}E_{ s^{\prime }s^{\prime }}\left ( \gamma M^{2} \right )^{2}$ at $y^{*}=2$ plotted against (d) ${\lambda}_{z}^{+}$, (e) ${\lambda}_{z}^{*}$ and (f) ${\lambda}_{z}/\delta $.}
    \label{fig: d24}
\end{figure}

\begin{figure}\centering
    \includegraphics[width=0.99\linewidth]{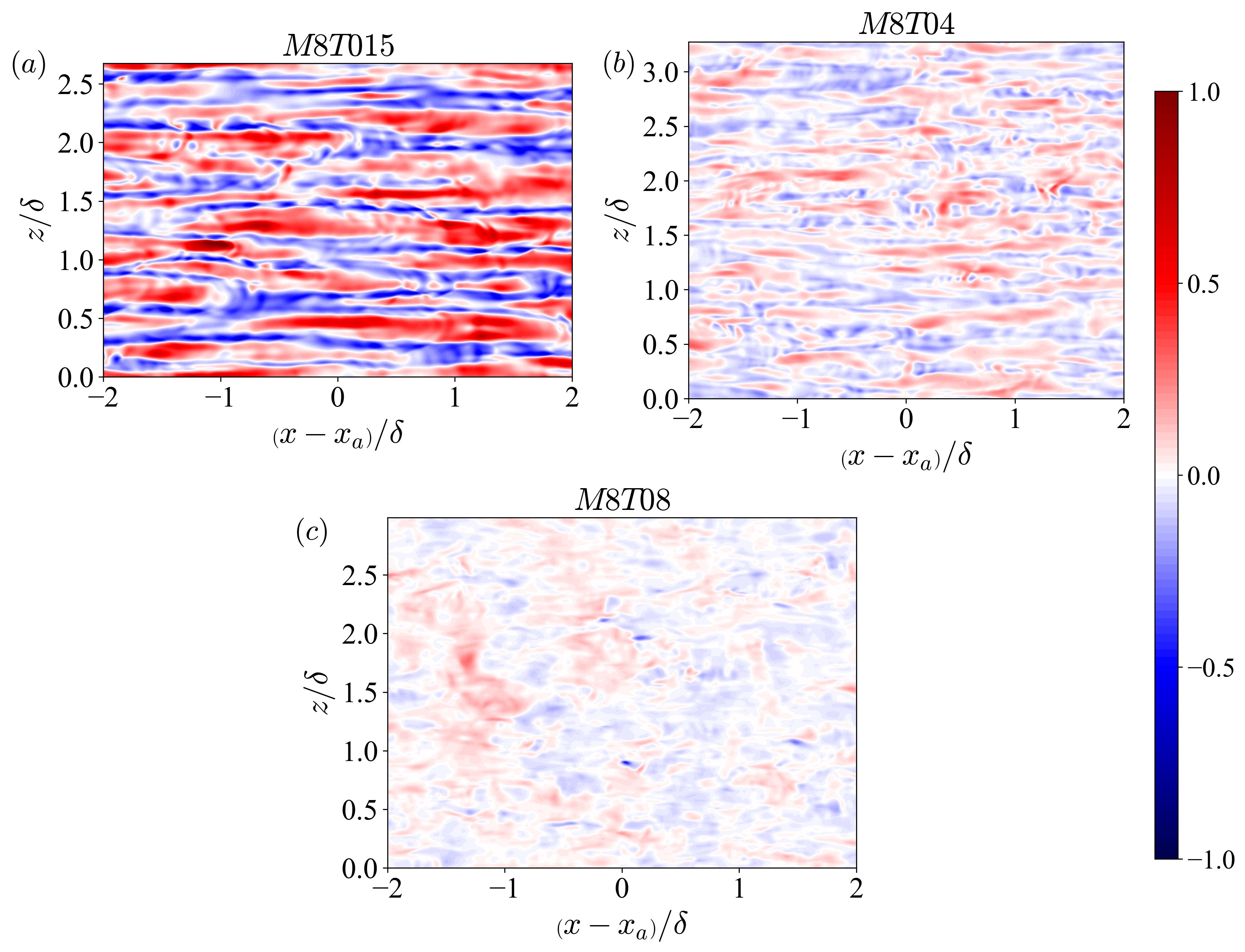}
    \caption{The instantaneous fields of the normalised fluctuating entropy $s^{\prime }\gamma M^{2}$ at $y^{*}=2$ in (a) ``M8T015'', (b) ``M8T04'' and (c) ``M8T08''.}
    \label{fig: d26}
\end{figure}

\begin{figure}\centering
    \includegraphics[width=0.99\linewidth]{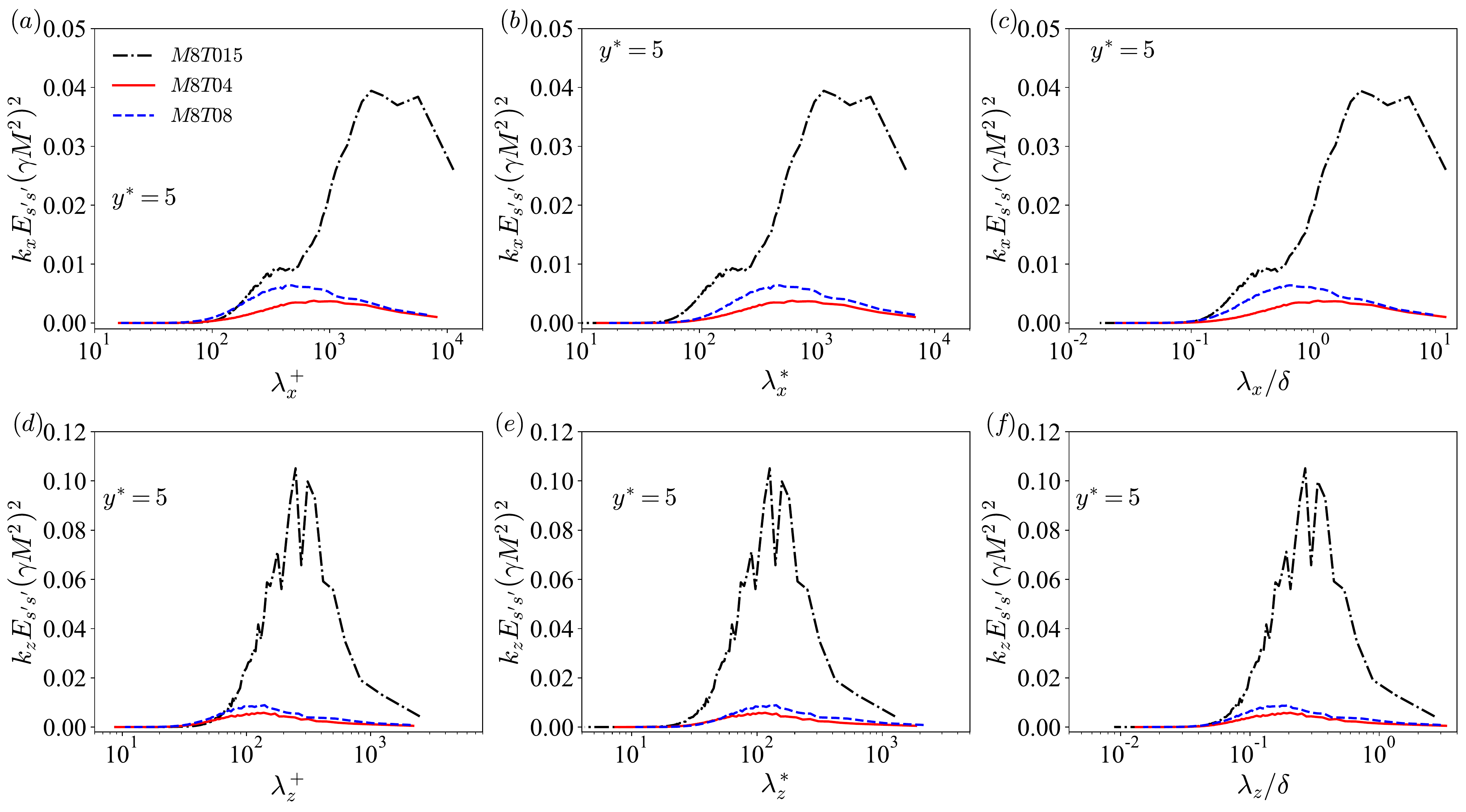}
    \caption{(a), (b) and (c): The normalised premultiplied streamwise spectra of the fluctuating entropy $k_{x}E_{ s^{\prime }s^{\prime }}\left ( \gamma M^{2} \right )^{2}$  at $y^{*}=5$ plotted against (a) ${\lambda}_{x}^{+}$, (b) ${\lambda}_{x}^{*}$ and (c) ${\lambda}_{x}/\delta $. (d), (e) and (f): The normalised premultiplied spanwise spectra of the fluctuating entropy $k_{z}E_{ s^{\prime }s^{\prime }}\left ( \gamma M^{2} \right )^{2}$ at $y^{*}=5$ plotted against (d) ${\lambda}_{z}^{+}$, (e) ${\lambda}_{z}^{*}$ and (f) ${\lambda}_{z}/\delta $.}
    \label{fig: d25}
\end{figure}

The normalised premultiplied streamwise and spanwise spectra of the fluctuating entropy $k_{x}E_{ s^{\prime }s^{\prime }}\left ( \gamma M^{2} \right )^{2}$ and $k_{z}E_{ s^{\prime }s^{\prime }}\left ( \gamma M^{2} \right )^{2}$ are depicted in figure \ref{fig: d21}. The $k_{x}E_{ s^{\prime }s^{\prime }}\left ( \gamma M^{2} \right )^{2}$ and $k_{z}E_{ s^{\prime }s^{\prime }}\left ( \gamma M^{2} \right )^{2}$ achieve their {primary} peaks near the edge of the boundary layer, which are consistent with the {primary} peak location of $s _{rms}^{\prime}\gamma M^{2}$ (figure \ref{fig: d6} (d)). The {secondary} peaks of entropy spectra appear at $y^{*}\approx 50$ in ``M8T08'' and ``M8T04'', which are coincident with the {secondary} peak location of $s _{rms}^{\prime}\gamma M^{2}$ (figure \ref{fig: d6} (a)). However, for the strongly cooled wall case ``M8T015'', the {secondary} peaks of entropy spectra appear at nearly $y^{*}=5$. The normalised premultiplied streamwise and spanwise spectra of the fluctuating entropy $k_{x}E_{ s^{\prime }s^{\prime }}\left ( \gamma M^{2} \right )^{2}$ and $k_{z}E_{ s^{\prime }s^{\prime }}\left ( \gamma M^{2} \right )^{2}$ at $y/\delta=0.7$, $y^{*}=50$, $y^{*}=2$ and $y^{*}=5$ are plotted in figure \ref{fig: d22}, figure \ref{fig: d23}, figure \ref{fig: d24} and figure \ref{fig: d25} respectively. It is shown in figure \ref{fig: d22} that $k_{x}E_{ s^{\prime }s^{\prime }}\left ( \gamma M^{2} \right )^{2}$ and $k_{z}E_{ s^{\prime }s^{\prime }}\left ( \gamma M^{2} \right )^{2}$ attain their peaks at ${\lambda}_{x}/\delta \approx 1.3$ and ${\lambda}_{z}/\delta \approx 1$ respectively. Furthermore, the peak values of $k_{x}E_{ s^{\prime }s^{\prime }}\left ( \gamma M^{2} \right )^{2}$ and $k_{z}E_{ s^{\prime }s^{\prime }}\left ( \gamma M^{2} \right )^{2}$ decrease as the wall temperature decreases. It can be seen in figure \ref{fig: d23} that the semi-local scaling can reduce the disparity between the peak locations of spectra. The $k_{x}E_{ s^{\prime }s^{\prime }}\left ( \gamma M^{2} \right )^{2}$ and $k_{z}E_{ s^{\prime }s^{\prime }}\left ( \gamma M^{2} \right )^{2}$ achieve their peaks at ${\lambda}_{x}^{*} \approx 700$ and ${\lambda}_{z}^{*} \approx 250$ respectively at $y^{*}=50$. {Furthermore, the peak values of the normalised premultiplied streamwise and spanwise spectra of the fluctuating entropy decrease as the wall temperature decreases, which are consistent with the observation that the values of $s _{rms}^{\prime}\gamma M^{2}$ decrease as the wall temperature decreases at $y^{*}=50$ (figure \ref{fig: d6} (a)).}

    {It is shown in figure \ref{fig: d6} (a) that a local secondary peak of $s _{rms}^{\prime}\gamma M^{2}$ appears near the wall in ``M8T015''. The underlying mechanism of this phenomenon is revealed as follows.} It is found in figure \ref{fig: d24} that the peak values of the entropy spectra in ``M8T04'' and ``M8T015'' are significantly larger than those in ``M8T08'', and the peak values increase as the wall temperature decreases. {Furthermore, the $k_{x}E_{ s^{\prime }s^{\prime }}\left ( \gamma M^{2} \right )^{2}$ in ``M8T015'' and ``M8T04'' attain their peaks at ${\lambda}_{x}/\delta \approx 2.3$ and 1.3 respectively, and the peak locations of the $k_{z}E_{ s^{\prime }s^{\prime }}\left ( \gamma M^{2} \right )^{2}$ in ``M8T015'' and ``M8T04'' are ${\lambda}_{z}/\delta  \approx 0.25$ and 0.2 respectively. The underlying structures of the characteristic scales of the entropy spectra are revealed by the instantaneous fields of the normalised fluctuating entropy $s^{\prime }\gamma M^{2}$ at $y^{*}=2$ shown in figure \ref{fig: d26}. It can be seen in figure \ref{fig: d26} (a) and (b) that the long and thin fluctuating entropy structures only appear when the wall is cooled (``M8T015'' and ``M8T04''). These special entropic structures exhibit the streaky patterns with alternating stripes of the high and low entropy, which is similar to the streaks of $u^{\prime }$ (shown in figure 5 (a) in \citet[]{Xu2021a}) with $R\left ( s ^{\prime},u ^{\prime} \right ) = 0.93$ at $y^{*}=2$ in ``M8T015''. However, the entropy structures in the nearly adiabatic wall case ``M8T08'' are pretty weak and fragmented. Here for the sake of description, the entropy structures in ``M8T08'' are named as ``the base entropic structures'', while the generated streaky patterns in ``M8T04'' and ``M8T015'' are called ``the streaky entropic structures'' (SES).  It is shown in figure \ref{fig: d21} that the characteristic length scale is significantly larger than the characteristic spanwise spacing scale (i.e. ${\lambda}_{x}/\delta  \gg  {\lambda}_{z}/\delta $) in ``M8T015'' and ``M8T04'', which is consistent with the long and thin nature of SES. Furthermore, the ${\lambda}_{x}/\delta$ and ${\lambda}_{z}/\delta $ in ``M8T015'' are larger than those in ``M8T04'', which is coincident with the observation that the SES become longer in the streamwise direction and fatter in the spanwise direction as the wall temperature decreases (figure \ref{fig: d26} (a) and (b)). Accordingly, it can be inferred that the characteristic streamwise length scale ${\lambda}_{x}/\delta $ and spanwise spacing scale ${\lambda}_{z}/\delta $ of the entropy spectra in ``M8T015'' and ``M8T04'' represent the scales of the SES. Furthermore, the aspect ratio ${\lambda}_{x}/{\lambda}_{z}$ of SES increases from 6.5 in ``M8T04'' to 9.2 in ``M8T015'', suggesting that the strongly cooled wall prefers to increase the streamwise length scale compared with the spanwise spacing scale of SES. Moreover, the intensity of the SES is significantly enhanced as the wall temperature decreases, which further leads to the significant enhancement of $s _{rms}^{\prime}\gamma M^{2}$ near the wall in the cooled wall cases (figure \ref{fig: d6} (a)). Specifically, the local secondary peak of $s _{rms}^{\prime}\gamma M^{2}$ near the wall in ``M8T015'' is attributed to the strong intensity of the SES. }

As the wall-normal location increases to $y^{*}=5$, it is seen in figure \ref{fig: d25} that the entropy spectra in ``M8T08'' and ``M8T04'' have pretty small peak values, and attain their peaks at ${\lambda}_{x}^{*} \approx 450$ and ${\lambda}_{z}^{*} \approx 140$ respectively at $y^{*}=5$, which have a similar characteristic aspect ratio ${\lambda}_{x}^{*}/{\lambda}_{z}^{*}$ with the characteristic scales at $y^{*}=50$. Furthermore, the peak values of the entropy spectra in ``M8T08'' and ``M8T04'' decrease as the wall temperature decreases, which are coincident with the behaviours at $y^{*}=50$ and $y/\delta=0.7$. {However, the peak values of the entropy spectra in strongly cooled wall case ``M8T015'' are significantly larger than those in other two cases. The entropy spectra in ``M8T015'' attain their peaks at ${\lambda}_{x}/\delta \approx 2.3$ and ${\lambda}_{z}/\delta  \approx 0.25$ respectively, which are similar to the characteristic scales at $y^{*}=2$. The above observations indicate that the SES exist in ``M8T015'', while disappear in ``M8T04'' at $y^{*}=5$. Therefore, it is implied that the SES can exist in a larger range of wall-normal distance $y^{*}$ as the wall temperature decreases.}

\begin{figure}\centering
    \includegraphics[width=0.99\linewidth]{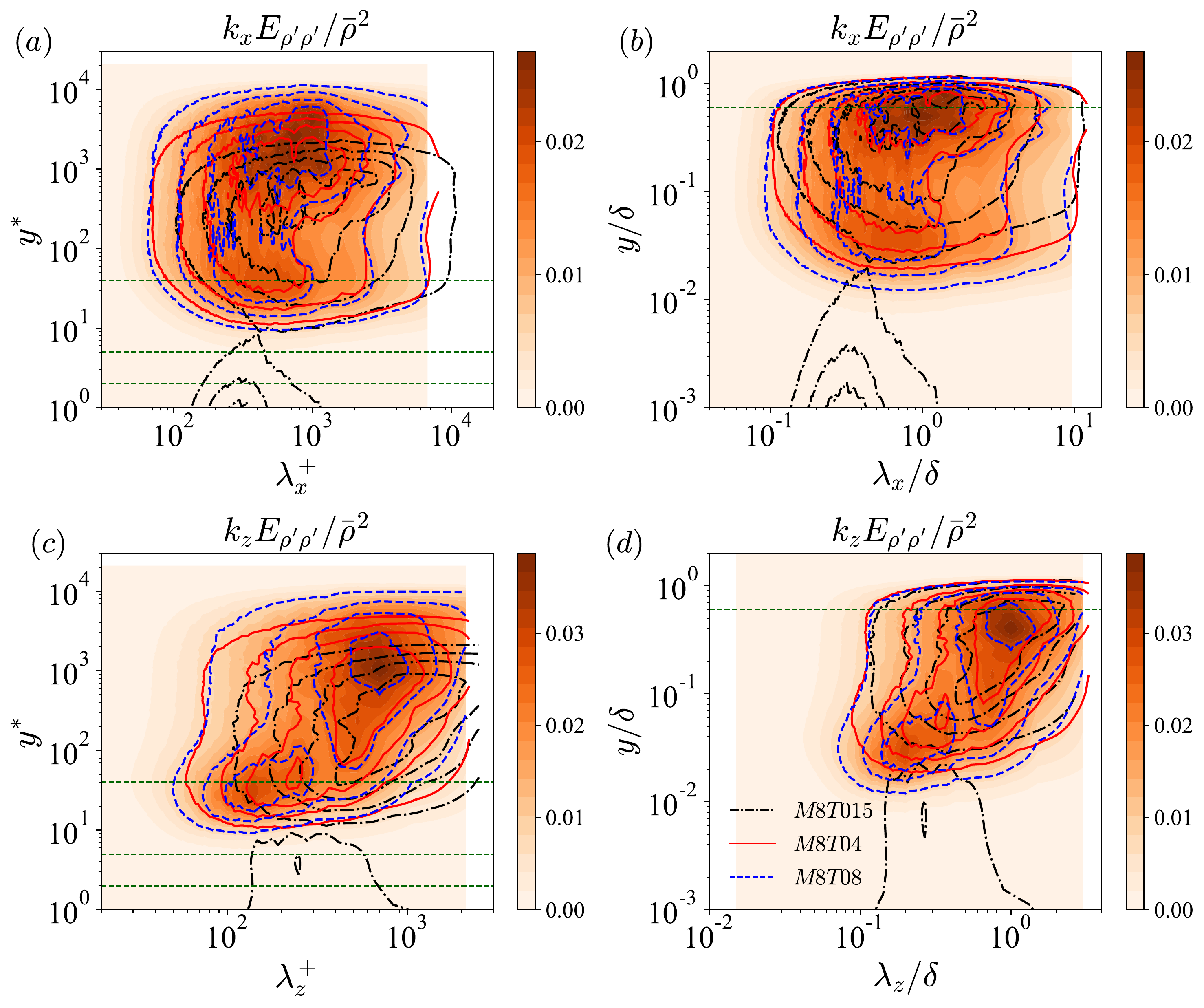}
    \caption{(a) and (b): The normalised premultiplied streamwise spectra of the fluctuating density $k_{x}E_{ \rho^{\prime }\rho^{\prime }}/\bar{\rho }^{2}$ in (a) inner scaling and (b) outer scaling. (c) and (d): The normalised premultiplied spanwise spectra of the fluctuating density $k_{z}E_{ \rho^{\prime }\rho^{\prime }}/\bar{\rho }^{2}$ in (c) inner scaling and (d) outer scaling. The {filled} contour represents the normalised premultiplied spectra in ``M8T08''. The line contour levels are (0.2, 0.4, 0.6, 0.8) times the peak values. The horizontal dashed lines represent $y^{*}=2, 5, 40$ in (a) (c) and $y/\delta=0.6$ in (b) (d) respectively.}
    \label{fig: d27}
\end{figure}

The normalised premultiplied streamwise and spanwise spectra of the fluctuating density $k_{x}E_{ \rho^{\prime }\rho^{\prime }}/\bar{\rho }^{2}$ and $k_{z}E_{ \rho^{\prime }\rho^{\prime }}/\bar{\rho }^{2}$ are shown in figure \ref{fig: d27}. It is found that the spectra of the fluctuating density are similar to those of the fluctuating entropy at $y^{*}>20$, which is consistent with the observation in figure \ref{fig: d12} (a) that the fluctuating density is dominated by its entropic mode at $y^{*}>20$. The $k_{x}E_{ \rho^{\prime }\rho^{\prime }}/\bar{\rho }^{2}$ and $k_{z}E_{ \rho^{\prime }\rho^{\prime }}/\bar{\rho }^{2}$ in three cases achieve their {primary} peaks at $y/\delta \approx 0.6$, and the {secondary} peak locations of the density spectra in ``M8T08'' and ``M8T04'' are  $y^{*} \approx 40$, which are coincident with peak locations of $\rho _{rms}^{\prime}/\bar{\rho }$ (figure \ref{fig: d6} (b) (e)). However, a complicated behaviour of the density spectra appears in ``M8T015'', mainly due to the strong variation of the relative contribution $\rho _{E,rms}^{\prime}/\left (\rho _{E,rms}^{\prime}+ \rho _{I,rms}^{\prime} \right )$ at $y^{*}<20$ with strongly cooled wall (figure \ref{fig: d12} (a)).

\begin{figure}\centering
    \includegraphics[width=0.99\linewidth]{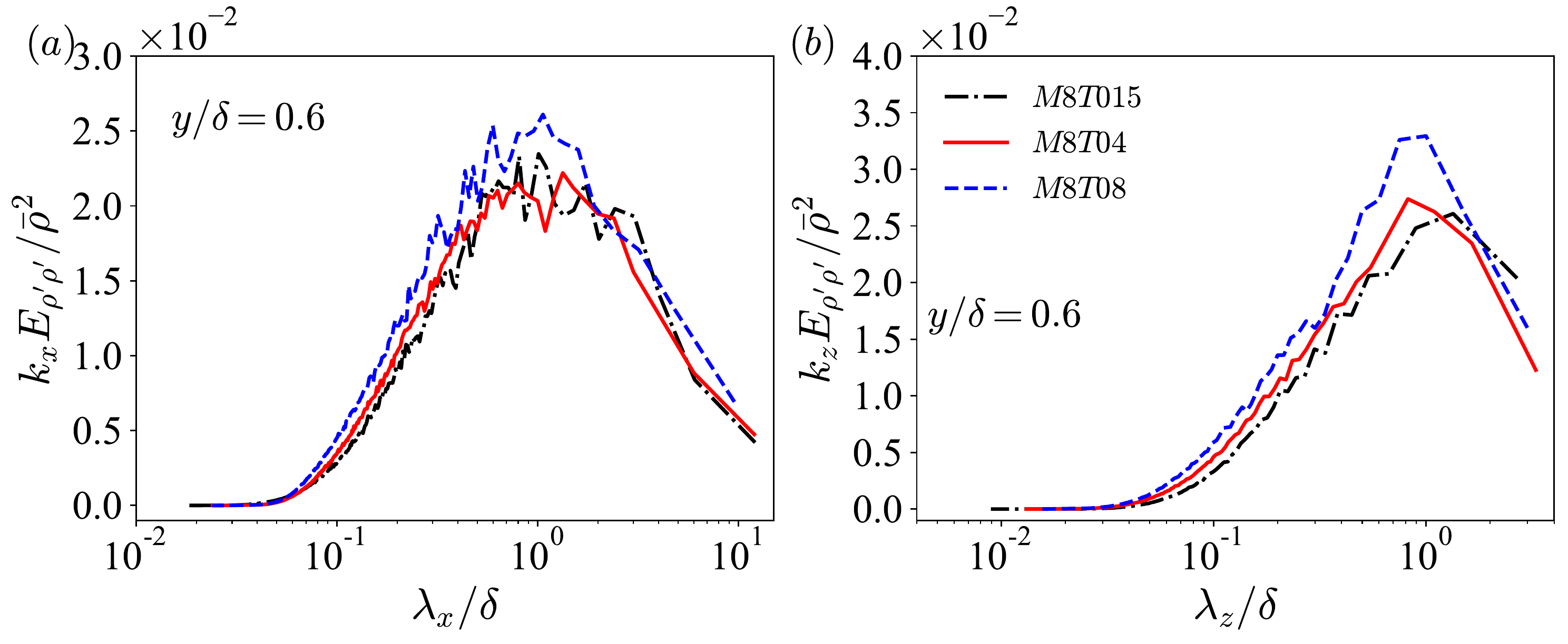}
    \caption{(a) The normalised premultiplied streamwise spectra of the fluctuating density $k_{x}E_{ \rho^{\prime }\rho^{\prime }}/\bar{\rho }^{2}$ at $y/\delta=0.6$ plotted against ${\lambda}_{x}/\delta $. (b) The normalised premultiplied spanwise spectra of the fluctuating density $k_{z}E_{ \rho^{\prime }\rho^{\prime }}/\bar{\rho }^{2}$ at $y/\delta=0.6$ plotted against ${\lambda}_{z}/\delta $.}
    \label{fig: d28}
\end{figure}

\begin{figure}\centering
    \includegraphics[width=0.99\linewidth]{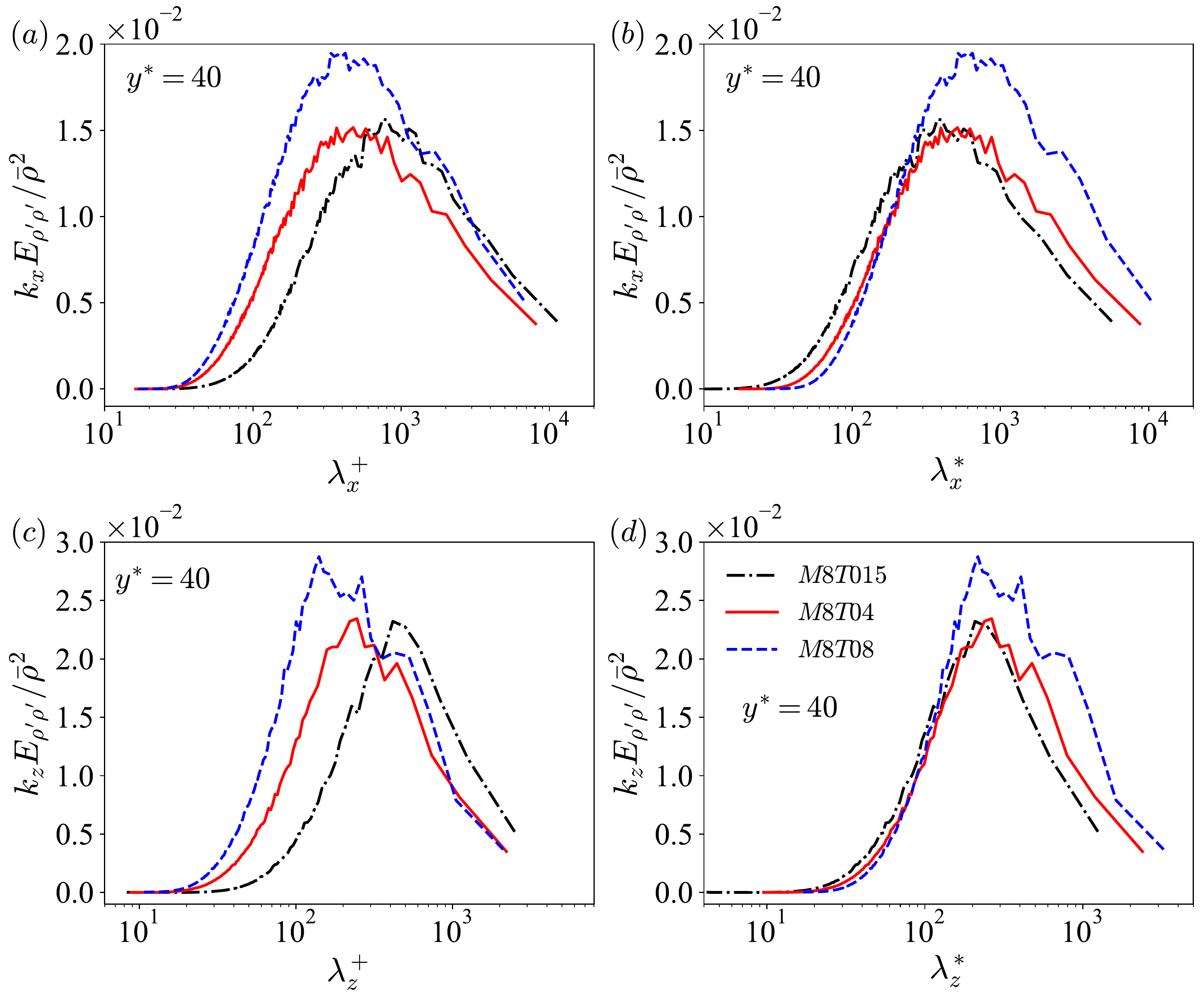}
    \caption{(a) and (b): The normalised premultiplied streamwise spectra of the fluctuating density $k_{x}E_{ \rho^{\prime }\rho^{\prime }}/\bar{\rho }^{2}$  at $y^{*}=40$ plotted against (a) ${\lambda}_{x}^{+}$ and (b) ${\lambda}_{x}^{*}$. (c) and (d): The normalised premultiplied spanwise spectra of the fluctuating density $k_{z}E_{ \rho^{\prime }\rho^{\prime }}/\bar{\rho }^{2}$ at $y^{*}=40$ plotted against (c) ${\lambda}_{z}^{+}$ and (d) ${\lambda}_{z}^{*}$.}
    \label{fig: d29}
\end{figure}

The normalised premultiplied streamwise and spanwise spectra of the fluctuating density $k_{x}E_{ \rho^{\prime }\rho^{\prime }}/\bar{\rho }^{2}$ and $k_{z}E_{ \rho^{\prime }\rho^{\prime }}/\bar{\rho }^{2}$ at $y/\delta=0.6$ and $y^{*}=40$ are depicted in figure \ref{fig: d28} and figure \ref{fig: d29} respectively. The $k_{x}E_{ \rho^{\prime }\rho^{\prime }}/\bar{\rho }^{2}$ and $k_{z}E_{ \rho^{\prime }\rho^{\prime }}/\bar{\rho }^{2}$ achieve their peaks at ${\lambda}_{x}/\delta \approx 1.3$ and ${\lambda}_{z}/\delta \approx 1$ respectively at $y/\delta=0.6$ (figure \ref{fig: d28}), and the peak locations of the density spectra are ${\lambda}_{x}^{*} \approx 700$ and ${\lambda}_{z}^{*} \approx 250$ respectively at $y^{*}=40$ (figure \ref{fig: d29}). It is noted that the peak locations of the density spectra are totally similar to those of the entropy spectra (figure \ref{fig: d22} and \ref{fig: d23}), which are mainly due to the reason that the fluctuating density is dominated by its entropic mode at $y^{*}>20$ (figure \ref{fig: d12} (a)).

\begin{figure}\centering
    \includegraphics[width=0.99\linewidth]{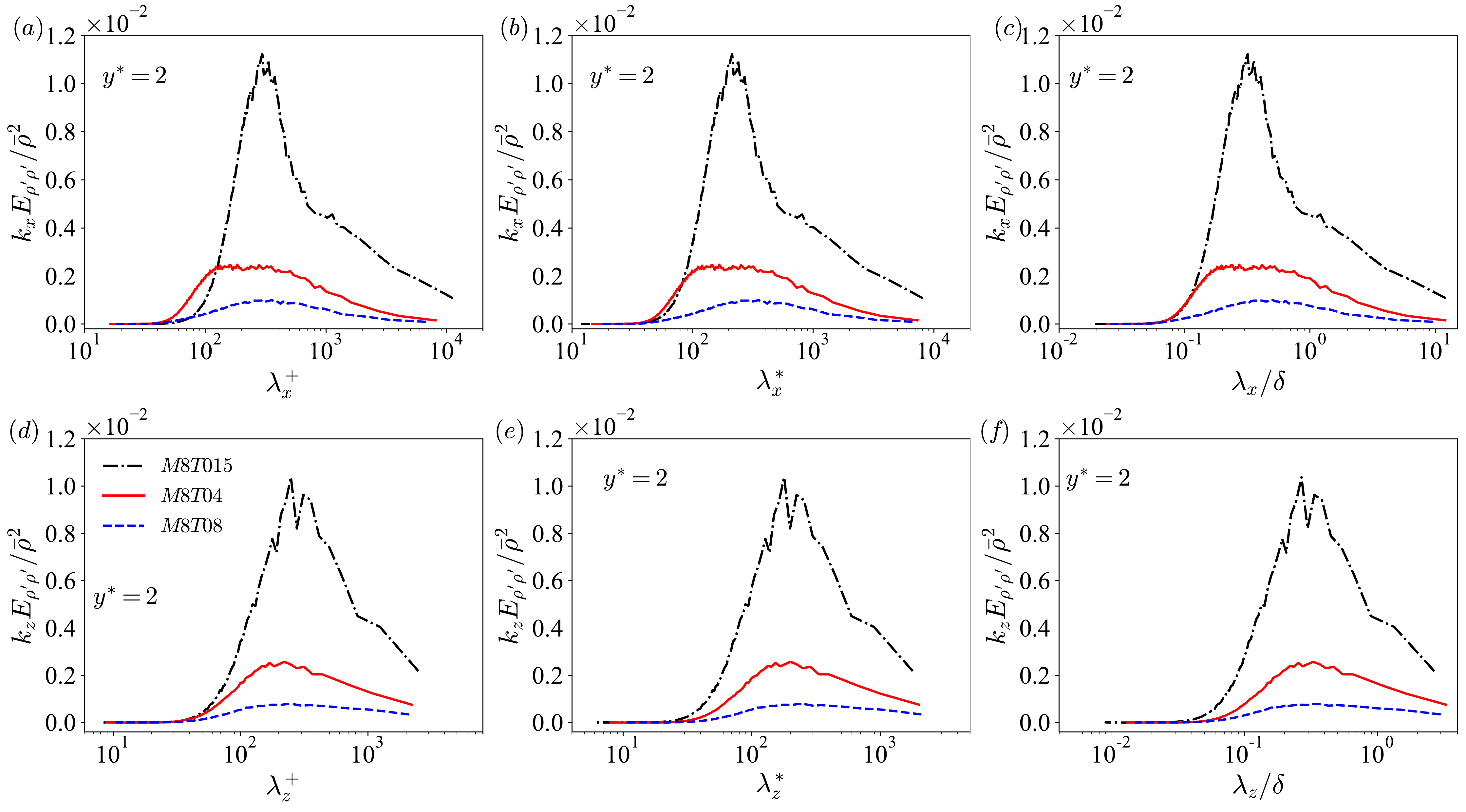}
    \caption{(a), (b) and (c): The normalised premultiplied streamwise spectra of the fluctuating density $k_{x}E_{ \rho^{\prime }\rho^{\prime }}/\bar{\rho }^{2}$  at $y^{*}=2$ plotted against (a) ${\lambda}_{x}^{+}$, (b) ${\lambda}_{x}^{*}$ and (c) ${\lambda}_{x}/\delta $. (d), (e) and (f): The normalised premultiplied spanwise spectra of the fluctuating density $k_{z}E_{ \rho^{\prime }\rho^{\prime }}/\bar{\rho }^{2}$ at $y^{*}=2$ plotted against (d) ${\lambda}_{z}^{+}$, (e) ${\lambda}_{z}^{*}$ and (f) ${\lambda}_{z}/\delta $.}
    \label{fig: d30}
\end{figure}

\begin{figure}\centering
    \includegraphics[width=0.99\linewidth]{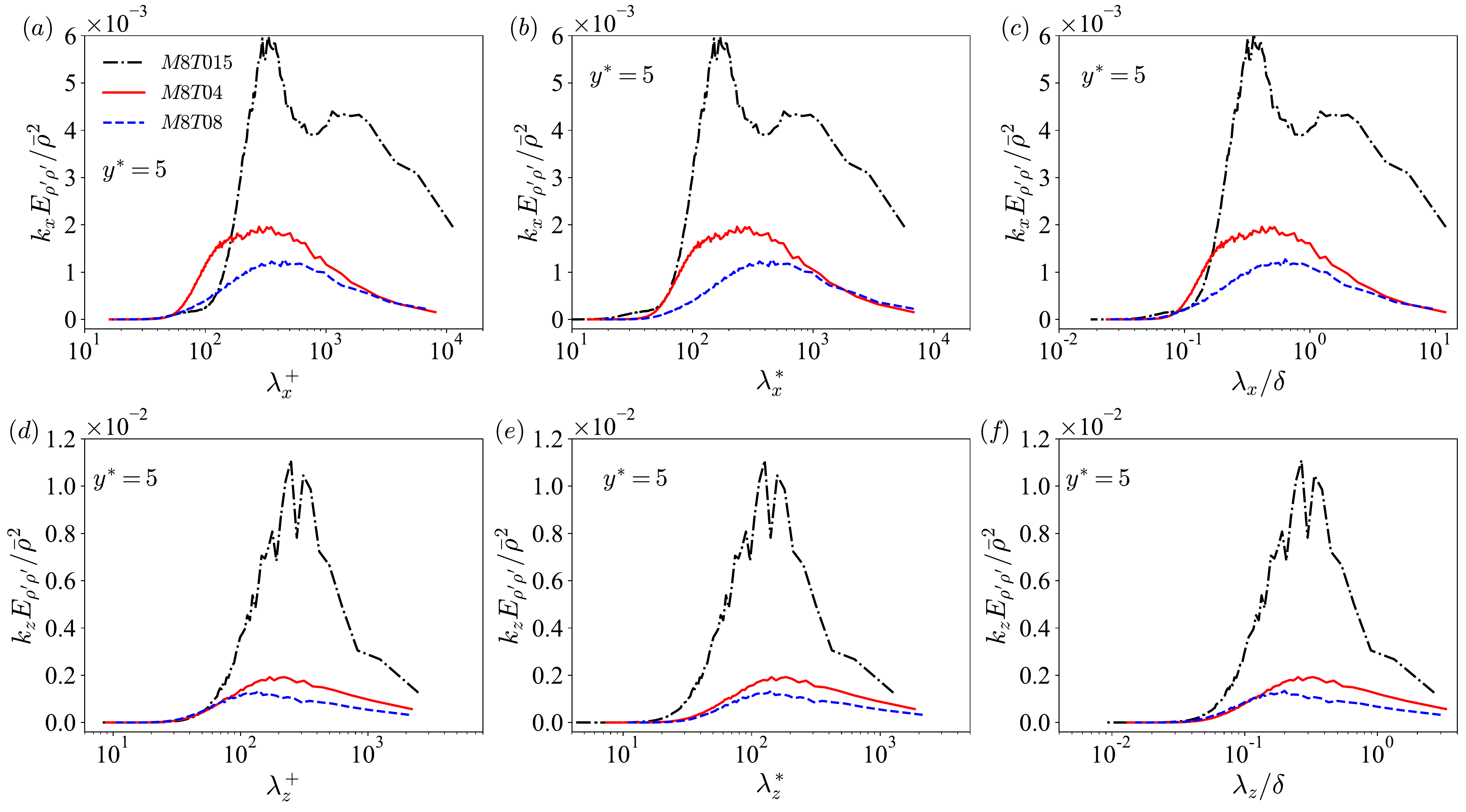}
    \caption{(a), (b) and (c): The normalised premultiplied streamwise spectra of the fluctuating density $k_{x}E_{ \rho^{\prime }\rho^{\prime }}/\bar{\rho }^{2}$  at $y^{*}=5$ plotted against (a) ${\lambda}_{x}^{+}$, (b) ${\lambda}_{x}^{*}$ and (c) ${\lambda}_{x}/\delta $. (d), (e) and (f): The normalised premultiplied spanwise spectra of the fluctuating density $k_{z}E_{ \rho^{\prime }\rho^{\prime }}/\bar{\rho }^{2}$ at $y^{*}=5$ plotted against (d) ${\lambda}_{z}^{+}$, (e) ${\lambda}_{z}^{*}$ and (f) ${\lambda}_{z}/\delta $.}
    \label{fig: d31}
\end{figure}

The normalised premultiplied streamwise and spanwise spectra of the fluctuating density $k_{x}E_{ \rho^{\prime }\rho^{\prime }}/\bar{\rho }^{2}$ and $k_{z}E_{ \rho^{\prime }\rho^{\prime }}/\bar{\rho }^{2}$ at $y^{*}=2$ and $y^{*}=5$ are depicted in figure \ref{fig: d30} and figure \ref{fig: d31} respectively.

{At $y^{*}=2$ (figure \ref{fig: d30}), the peak locations of the density spectra are similar to those of the pressure spectra at $y^{*}=2$ in three cases, suggesting that the acoustic mode of density is dominant in the fluctuating density at $y^{*}=2$. However, the values of the density spectra at large values of ${\lambda}_{x}/\delta $ and ${\lambda}_{z}/\delta $ are much larger than those of the pressure spectra at $y^{*}=2$ in ``M8T04'' and ``M8T015'', which indicate that the relative contribution of the entropic mode to the fluctuating density is enhanced at $y^{*}=2$ when the wall is cooled.}

{The enhancement of the relative contribution of the entropic mode to the fluctuating density in strongly cooled wall case becomes much more significant at $y^{*}=5$. It is found in figure \ref{fig: d31} that the peak locations of the density spectra are also similar to those of the pressure spectra at $y^{*}=5$ in three cases, indicating that the acoustic mode of density still has a major contribution to the fluctuating density at $y^{*}=5$. However, a secondary peak of $k_{x}E_{ \rho^{\prime }\rho^{\prime }}/\bar{\rho }^{2}$ appears at ${\lambda}_{x}/\delta \approx 2.3$ in ``M8T015'', suggesting that the relative contribution of the entropic mode to the fluctuating density becomes significantly larger in the strongly cooled wall case ``M8T015''. It is shown in figure \ref{fig: d12} (a) that the relative contribution $\rho _{E,rms}^{\prime}/\left (\rho _{E,rms}^{\prime}+ \rho _{I,rms}^{\prime} \right )$ in ``M8T015'' has a hump marked by the green dashed box. Here, the density spectra at $y^{*}=5$ in ``M8T015'' reveal that the hump is mainly due to the appearance of SES when the wall is strongly cooled. The SES significantly enhance the intensity of the entropic mode of density, which further enhance the relative contribution of the entropic mode to the fluctuating density.
}

\begin{figure}\centering
    \includegraphics[width=0.99\linewidth]{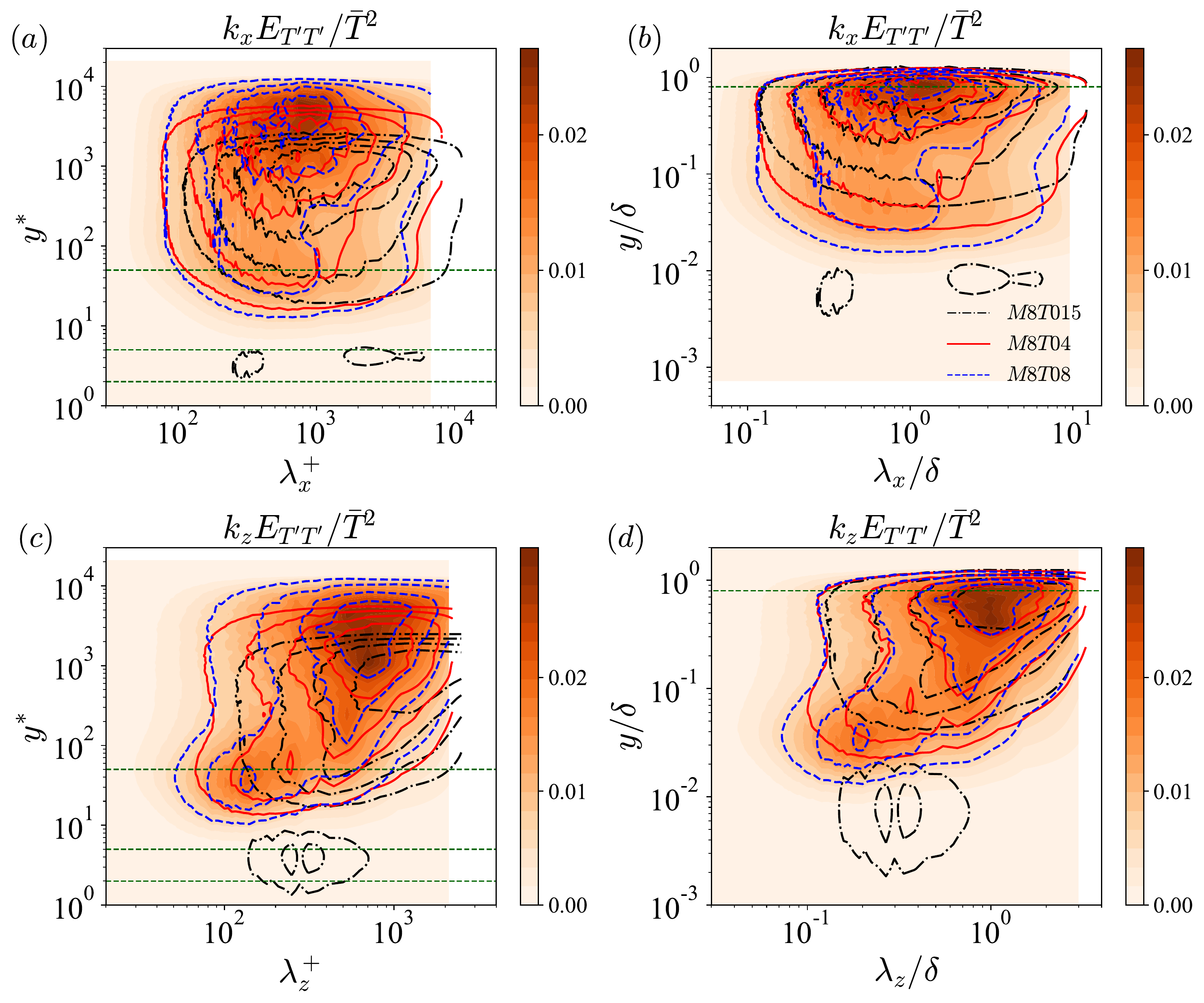}
    \caption{(a) and (b): The normalised premultiplied streamwise spectra of the fluctuating temperature $k_{x}E_{ T^{\prime }T^{\prime }}/\bar{T}^{2}$ in (a) inner scaling and (b) outer scaling. (c) and (d): The normalised premultiplied spanwise spectra of the fluctuating temperature $k_{z}E_{ T^{\prime }T^{\prime }}/\bar{T}^{2}$ in (c) inner scaling and (d) outer scaling. The {filled} contour represents the normalised premultiplied spectra in ``M8T08''. The line contour levels are (0.2, 0.4, 0.6, 0.8) times the peak values. The horizontal dashed lines represent $y^{*}=2, 5, 50$ in (a) (c) and $y/\delta=0.8$ in (b) (d) respectively.}
    \label{fig: d32}
\end{figure}

The normalised premultiplied streamwise and spanwise spectra of the fluctuating temperature $k_{x}E_{ T^{\prime }T^{\prime }}/\bar{T}^{2}$ and $k_{z}E_{ T^{\prime }T^{\prime }}/\bar{T}^{2}$ are shown in figure \ref{fig: d32}. It is found that the temperature spectra are similar to the entropy spectra (figure \ref{fig: d21}), indicating the dominant contribution of the entropic mode of temperature as shown in figure \ref{fig: d12} (b).

\begin{figure}\centering
    \includegraphics[width=0.99\linewidth]{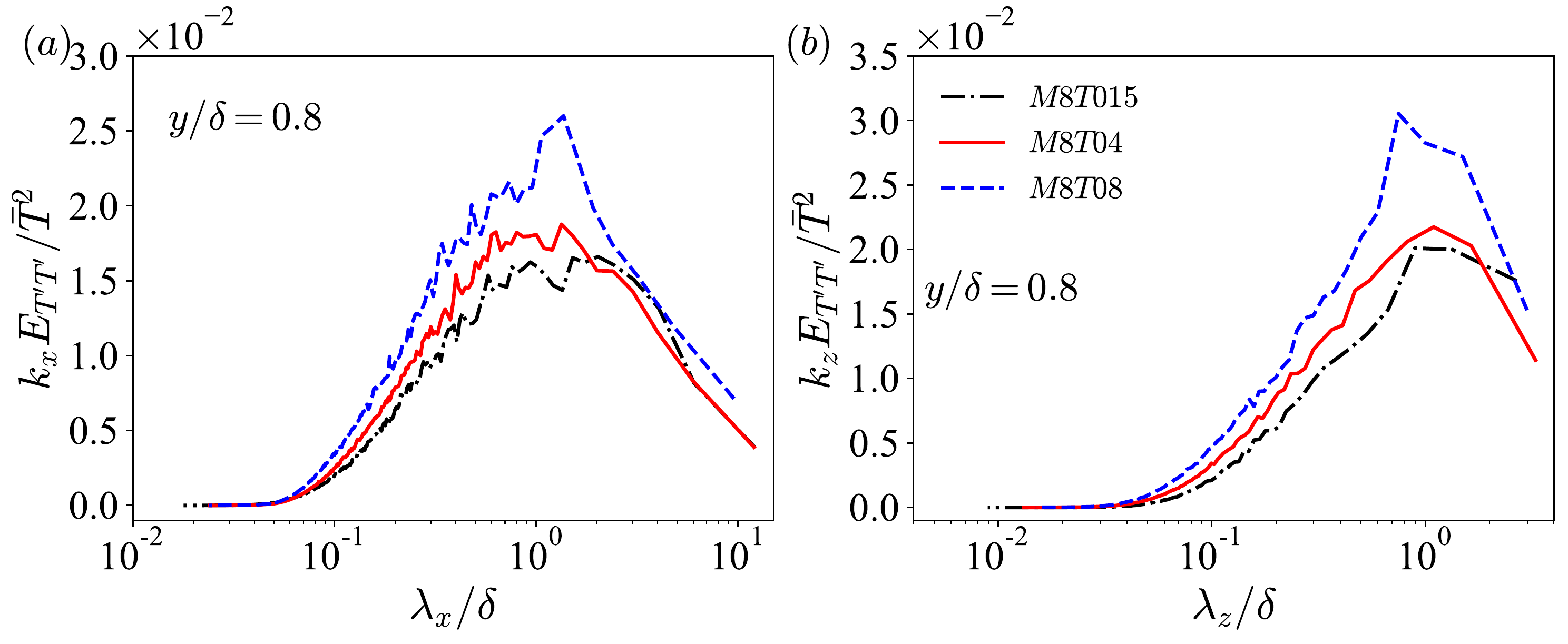}
    \caption{(a) The normalised premultiplied streamwise spectra of the fluctuating temperature $k_{x}E_{ T^{\prime }T^{\prime }}/\bar{T}^{2}$ at $y/\delta=0.8$ plotted against ${\lambda}_{x}/\delta $. (b) The normalised premultiplied spanwise spectra of the fluctuating temperature $k_{z}E_{ T^{\prime }T^{\prime }}/\bar{T}^{2}$ at $y/\delta=0.8$ plotted against ${\lambda}_{z}/\delta $.}
    \label{fig: d33}
\end{figure}

\begin{figure}\centering
    \includegraphics[width=0.99\linewidth]{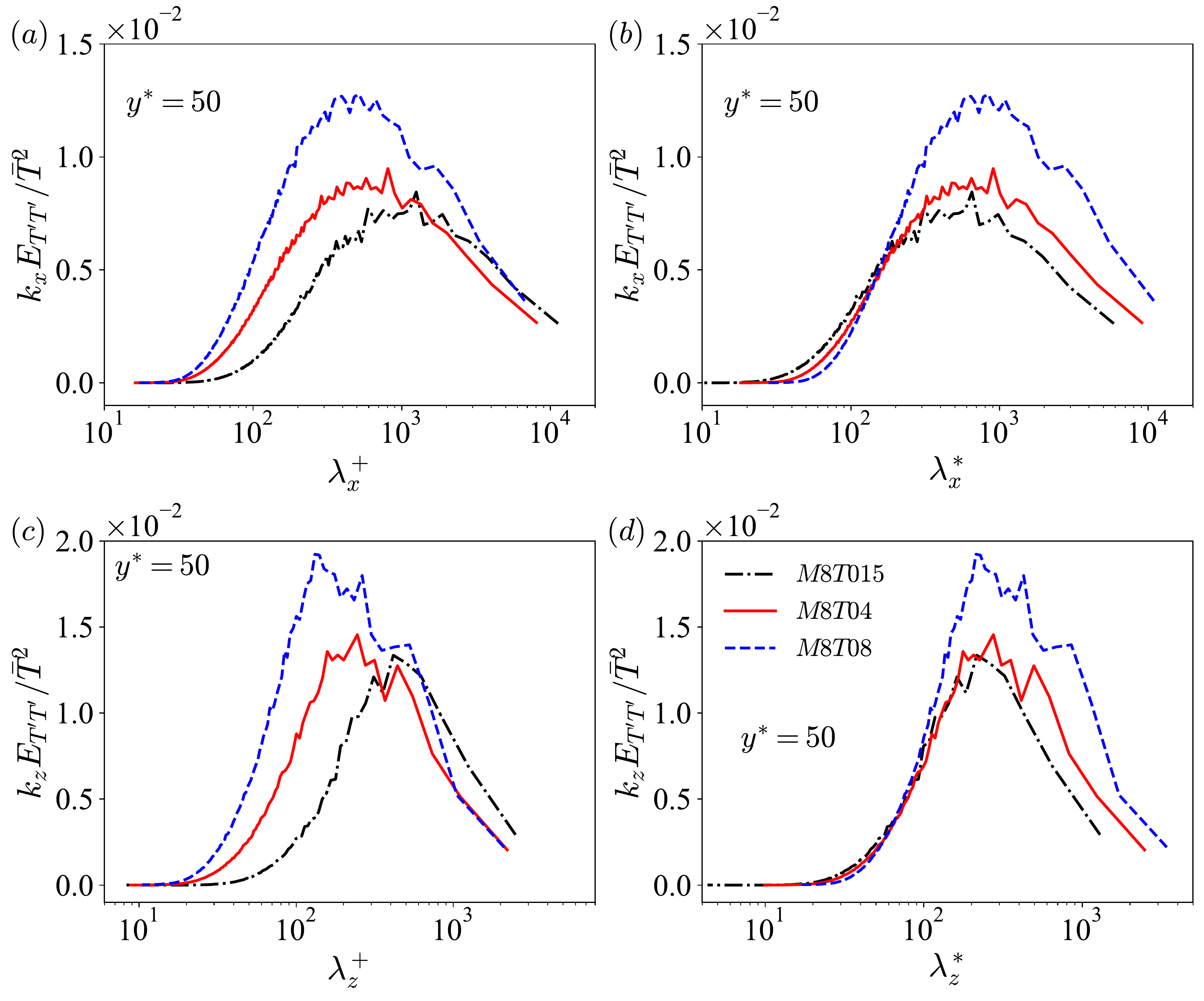}
    \caption{(a) and (b): The normalised premultiplied streamwise spectra of the fluctuating temperature $k_{x}E_{ T^{\prime }T^{\prime }}/\bar{T}^{2}$  at $y^{*}=50$ plotted against (a) ${\lambda}_{x}^{+}$ and (b) ${\lambda}_{x}^{*}$. (c) and (d): The normalised premultiplied spanwise spectra of the fluctuating temperature $k_{z}E_{ T^{\prime }T^{\prime }}/\bar{T}^{2}$ at $y^{*}=50$ plotted against (c) ${\lambda}_{z}^{+}$ and (d) ${\lambda}_{z}^{*}$.}
    \label{fig: d34}
\end{figure}

The normalised premultiplied streamwise and spanwise spectra of the fluctuating temperature $k_{x}E_{ T^{\prime }T^{\prime }}/\bar{T}^{2}$ and $k_{z}E_{ T^{\prime }T^{\prime }}/\bar{T}^{2}$ at $y/\delta=0.8$ and $y^{*}=50$ are shown in figure \ref{fig: d33} and figure \ref{fig: d34} respectively. {The temperature spectra attain their peaks at ${\lambda}_{x}/\delta \approx 1.3$ and ${\lambda}_{z}/\delta \approx 1$ at $y/\delta=0.8$, which are similar to the observation in \citet[]{Cogo2022}. Furthermore, the peak locations of the temperature spectra are ${\lambda}_{x}^{*} \approx 700$ and ${\lambda}_{z}^{*} \approx 250$ respectively at $y^{*}=50$. It is noted that the peak locations of the temperature spectra far from the wall are similar to the behaviours of the entropy spectra, suggesting the dominance of the entropic mode in fluctuating temperature at $y^{*}>20$ (figure \ref{fig: d12} (b)).}

\begin{figure}\centering
    \includegraphics[width=0.99\linewidth]{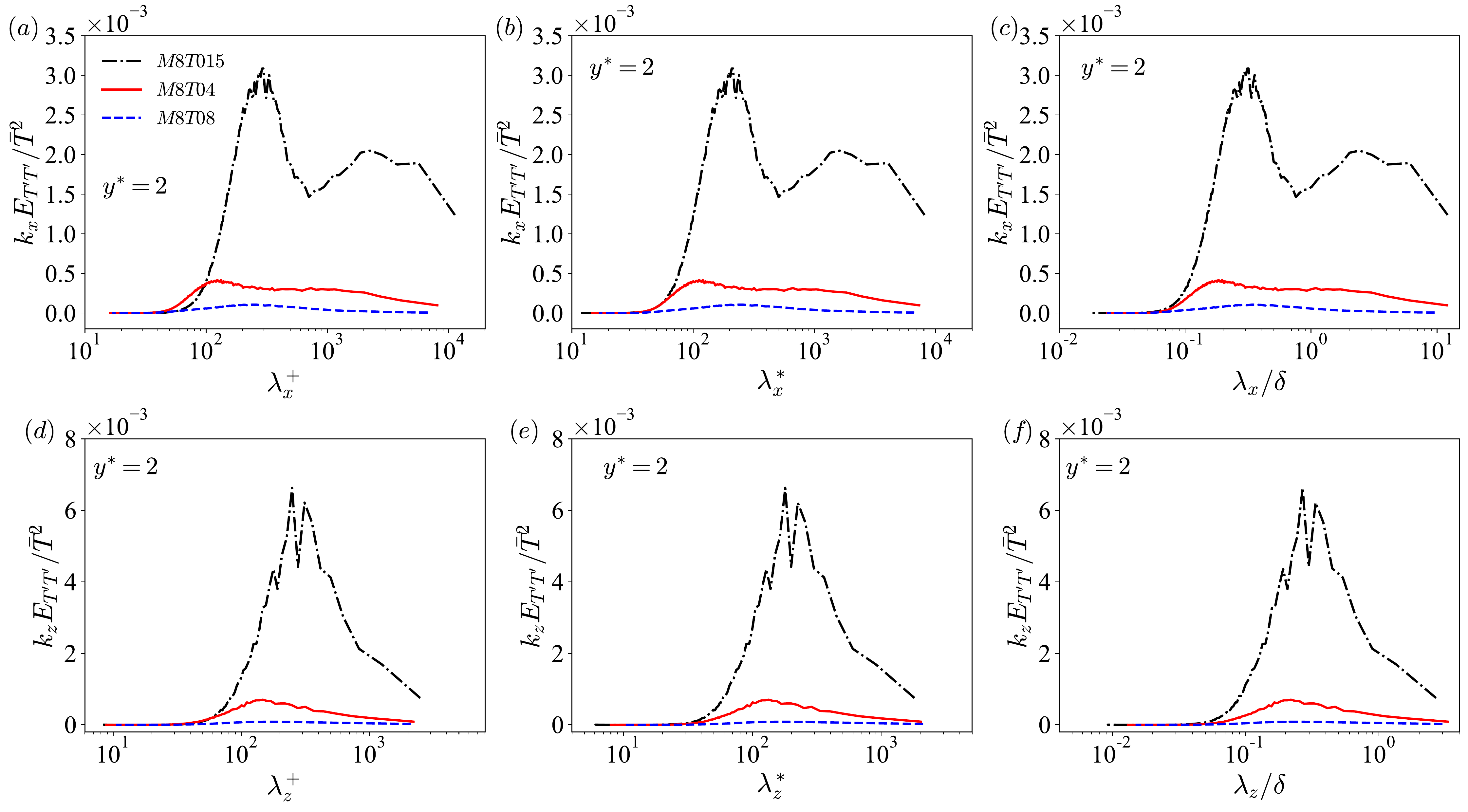}
    \caption{(a), (b) and (c): The normalised premultiplied streamwise spectra of the fluctuating temperature $k_{x}E_{ T^{\prime }T^{\prime }}/\bar{T}^{2}$  at $y^{*}=2$ plotted against (a) ${\lambda}_{x}^{+}$, (b) ${\lambda}_{x}^{*}$ and (c) ${\lambda}_{x}/\delta $. (d), (e) and (f): The normalised premultiplied spanwise spectra of the fluctuating temperature $k_{z}E_{ T^{\prime }T^{\prime }}/\bar{T}^{2}$ at $y^{*}=2$ plotted against (d) ${\lambda}_{z}^{+}$, (e) ${\lambda}_{z}^{*}$ and (f) ${\lambda}_{z}/\delta $.}
    \label{fig: d35}
\end{figure}

\begin{figure}\centering
    \includegraphics[width=0.99\linewidth]{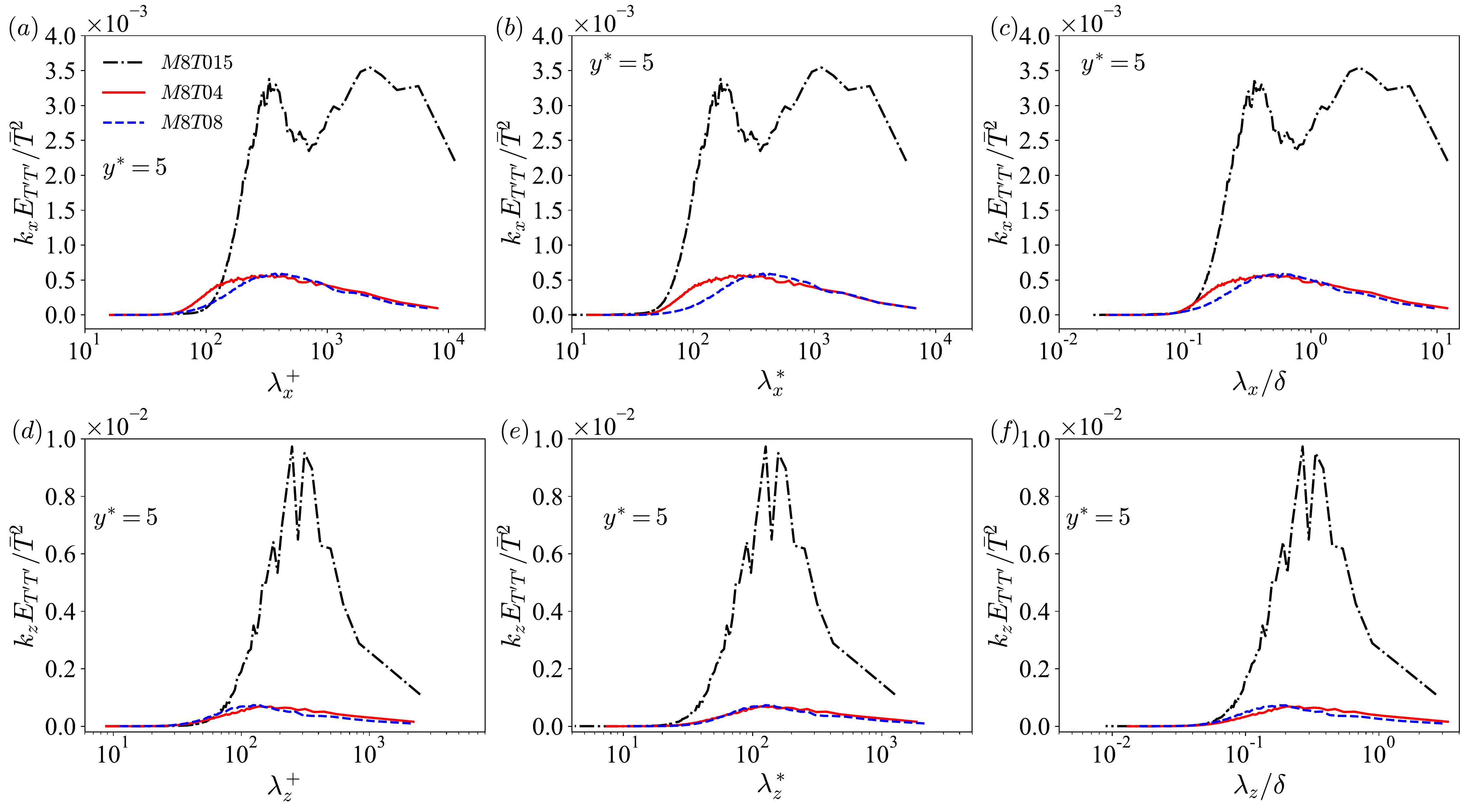}
    \caption{(a), (b) and (c): The normalised premultiplied streamwise spectra of the fluctuating temperature $k_{x}E_{ T^{\prime }T^{\prime }}/\bar{T}^{2}$  at $y^{*}=5$ plotted against (a) ${\lambda}_{x}^{+}$, (b) ${\lambda}_{x}^{*}$ and (c) ${\lambda}_{x}/\delta $. (d), (e) and (f): The normalised premultiplied spanwise spectra of the fluctuating temperature $k_{z}E_{ T^{\prime }T^{\prime }}/\bar{T}^{2}$ at $y^{*}=5$  plotted against (d) ${\lambda}_{z}^{+}$, (e) ${\lambda}_{z}^{*}$ and (f) ${\lambda}_{z}/\delta $.}
    \label{fig: d36}
\end{figure}

The normalised premultiplied streamwise and spanwise spectra of the fluctuating temperature $k_{x}E_{ T^{\prime }T^{\prime }}/\bar{T}^{2}$ and $k_{z}E_{ T^{\prime }T^{\prime }}/\bar{T}^{2}$ at $y^{*}=2$ and $y^{*}=5$ are shown in figure \ref{fig: d35} and figure \ref{fig: d36} respectively. {The values of the temperature spectra at $y^{*}=2$ in ``M8T08'' are pretty small, which is consistent with the small values of $T _{rms}^{\prime}/\bar{T}$ at $y^{*}=2$ in nearly adiabatic wall case (figure \ref{fig: d6} (c)). The primary peak locations of $k_{x}E_{ T^{\prime }T^{\prime }}/\bar{T}^{2}$ in ``M8T04'' and ``M8T015'' are ${\lambda}_{x}^{*} \approx 120$ and 210 at $y^{*}=2$ respectively, which are similar to the peak locations of $k_{x}E_{ p^{\prime }p^{\prime }}/\bar{p}^{2}$ at $y^{*}=2$ (figure \ref{fig: d19} (b)). Moreover, the $k_{x}E_{ T^{\prime }T^{\prime }}/\bar{T}^{2}$ in ``M8T04'' and ``M8T015'' attain their secondary peaks at ${\lambda}_{x}/\delta \approx 1.3$ and 2.3 respectively, which are also coincident with the peak locations of $k_{x}E_{ s^{\prime }s^{\prime }}\left ( \gamma M^{2} \right )^{2}$ at $y^{*}=2$ (figure \ref{fig: d24} (c)). These observations indicate that both the acoustic and entropic modes have significant contributions to the fluctuating temperature at $y^{*}=2$ when the wall is cooled.}

{At $y^{*}=5$, it is found in figure \ref{fig: d36} that the temperature spectra in ``M8T08'' attain their peaks at ${\lambda}_{x}^{*} \approx 450$ and ${\lambda}_{z}^{*} \approx 140$ respectively, which are consistent with the peak locations of the entropy spectra at $y^{*}=5$ in ``M8T08''. This observation indicates that the fluctuating temperature is dominated by its entropic mode at $y^{*}=5$ in the nearly adiabatic wall case. In ``M8T04'', the $k_{x}E_{ T^{\prime }T^{\prime }}/\bar{T}^{2}$ has a relative wide peak at $y^{*}=5$, indicating that both the acoustic and entropic modes have strong contributions to the fluctuating temperature. This observation is consistent with the fact that $T _{E,rms}^{\prime}/\left (T _{E,rms}^{\prime}+T _{I,rms}^{\prime}  \right ) \approx 0.53$ at $y^{*}=5$ in ``M8T04'' (figure \ref{fig: d12} (b)). However, the behaviours of the temperature spectra in ``M8T015'' are quite different. To be specific, the $k_{x}E_{ T^{\prime }T^{\prime }}/\bar{T}^{2}$ in ``M8T015'' attains its primary peak at ${\lambda}_{x}/\delta \approx 2.3$, and this primary peak location represents the characteristic length scale of the SES. Furthermore, the secondary peak location of the streamwise temperature spectra is similar to that of the streamwise pressure spectra, indicating that the TAPNS also appear in the fluctuating temperature when the wall is strongly cooled. Similar to the fluctuating density, it is shown in figure \ref{fig: d12} (b) that the relative contribution $T _{E,rms}^{\prime}/\left (T _{E,rms}^{\prime}+T _{I,rms}^{\prime}  \right )$ in ``M8T015'' also has a hump marked by the green dashed box. This hump is mainly attributed to the strong intensity of SES near the wall in strongly cooled wall case ``M8T015''.}

\section{Discussions}\label{sec: n4}
According to the above numerical results, some discussions are made in this section.

\subsection{Discussion about the relative contributions of the acoustic and entropic modes of density and temperature}
As the wall temperature decreases, the intensities of the pressure and the acoustic modes of density and temperature significantly increase. Specifically, when the wall is strongly cooled, the TAPNS appear in the pressure and the acoustic modes of density and temperature at the wall and in the vicinity of the wall.

When the wall is strongly cooled, the SES appear in the entropy and the entropic modes of density and temperature near the wall. Different from the TAPNS locating at the wall and in the vicinity of the wall, the SES are relatively weak at the wall, and have the largest intensities slightly away from the wall (such as $y^{*}\approx 5$ in ``M8T015''). It is shown above that the intensities of the entropy and the entropic modes of density and temperature near the wall are enhanced as the wall temperature decreases, and this enhancement is mainly caused by the SES. However, the intensities of the entropy and the entropic modes of density and temperature decrease as the wall temperature decreases in the far-wall region.

As shown in figure \ref{fig: d12}, the entropic modes of density and temperature are dominant far from the wall ($y^{*}>20$).  It is found above that the intensities of the pressure and the acoustic modes of density and temperature attain the peak values near the wall and then monotonically decrease away from the wall, while the intensities of the entropy and the entropic modes of density and temperature achieve their {primary} peaks at the edge of the boundary layer. These observations further lead to the result that the relative contributions of the entropic modes of density and temperature increase as $y^{*}$ increases among the boundary layer. Moreover, the relative contributions of the entropic modes of density and temperature at $y^{*}>20$ decrease as the wall temperature decreases, which can be ascribed to the enhancement of the pressure and the acoustic modes of density and temperature, as well as the decrement of the entropy and the entropic modes of density and temperature as the wall temperature decreases.

However, the variations of the relative contributions of the entropic modes of density and temperature are rather complicated at $y^{*}<20$. In the nearly adiabatic wall case ``M8T08'', {the intensities of the acoustic modes of density and temperature are much larger than those of the entropic modes near the wall. Therefore, the acoustic modes of density and temperature are dominant in the near-wall region. When the wall is cooled, especially in ``M8T015'', both the TAPNS and the SES appear near the wall. The intensities of the TAPNS are weaker than those of the SES at the wall and in the vicinity of the wall. Accordingly, the relative contributions of the entropic modes of density and temperature become larger in this region as the wall temperature decreases. As $y^{*}$ further increases, the intensities of the TAPNS decrease while those of the SES increase, leading to the enhancement of the relative contributions of the entropic modes of density and temperature.} At $y^{*} \approx 5$ in ``M8T015'', the SES have the strongest intensities, resulting in the local maximum values of the relative contributions of the entropic modes of density and temperature. As $y^{*}$ further increases, the intensities of the SES decrease, giving rise to the decrease of the relative contributions of the entropic modes of density and temperature in ``M8T015''. {Therefore, the humps marked by the green dashed boxes in figure \ref{fig: d12} are mainly attributed to the strong intensities of the SES in ``M8T015''.}

\subsection{Discussion about the generating mechanism of the streaky entropic structures (SES)}
The generating mechanism of the streaky entropic structures (SES) appeared near the wall in the cooled wall cases is illustrated in this subsection.

It has been shown that the vortices lead to the streaky structures of the fluctuating streamwise velocity by advecting the mean velocity gradient \citep[]{Blackwelder1979,Jimenez1999}. Therefore, similar to the analysis of the fluctuating streamwise velocity, the quadrant analysis \citep[]{Wallace2016} is introduced to investigate the generating mechanism of the streaky entropic structures (SES).

Based on the quadrant analysis, four quadrants are created by the fluctuating temperature and the wall-normal fluctuating velocity, and the instantaneous turbulent heat flux ${T}^{\prime}{v}^{\prime}$ located in these four quadrants are called four events \citep[]{Wallace2016}, that is, (1) $Q_{1}$ : ${T}'>0,\,{v}'>0$; (2) $Q_{2}$ : ${T}'<0,\,{v}'>0$; (3) $Q_{3}$ : ${T}'<0,\,{v}'<0$; (4) $Q_{4}$ : ${T}'>0,\,{v}'<0$. Similarly, the instantaneous turbulent entropy flux ${s}^{\prime}{v}^{\prime}$ can be divided into four events: (1) $Q_{1}$ : ${s}'>0,\,{v}'>0$; (2) $Q_{2}$ : ${s}'<0,\,{v}'>0$; (3) $Q_{3}$ : ${s}'<0,\,{v}'<0$; (4) $Q_{4}$ : ${s}'>0,\,{v}'<0$. Q2 and Q4 events represent the ejection and sweep events, which are gradient-type motions; while Q1 and Q3 events denote the outward and inward interactions, which are countergradient-type motions \citep[]{Wallace2016}. Q2 event describes the motion that the near-wall low-temperature or low-entropy streaks rise up to the far-wall fluid, while Q4 event implies that the high-temperature or high-entropy streaks in the outer layer sweep down to the near-wall fluid.

\begin{figure}\centering
    \includegraphics[width=0.99\linewidth]{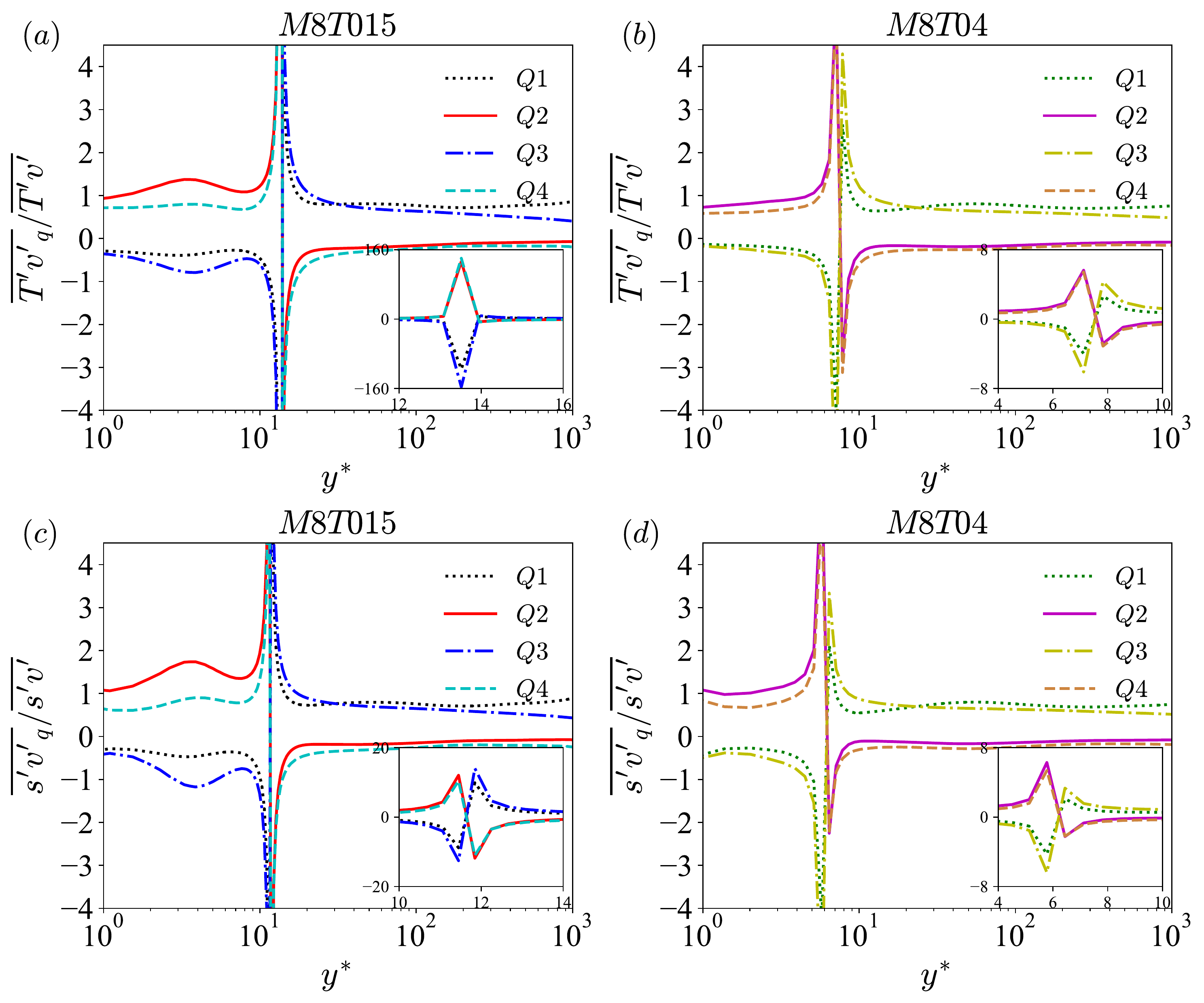}
    \caption{(a) and (b): Quadrant contributions to the turbulent heat flux $\overline{{T}^{\prime}{v}^{\prime}}$ along wall-normal direction in (a) ``M8T015'' and (b) ``M8T04''. (c) and (d): Quadrant contributions to the turbulent entropy flux $\overline{{s}^{\prime}{v}^{\prime}}$ along wall-normal direction in (c) ``M8T015'' and (d) ``M8T04''. The insets shown the sharp peaks of the quadrant contributions.}
    \label{fig: d37}
\end{figure}

Quadrant contributions to the turbulent heat flux $\overline{{T}^{\prime}{v}^{\prime}}$ and the turbulent entropy flux $\overline{{s}^{\prime}{v}^{\prime}}$ are shown in figure \ref{fig: d37}. It is shown that in the near-wall region where the wall-normal gradient of the mean temperature $\partial \bar{T}/\partial y$ is positive (figure \ref{fig: d2} (e)), the Q2 (ejection) and Q4 (sweep) events give strong positive contributions to $\overline{{T}^{\prime}{v}^{\prime}}$ and $\overline{{s}^{\prime}{v}^{\prime}}$, while the Q1 (outward) and Q3 (inward) events give small negative contributions to $\overline{{T}^{\prime}{v}^{\prime}}$ and $\overline{{s}^{\prime}{v}^{\prime}}$. As the wall-normal location $y^{*}$ increases, the sharp peaks appear at the wall-normal location of the  ``turning points'' where the wall-normal gradient of the mean temperature $\partial \bar{T}/\partial y$ is zero (shown by the green circles in figure \ref{fig: d2} (e)). As the wall-normal location $y^{*}$ further increases, the wall-normal gradient of the mean temperature $\partial \bar{T}/\partial y$ changes from positive to negative (figure \ref{fig: d2} (e)). Consequently, the Q1 (outward) and Q3 (inward) events give strong positive contributions to $\overline{{T}^{\prime}{v}^{\prime}}$ and $\overline{{s}^{\prime}{v}^{\prime}}$, while the Q2 (ejection) and Q4 (sweep) events give small negative contributions to $\overline{{T}^{\prime}{v}^{\prime}}$ and $\overline{{s}^{\prime}{v}^{\prime}}$. Therefore, it can be concluded that the streaky entropic structures (SES) appeared near the wall in the cooled wall cases are mainly caused by the advection effect of the strong positive wall-normal gradient of the mean temperature. When $\partial \bar{T}/\partial y >0$, the ejection and sweep events give dominant positive contributions, and lead to the streaky entropic structures. As the wall temperature decreases, the positive values of $\partial \bar{T}/\partial y $ become larger, and the ejection and sweep events become stronger, which further lead to the stronger streaky entropic structures. Furthermore, the wall-normal range of the $\partial \bar{T}/\partial y >0$ region also increases as the wall temperature decreases (figure \ref{fig: d2} (e)), which further leads to the larger wall-normal region where the SES exist in colder wall case.

\section{Summary and conclusion}\label{sec: n5}
In this paper, the wall cooling effect on the spectra and structures of the thermodynamic variables is systematically investigated in hypersonic turbulent boundary layers by direct numerical simulations. The turbulent intensities and the streamwise and spanwise spectra of the fluctuating streamwise velocity and thermodynamic variables, including the density, temperature, pressure and entropy, are meticulously studied. It is found that the wall cooling effect has a significantly larger influence on the thermodynamic variables compared with the fluctuating streamwise velocity, which leads to a great challenge to the accurate modelling of the thermodynamic variables.

The fluctuating density and temperature can be divided into the acoustic and entropic modes based on the Kovasznay decomposition. The fluctuating pressure is positively linearly correlated with the acoustic modes of density and temperature with $R\left ( \rho _{I} ^{\prime},p^{\prime} \right )=1$ and $R\left ( \rho _{I} ^{\prime},T _{I}^{\prime} \right )=1$. {Furthermore, the entropic mode of density is almost negatively linearly correlated with the fluctuating entropy and the entropic mode of temperature with $R\left ( \rho _{E} ^{\prime},T _{E}^{\prime} \right ) \approx -1$ and $R\left ( \rho _{E} ^{\prime},s^{\prime} \right ) \approx -1$.}

    {It is found that the intensities of the fluctuating pressure and the acoustic modes of density and temperature are significantly enhanced as the wall temperature decreases, especially at the wall and in the vicinity of the wall. When the wall is cooled, the travelling-wave-like alternating positive and negative structures (TAPNS) appear at the wall and in the vicinity of the wall. These TAPNS give rise to the fact that the intensities of the pressure and the acoustic modes of density and temperature in cooled wall cases (i.e. ``M8T04'' and ``M8T015'') achieve their primary peaks at the wall.} The TAPNS are short and fat (i.e. ${\lambda}_{x}^{*}<{\lambda}_{z}^{*}$). As the wall temperature decreases, the intensities and the characteristic streamwise length and spanwise spacing scales of the TAPNS increase, and the wall-normal range where the TAPNS exist also increases.

    {It is also shown that the entropy and the entropic modes of density and temperature achieve their primary peaks near the edge of the boundary layer. Furthermore, as the wall temperature decreases, the intensities of the entropy and the entropic modes of density and temperature decrease far from the wall, while are significantly enhanced in the near-wall region. The enhancement of the intensities of the entropy and the entropic modes of density and temperature near the wall in the cooled wall cases can be attributed to the appearance of the streaky entropic structures (SES). Specifically, the interesting phenomena are observed in figure \ref{fig: d6} (a)-(c) that the $s _{rms}^{\prime}\gamma M^{2}$, $\rho _{rms}^{\prime}/\bar{\rho }$ and $T _{rms}^{\prime}/\bar{T}$ in ``M8T015'' have local secondary peaks at $y^{*}=5$, and these phenomena are mainly due to the strong intensities of the SES. The SES are long and thin (i.e. ${\lambda}_{x}/\delta \gg {\lambda}_{z}/\delta $). As the wall temperature decreases, the intensities and the characteristic streamwise length and spanwise spacing scales of the SES are enhanced, and the wall-normal range where the SES exist also increases.}

    {It is shown in figure \ref{fig: d6} (a)-(c) that the profiles of $s _{rms}^{\prime}\gamma M^{2}$, $\rho _{rms}^{\prime}/\bar{\rho }$ and $T _{rms}^{\prime}/\bar{T}$ are similar to each other, which are mainly due to the observation that the density and temperature are dominated by their entropic modes far from the wall ($y^{*}>20$).} Moreover, the relative contributions of the entropic modes become weaker as the wall temperature decreases. However, in the near-wall region $y^{*}<20$, the wall temperature has a significant effect on the relative contributions of the entropic modes of density and temperature. When the wall is nearly adiabatic (i.e. ``M8T08''), the acoustic modes of density and temperature are dominant in the vicinity of the wall. When the wall is cooled, the intensities of SES are larger than those of TAPNS, which further result in the enhancement of the relative contributions of the entropic modes of density and temperature near the wall.

Furthermore, the quadrant analysis shows that streaky entropic structures (SES) are mainly caused by the advection effect of the strong positive wall-normal gradient of the mean temperature associated with ejection and sweep events. As the wall temperature decreases, the magnitudes and the wall-normal range of the positive $\partial \bar{T}/\partial y $ significantly increase, which further lead to the stronger intensities and larger wall-normal range of SES.

    {In conclusion, the wall cooling effect on the multi-scale properties of the thermodynamic variables in hypersonic boundary layers is systematically investigated. Two special structures TAPNS and SES are revealed in the near-wall region when the wall is cooled, and should be specially considered in the accurate modelling of the thermodynamic variables.}

\acknowledgments{
    \noindent\textbf{Funding.} This work was supported by the NSFC Basic Science Center Program (Grant No. 11988102), by National Natural Science Foundation of China (NSFC Grants No. 91952104, 92052301, 12172161 and 91752201), by the Technology and Innovation Commission of Shenzhen Municipality (Grant Nos. KQTD20180411143441009 and JCYJ20170412151759222), and by Department of Science and Technology of Guangdong Province (Grant No. 2019B21203001). This work was also supported by Center for Computational Science and Engineering of Southern University of Science and Technology.

    \noindent\textbf{Declaration of Interests.} The authors report no conflict of interest.}

\section*{{Appendix A. Validation of the DNS cases}}
\setcounter{equation}{0}\setcounter{subsection}{0}
\renewcommand{\theequation}{A.\arabic{equation}}
\renewcommand{\thesubsection}{A.\arabic{subsection}}

{In this Appendix, the DNS cases in this study are validated via comparisons with the available DNS database in \citet[]{Zhang2018}.}

\begin{figure}\centering
    \includegraphics[width=0.99\linewidth]{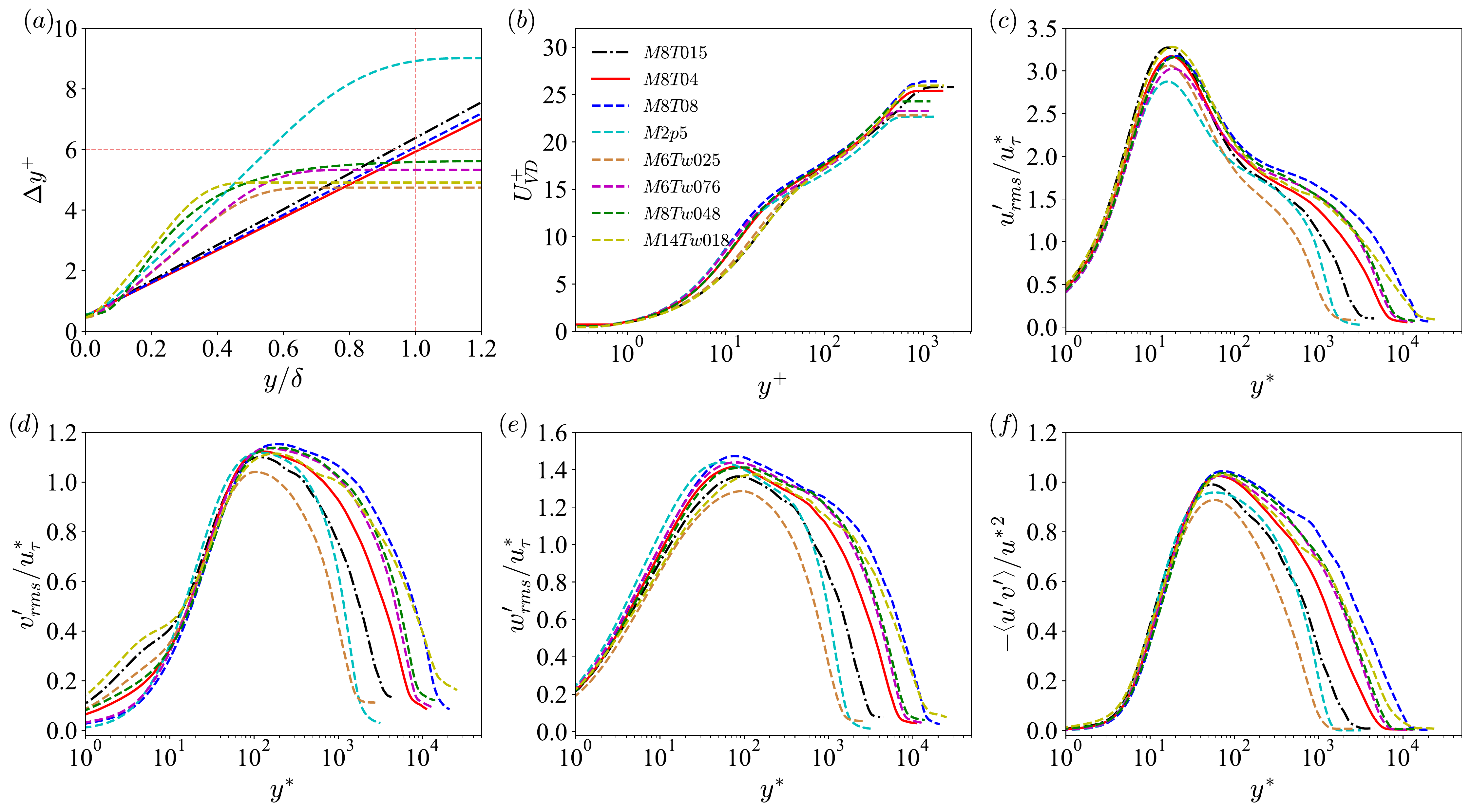}
    \caption{The comparisons between the DNS cases in this study and the DNS database in \citet[]{Zhang2018}: (a) the wall-normal grid spacing $\Delta y^{+}$, (b) the van Driest transformed mean velocity $U^{+}_{VD}$, (c) the intensities of the streamwise fluctuating velocity $u _{rms}^{\prime}/u_{\tau}^{*}$, (d) the intensities of the wall-normal fluctuating velocity $v _{rms}^{\prime}/u_{\tau}^{*}$, (e) the intensities of the spanwise fluctuating velocity $w _{rms}^{\prime}/u_{\tau}^{*}$, (f) the Reynolds shear stress $-\left \langle u^{\prime}v^{\prime}\right \rangle/{u^{*}}^{2}$.}
    \label{fig: app1}
\end{figure}

{It is noted that the van Driest transformed mean velocity $U^{+}_{VD}$ is defined as \citep[]{VanDriest1951}}
\begin{equation}
    {U^{+}_{VD}=\int_{0}^{U^{+}}\left ( \bar{\rho }/\bar{\rho }_{w} \right )^{1/2}dU^{+},}
\end{equation}
{where $U^{+}=U/u_{\tau}$ and $U$ is the mean streamwise velocity. The comparisons of the wall-normal grid spacing $\Delta y^{+}$, the van Driest transformed mean velocity $U^{+}_{VD}$, the intensities of the streamwise, wall-normal, spanwise fluctuating velocities and the Reynolds shear stress between the DNS cases in this study and the DNS database in \citet[]{Zhang2018} are shown in figure \ref{fig: app1}.}

{It is found in figure \ref{fig: app1} (a) that the wall-normal grid spacing $\Delta y^{+}$ in the DNS cases are smaller than those of the DNS database in \citet[]{Zhang2018} at $y/\delta < 0.8$, and $\Delta y^{+} \approx 6$ at $y/\delta =1.0$, indicating that the wall-normal resolutions of the DNS cases in this study are fine enough. Furthermore, it is shown in figure \ref{fig: app1} (b)-(f) that the profiles of $U^{+}_{VD}$, $u _{rms}^{\prime}/u_{\tau}^{*}$, $v _{rms}^{\prime}/u_{\tau}^{*}$, $w _{rms}^{\prime}/u_{\tau}^{*}$ and $-\left \langle u^{\prime}v^{\prime}\right \rangle/{u^{*}}^{2}$ in ``M8T04'' are similar to those in ``M8Tw048'' in \citet[]{Zhang2018}. When the wall is strongly cooled, the van Driest transformed mean velocity, the intensities of the fluctuating velocites and the Reynolds shear stress in ``M8T015'' have similar behaviours with those in ``M14Tw018'' in \citet[]{Zhang2018}. These observations validate the accuracy of the DNS cases in this study.}

\bibliographystyle{jfm}
\bibliography{jfm-instructions}

\end{document}